%% file: main.tex
\documentclass[preprint2]{aastex}
\usepackage{subfigure,lscape,amsmath}
\usepackage[]{geometry}
\input{def}
\shorttitle{Swift Quasar Catalog}
\shortauthors{Wu J. et al.}
\begin{document}

\title{A Quasar Catalog with Simultaneous UV, Optical and X-ray Observations by \emph{Swift}}

\author{Jian Wu\altaffilmark{1,4}, Daniel Vanden Berk\altaffilmark{2}, Dirk Grupe\altaffilmark{1}, Scott Koch\altaffilmark{1}, Jonathan Gelbord\altaffilmark{1}, Donald P. Schneider\altaffilmark{1,3}, Caryl Gronwall\altaffilmark{1,3}, Sarah Wesolowski\altaffilmark{2} and Blair L. Porterfield\altaffilmark{1}}

\email{jwu@astro.psu.edu}

\altaffiltext{1}{Department of Astronomy \& Astrophysics, Pennsylvania State Univesity, 525 Davey Lab, University Park, PA, 16802, USA}
\altaffiltext{2}{Department of Physics, Saint Vincent College, 300 Fraser Purchase Road, Latrobe, PA, 15650}
\altaffiltext{3}{Institute for Gravitation and the Cosmos, The Pennsylvania State University, University Park, PA 16802}
\altaffiltext{4}{College of Information Sciences and Technology, The Pennsylvania State University, University Park, PA 16802}

\begin{abstract}
We have compiled a catalog of optically-selected quasars with simultaneous observations in UV/optical and X-ray bands by the \emph{Swift} Gamma Ray Burst Explorer. Objects in this catalog are identified by matching the \emph{Swift} pointings with the Sloan Digital Sky Survey Data Release 5 quasar catalog. The final catalog contains 843 objects, among which 637  have both UVOT and XRT observations and 354  of which are detected by both instruments. The overall X-ray detection rate  is $\sim60\%$ which rises to $\sim85\%$ among sources with at least 10~ks of XRT exposure time. We construct the time-averaged spectral energy  distribution for each of the 354 quasars using UVOT photometric measurements and XRT spectra. From model fits to these SEDs, we find that the big blue bump contributes about $\sim0.3$~dex to the quasar luminosity. We re-visit the \aox-\luv\ relation  by selecting a clean sample with only type 1 radio-quiet quasars; the  dispersion of this relation is reduced by at least 15\% compared to studies that use non-simultaneous UV/optical and X-ray data. We only found a weak correlation between \lle\ and \auv. We do not find significant correlations between \ax\ and \aox,  \aox\ and \auv,  and \ax\ and \loglxint. The correlations between \auv\ and \ax,  \aox\ and \ax, \aox\ and \auv, \lle\ and \ax, and \lle\ and \aox\ are stronger amongst low-redshift quasars, indicating that these correlations are  likely driven by the changes of SED shape with accretion state.

\end{abstract}


\input{sec1}

\input{sec2}

\input{sec3}

\input{sec4}

\input{sec5}

\input{sec6}

\acknowledgments

We received much assistance for this project from people at the
\swift\ Mission Operation Center. Jamie Kennea helped us
understand the XRT exposure maps; Pete Roming offered comments
on the UVOT photometric calculations. We also received some 
helpful comments and discussions from Kim Page and Phil Evans. 
We also thank Bin Luo, Patrick Broos, Richard Wade, and Cristian Saez for
their suggestions on this project. We also thank the referee for his/her
useful comments which improves the quality of this paper. This project is 
financially supported by NSF grant AST06-07634 and NASA ADAP 
grant NNX09AC87G.  

Funding for the SDSS and SDSS-II has been provided by the Alfred P. Sloan
Foundation, the Participating Institutions, the National Science Foundation,
the U.S. Department of Energy, the National Aeronautics and Space
Administration, the Japanese Monbukagakusho, the Max Planck Society, and the
Higher Education Funding Council for England. The SDSS website is
\url{http://www.sdss.org/}.

The Institute for Gravitation and the Cosmos is supported by the 
Eberly College of Science and the Office of the Senior Vice President
for Research at the Pennsylvania State University.

Swift is supported at PSU by NASA contract NAS5-00136.
\include{app}

\bibliographystyle{apj}
\bibliography{ref}

\clearpage

\include{fig}
\clearpage

\include{table}
\clearpage
\end{document}

%% file: def.tex
\newcommand{\civfull}{C~{\sc iv}~$\lambda$1549}

\newcommand{\civ}{C~{\sc iv}}

\newcommand{\ewciv}{EW(C~{\sc iv})}

\newcommand{\feii}{Fe~{\sc ii}}
\newcommand{\luv}{$L_{\rm 2500~\AA}$} 
\newcommand{\logluv}{$\log{L_{\mbox{2500~\AA}}}$}
\newcommand{\logluvph}{$\log{L_{\mbox{2500~\AA}}}$}
\newcommand{\elogluvph}{$\delta\log{L_{\mbox{2500~\AA}}}$}
\newcommand{\lx}{$L_{\mbox{2~keV}}$}
\newcommand{\loglx}{$\log{L_{\mbox{2~keV}}}$}
\newcommand{\eloglx}{$\delta\log{L_{\mbox{2~keV}}}$}
\newcommand{\lradio}{$L_{\mbox{5~GHz}}$}
\newcommand{\aox}{$\alpha_{\rm ox}$}
\newcommand{\aoxph}{$\alpha_{\rm ox,ph}$}
\newcommand{\eaoxph}{$\delta\alpha_{\rm ox,ph}$}

\newcommand{\luvunit}{erg~s$^{-1}$~Hz$^{-1}$}
\newcommand{\fwunit}{erg~s$^{-2}$~cm$^{-2}$~\AA$^{-1}$}

\newcommand{\mgiifull}{Mg~{\sc ii}~$\lambda2798$}
\newcommand{\mgii}{Mg~{\sc ii}}
\newcommand{\ha}{H$\alpha$}
\newcommand{\hb}{H$\beta$}

\newcommand{\hi}{H~{\sc i}}

\newcommand{\lambdauvo}{$5600$~\AA}

\newcommand{\ciiiuv}{C~{\sc iii}]~$\lambda1908$}

\newcommand{\oiii}{[O~{\sc iii}]}

\newcommand{\oiiidoublet}{[O~{\sc iii}]~$\lambda\lambda4960,5008$}

\newcommand{\lya}{Ly$\alpha$}

\newcommand{\ciiiuvfull}{C{\,\sc iii}~$\lambda1908$}

\newcommand{\kms}{km~s$^{-1}$}

\newcommand{\hbeta}{H$\beta$}
\newcommand{\halpha}{H$\alpha$}
\newcommand{\ergs}{erg~s$^{-1}$}

\newcommand{\hgamma}{H~$\gamma$}
\newcommand{\xgamma}{$\Gamma$}
\def\figurenum#1{\def\thefigure{#1}\let\@currentlabel\thefigure
    \addtocounter{figure}{\count22}}
\newcommand{\loglbolexp}{$\log{L_{\rm bol,EXP}}$} 
\newcommand{\loglboltpl}{$\log{L_{\rm bol,TPL}}$} 
\newcommand{\lle}{$L_{\rm bol}/L_{\rm Edd}$}
\newcommand{\lbol}{$L_{\rm bol}$}

\newcommand{\mbh}{$M_{\rm BH}$}

\newcommand{\swift}{\emph{Swift}}
\newcommand{\auv}{$\alpha_{\rm UV}$}
\newcommand{\auvph}{$\alpha_{\rm UV,ph}$} 
\newcommand{\eauvph}{$\delta\alpha_{\rm UV,ph}$} 
\newcommand{\ax}{$\alpha_{\rm x}$}

\newcommand{\onesig}{1$\sigma$}
\newcommand{\twosig}{2$\sigma$}
\newcommand{\threesig}{3$\sigma$}
\newcommand{\psqcm}{cm$^{-2}$} 
\newcommand{\logfxabs}{$\log{F_{\rm obs}(\mbox{0.3--10~keV)}}$}
\newcommand{\logfxunabs}{$\log{F_{\rm unobs}(\mbox{0.3--10~keV)}}$}
\newcommand{\fx}{$F(\mbox{0.3--10~keV})$}
\newcommand{\ftwokev}{$f(\mbox{2~keV})$}
\newcommand{\loglxint}{$\log{L(\mbox{0.3--10~keV})}$}
\newcommand{\cps}{cnt~s$^{-1}$}
\newcommand{\nhg}{$N_{\rm H,G}$}
\newcommand{\nhi}{$N_{\rm H,i}$}
\newcommand{\nhim}{$\delta_{-}(N_{\rm H,i})$} 
\newcommand{\nhip}{$\delta_{+}(N_{\rm H,i})$} 

\newcommand{\nxph}{$N_{\rm Xph}$}
\newcommand{\modela}{Model~A}
\newcommand{\modelb}{Model~B}
\newcommand{\modelc}{Model~C}

\newcommand{\tbb}{$T_{\rm B}$}
\newcommand{\auvx}{$\alpha_{\rm UVX}$}
\newcommand{\aoxluv}{\aox--\luv}
\newcommand{\aoxluvbf}{$\mathbf{\alpha_{\rm ox}}$--$\mathbf{L_{2500~\mbox{\AA}}}$}
\newcommand{\xspec}{\emph{XSPEC}}

%% file: sec1.tex
\section{Introduction}
\label{swiftq-intro}
Variability is a ubiquitous phenomenon of quasars 
\citep[e.g.,][]{mat63,smi63,gas03,van04,wil05,wil06,meu11}
and has been observed in radio, infrared, UV/optical, 
X-ray and even $\gamma$-ray bands 
\citep[e.g.,][]{kov02,rie08,sak10,par10,gru10}. While
variability provides considerable information on the size
scales of the quasar central engine and can be utilized
to estimate the mass of the central super massive black hole 
\citep[e.g.,][]{pet04}, it is a significant source of scatter 
in multi-wavelength correlations. Clearly, when fluxes of variable sources 
measured in different wavebands at different times are combined, the combined 
spectral shape may not be representative of the spectral shape 
at a specific time. It is therefore essential to take simultaneous 
observations when determining multi-waveband properties 
in order to understand the true spectral shape.

The dispersion in multi-waveband correlations produced by variability
can be estimated. For example,
there is evidence that at least 60\%\ of the dispersion of the 
Baldwin Effect and at least 75\%\ of the \ewciv-\aox\ relation
can be attributed to variability \citep{jwu09}.
However, a large multi-waveband program of simultaneous observations
to verify that the dispersions of various relations can be 
reduced has not yet been performed. In this work, we will use 
simultaneously observed UV/optical and X-ray data to study quasar SEDs.

The \swift\ Gamma-ray Burst (GRB) Explorer \citep{geh04} includes co-aligned
X-ray and UV/optical detectors. The X-ray telescope (XRT; Burrows et al. 
2005\nocite{bur05}) is an imaging spectrometer that covers the 0.3--10~keV
band. The Ultraviolet Optical Telescope (UVOT; Roming et al. 2005\nocite{rom05}) 
provides photometry in six bands from $\sim1928$~\AA\ to $\sim5468$~\AA. By
default, XRT and UVOT are operated simultaneously. Because \swift\
is primarily a GRB mission, there are few quasars 
that were specifically targeted
for observations. However, due to the relatively large fields of view 
(FOVs) of both the UVOT and the XRT,
a large number of serendipitous sources, including quasars, are observed.
The Sloan Digital Sky Survey (SDSS; York et al. 2000\nocite{yor00})
Data Release 5 (DR5; Adelman-McCarthy et al. 2007\nocite{ade07}) quasar catalog
\citep{sch07} contains 77,429 optically-selected quasars in 5,740~deg$^2$; 
the UVOT FOV is $17^\prime\times17^\prime$, so there is a density of 
approximately one SDSS quasar per UVOT field. We have matched
the \swift\ pointings from launch to June 2008 ($\sim3.5$~years) with the SDSS
DR5 quasar catalog, to examine the properties of quasars observed 
simultaneously in UV and X-ray light.

Catalogs using similar strategies were constructed by 
\citet{tue08} and \citet{gru10}. 
Our work is unique in terms of the selection criteria and sample size. 
The sample in \citet{tue08} contains 153 hard 
X-ray (14--195~keV) selected local AGNs with a mean redshift of $\sim0.03$. 
Because hard X-ray photons more easily penetrate gas and dust than UV/optical
and soft X-ray photons, this study provides a more homogeneous sample of 
quasars than those based on UV/optical and soft X-ray bands. 
Using this sample, \citet{win09} estimated the fraction of ``Hidden" AGNs 
\citep{ued07} in the local universe to be $\sim24\%$. 
The sample by \citet{gru10} contains 92 soft X-ray selected 
AGNs with redshifts ranging from $0.002$ to $0.349$. Due to their selection
criteria, objects in this sample are X-ray bright
Type~1 AGNs with few Seyfert~1.5 objects 
(e.g., Mkn~841; see Wilkes et al. 1999\nocite{wil99}).
Their selection criteria are not biased for or against RL AGNs (about
$10\%$ of AGNs are RL). 
Using this sample, \citet{gru10} constructed composite AGN SEDs 
with simultaneous observations from \swift. By fitting these SEDs 
with two different models,
they attempted to constrain AGN bolometric corrections (BCs).
They found a significant correlation between UV and X-ray spectral indices, 
namely \auv\ and \ax\footnote{The spectral indices are
defined as $f_\nu\propto\nu^{+\alpha}$ in this work.} for 
AGNs with $\alpha_{\rm X}>-1.6$.
Because their sample contains a large fraction of narrow line Seyfert~1 
(NLS1) galaxies, they were also able to examine differences between broad line
Seyfert~1 (BLS1) galaxies and NLS1 galaxies in terms of the \auv--\lle\
relation. 

As a result of the detection limits in their selected wavebands, both
of the above studies are limited to nearby AGNs and relatively small
sample sizes. We find 1034 SDSS quasars within $20^\prime$ of \swift\
pointings, which is almost an order of magnitude larger than either of 
these samples. 

One of the prominent features in the AGN SED is the big blue bump
(BBB) in the extreme UV (EUV) energy band, which is believed to 
be primarily produced by thermal emission from an accretion disk 
\citep[e.g.,][]{shi78}. Because of strong
Galactic and (possibly) intrinsic extinction, it is almost impossible to
observe this wavelength region. 
This feature is likely, however, to be an important contributor
to quasar bolometric luminosity. Based on the observed 
data in UV/optical and X-ray bands, we can place some constraints on the 
flux contribution of this feature. \citet{gru10} were able to constrain the 
BBB using their sample of $\sim100$ AGNs at low redshift $z<0.4$.
In contrast, over 50\% of the quasars 
in our sample are at $z\gtrsim1$, thus, we will be able to better constrain
the BBB feature.

It should be emphasized that we are not repeating the 
global SED work of \citet{elv94} and \citet{ric06}. 
We are attempting to use simultaneously acquired data to constrain 
the BBB. This component, because of the lack of observational data, 
is usually represented as a power-law connecting a UV flux 
point, such as 2500~\AA, to an X-ray point, such as 2~keV 
\citep[e.g.,][]{ric06}. The slope determined by these two points 
is defined as \aox\ \citep{tan79},
\begin{equation}
\label{aoxdefinition}
\alpha_{\rm ox}=0.3838\log{\left[L_{2~\mbox{keV}}/L_{2500~\mbox{\AA}}\right]}
\end{equation}
which is used to characterize the spectral hardness between the UV and X-ray bands
\citep[e.g.,][]{avn82,avn86,and87,wil94,vig03,str05,ste06,jus07}. 
The expression above follows the notatioin convention in \citet{jus07},
so $L_{2500~\mbox{\AA}}$ is the monochromatic luminosity at 2500~\AA\ in
ergs~s$^{-1}$~Hz$^{-1}$.
A long-standing problem with the measurement of \aox\ is that it is
difficult to obtain simultaneous measurements of an object in the
X-ray and UV/optical bands. Source variability introduces scatter into 
measurements of \aox. Our simultaneous observations remove this 
noise.

This paper is organized as follows. In Section~\ref{swiftq-obsanddata}, 
we describe the observations and data processing; in 
Section~\ref{swiftq-results}, we present our data processing results, 
including the UV/optical light curves,
and the composite SED for each quasar; in Section~\ref{swiftq-catalog}, we 
present our final catalog; in Section~\ref{swiftq-analyses}, we select a 
sample of quasars from this catalog to study the dispersion of the 
\aoxluv\ relation and other correlations. 
In Section~\ref{swiftq-conclusion}, we summarize
this catalog and our conclusions. Throughout this work, we adopt the 
following cosmology: $\Omega_{\rm M}=0.3$, $\Omega_\Lambda=0.7$, $H_0=70$
km~s$^{-1}$~Mpc$^{-1}$.

%% file: sec2.tex
\section{Observations and Data Processing}
\label{swiftq-obsanddata}
Our quasar sample was compiled in the following steps.
\begin{enumerate}
  \item Candidate objects for our catalog were selected as any
        SDSS DR5 quasar that lie within $20^\prime$ of the center of 
        the \swift\ FOV in any pointing from launch through June 2008. 
  \item XRT data were processed to obtain X-ray count rates, 
        spectra and spectral parameters.
  \item UVOT data were processed to obtain UV and optical photometry. 
  \item UVOT photometry were supplemented with measurements at other
        wavelengths from published catalogs.
  \item Quasar SEDs were constructed.
  \item Additional parameters were calculated based on the SEDs of each quasar.
\end{enumerate}
The raw sample is constructed by matching 3.5~years \swift\ pointings and 
the SDSS DR5 quasar catalog and contains 1034 objects. Fig.~\ref{fig-miz}
shows the distribution of our quasar sample in the luminosity 
(represented by absolute magnitude $M_i$) -- redshift diagram. The 
distribution of our sample in this diagram is consistent with the 
SDSS DR5 quasar catalog. 
\subsection{XRT Data}
\label{processxrt}
We processed the XRT data using the task \emph{xrtpipeline} 
(HEADAS version 6.10) from \emph{FTOOLS} 
\citep{blackburn95}. For each observation identified by an 
observational ID (hereafter OBSID), this routine stacks all available
XRT snapshots (individual exposures with the same segment number) 
observed in photon counting mode and generates a composite 
\emph{sky image} for each OBSID with an associated exposure map. 

Source and background extraction regions are then defined for 
each quasar in the candidate list. The quasar coordinates are adopted from SDSS 
DR5 quasar catalog, which are accurate to $\approx1^{\prime\prime}$. 
The source and background regions are circles with default radii 
of $30^{\prime\prime}$ and $120^{\prime\prime}$, respectively. These circles 
are visually inspected in the deepest sky image. This ensures 
that the background does not include serendipitous X-ray sources.
The position and size of the background region circle is adjusted so that
it lies entirely within the field of view in all sky images
covering this object.
We exclude sky images in which the source region is
not fully within the FOV or not covered by 
all individual exposures. We also discard sky images in which the 
source is severely contaminated by nearby X-ray sources. 
Depending on the brightness of the X-ray source, the 
source region radius is adjusted to be large enough to ensure that the photon
density at the border is at the same level as the background. 
We then enlarge the background region so that its radius is at 
least four times as large as the source region radius.

XRT data are prepared using standard \emph{FTOOLS} packages. 
Event files for each OBSID are extracted and cleaned using 
\emph{XSelect} (version V2.4a). \emph{xrtmkarf} (version 0.5.6) 
is used to create an auxiliary response file (ARF) for each OBSID. 
The ARF file includes corrections for filter transmission,
vignetting, effective area and point spread function (PSF). 
It also accounts for hot pixels and hot columns, which are masked out 
from data and decrease the effective exposure time. The 
response matrix file (RMF) we use is
\emph{swxpc0to12s0\_20010101v011.rmf}\footnote{We noticed that this
RMF file was intended for use with data taken from launch through 
28 February 2007 only and that other RMFs are appropriate for
data taken from March through August 2007 and for data taken since
30 August 2007. To test the sensitivity of our simple spectral
models (described below) to the choice of response matrix, we fitted
models to several high signal to noise sources using each RMF and
found the impact upon the model parameters to be small: the spectral
index changes by less than 0.02 and the intrinsic column density 
 changes by no more than a few tens of percent. We therefore 
use a single RMF for simplicity and we choose the one for data 
taken prior to 28 February 2007 as the majority of our data is 
from this time interval and most of our sample members were 
observered at least in part before this date.}

The total number of background-subtracted X-ray counts, \nxph, 
is determined for each quasar by combining the counts measured
in each sky image. This value is used to assgin a quality
flag to each object: \emph{g} (good with $N_{\rm Xph}>100$), 
\emph{a} (acceptable with $10<N_{\rm Xph}\leq100$), 
\emph{w} (weak with $N_{\rm Xph}\leq10$), and \emph{o} (out of FOV). 
In the candidate sample of 1034 quasars, 103 objects are flagged as 
good (\emph{g}), 296 are flagged as acceptable (\emph{a}) and 406 are 
flagged as weak (\emph{w}). The rest (229 objects) are flagged as \emph{o} 
because they lack useful sky images. 

Next, we produce summed event lists for the source and background regions,
including all observations. From these we extract time-averaged spectra.
We also generate the ARF file for the composite source spectra
using \emph{addarf} (version 1.2.6).
The X-ray energy spectra are binned using \emph{FTOOLS} task 
\emph{grppha} (version 3.0.1). 

The X-ray energy spectra are fit using \xspec\ (Arnaud
1996\nocite{arn96}; version 12.5.1n). The binning strategy and 
the statistical method we use are described in Table~\ref{t-xrtbin}. 
Basically, we use $\chi^2$ statistics and have at least ten 
spectral bins if \nxph$>100$. If \nxph$<100$ we apply minimal
binning to eliminate unoccupied spectral bins (this avoids 
having bins with negative background-subtracted fluxes) and
use Cash statistics \citep{cas79}. If \nxph$<10$, we 
only estimate flux or flux upper limits. 
For each object with 
\nxph$\gtrsim10$, we fit each X-ray spectrum with an absorbed 
power-law model. This model includes Galactic column density 
along the line of sight to the quasar \nhg\ and possibly
additional absorption intrinsic to it \nhi. The
values of \nhg\ are fixed based upon the 
Leiden/Argentine/Bonn (LAB) survey of Galactic 
\hi\ \citep{har97,kal05}. We define four variations of 
this model, depending upon whether the photon spectral index (\xgamma)
or \nhi\ are allowed to vary as a fixed parameter. We define 
four variations of this model, depending upon whether this spectral index
\ax\ or \nhi\ are allowed to vary (models A-D; Table~\ref{tab-models}). 
 We visually inspect the fitting 
quality and apply the following rules to select 
the model that produces the best constraint and quality: 
\begin{enumerate}
  \item If $N_{\rm Xph}<30$, we use Model~A. 
        In this case, it is impossible to 
        constrain the intrinsic absorption and the constraint on 
        \ax\ is poor (\onesig\ uncertainty of \ax\ is larger than 
        $\sim0.5$). Therefore, we fit only the flux levels and fix 
        \ax$=-1$, which is the average value of X-ray spectral 
        index for RQ type 1 quasars \citep[e.g.,][]{nandra94,page03,pag04,you09}. 
  \item If $30\leq N_{\rm Xph}<100$, we allow one additional free parameter, 
        either \nhi\ or \ax, but not both:
         \label{modelselruleii}
        \begin{itemize}
           \item If the intrinsic column density \nhi\ is 
                inconsistent with zero and well constrained,
                 we choose Model~D. 
           \item If \nhi\ is consistent with zero, and \aox\ is reasonably
                 well constrained ($|\delta\alpha_{\rm x}|<0.5$), we choose \modelb. 
           \item If neither of these criteria are met, we choose \modela.
        \end{itemize}
  \item In the case that $N_{\rm Xph}\geq100$, if \ax\ is well
        constrained and the \onesig\ lower bound of \nhi\ is not consistent 
        with zero, we use \modelc. Otherwise, we follow Rule~2.
\end{enumerate}
The visual inspection process is performed by three people and a consensus 
is reached to ensure objectivity. The number of 
quasars selected for each fitting model is shown in Table~\ref{tab-models}. 
Examples of the fitting are shown in Fig.~\ref{fig-xrtexample}. 

Based on the spectral fitting results, we calculate a number of
parameters, including the photon count rate, the observed and
unabsorbed X-ray flux between 0.3 and 10~keV, and the monochromatic 
luminosity at 2~keV in the emitted frame. These
values are included in the final catalog (see Section~\ref{swiftq-catalog}). 
\subsection{Weak X-ray Sources}
We process weak X-ray sources (flagged as \emph{w} in the final catalog)
with $N_{\rm Xph}<10$ separately because some are not significantly 
detected by XRT and we can only estimate their flux upper limits. 

We apply the Bayesian method of \citep{kraft91} to determine whether
an X-ray source is detected or not. We define a source to be a 
non-detection if \threesig\ confidence level lower limit is consistent
with zero source counts. 98 out of 406 weak X-ray sources are detected by XRT. 

For the detected sources, we fit Model~A to 
determine \fx\ and \ftwokev, which is the obsersved flux between 
0.3 and 10~keV and the observed flux density at 2~keV. For undetected 
sources, we consider two cases. If the background-subtracted count 
rate is positive, we can use the \xspec\ to apply Model~A. 
This has the advantage of including
the calibrations present in the ARF file, but the model will be fitted
to the measured counts and not the upper limit and therefore must be 
rescaled. The flux upper limit is determined using 
\begin{equation}
  F_{\rm uplim}=F_{\rm Xspec}\frac{N_{\rm uplim}}{N_{\rm Xph}}.
\end{equation}
in which $N_{\rm uplim}$ is the \threesig\ source photon count upper 
limit. 

If an undetected X-ray source has too few counts (typically 
\nxph$<2$), \xspec\ is unable to apply models to the data.  
In these cases we manually apply two corrections to convert 
the count limits into limiting count rates.  The first correction 
is for vignetting.  Vignetting is a reduction of the effective 
area of the telescope at off-axis positions.  We adopt a 
vignetting function of $V(\theta)=1-C\theta^2$, where $\theta$ 
is the off-axis angle in arcminute.  The coefficient $C$ varies 
with energy (Cusumano \& the XRT Calibraqtion Team, 2006; Kennea, 
private communication); here we adopt the value appropriate for 2~keV 
as it is representative and the uncertainty in our measurements is 
dominated by small number statistics.  This correction is applied by 
reducing the effective exposure time by the vignetting factor, 
thereby increasing the count rate limit.  The second correction is 
to account for the finite size of the aperture used to measure 
source counts.  We generate an ARF file appropriate for the size 
and position of the source extraction region.  The output of this
process includes a report of the fraction of the source fluence 
enclosed within the region; we divide the count limit by this 
fraction to apply the PSF correction. Finally, the vignetting- 
and PSF-corrected count rate limit is converted to limiting observed 
flux by using the tool \emph{pimms} \citep{mukai93} together with 
the assumption of \modela. From this model we determine the 
absorption-corrected flux, flux density and monochromatic luminosity at 2~keV.
\subsection{UVOT Data}
\label{swiftq-uvotdata}
Instead of using a pipeline to obtain a co-added event file as was done
with the XRT data, 
we process each individual UVOT sky image for each object. The 
composite photometry is obtained by summing over photon 
counts from each individual image and dividing the sum 
by the total exposure time.

First, we must identify suspect and defective sky images. Because we 
are using serendipitous observations, it is inevitable that some sources
are too close to the edge of the UVOT FOV\footnote{Objects are not necessarily 
always within the FOV because the matching radius used to select the
raw sample is larger than
UVOT FOV and/or because UVOT was working under other modes.}. 
The goal of this process is to exclude the low quality sky 
images so that the
default source and background regions can be directly applied on good ones.

The source region is defined using a circle with a radius of
$3^{\prime\prime}$, which is recommended in the UVOT
photometric calibration \citep{poo08}. Since a
$5^{\prime\prime}$ radius aperture, which contains
$85.8\pm3.8\%$ of the PSF, was used for calibrating the UVOT, an
aperture correction is applied to the data when running
\emph{uvotsource} \citep{poo08}. 
The inner and outer radii of the background regions are
$r=27_.^{\prime\prime}5$ and $R=35_.^{\prime\prime}0$ respectively;
this is the standard background region used to construct the first GRB
afterglow catalog \citep{rom09}. The large difference between the
source and background radii ensure that the background area is at least 
$50$ times as large as the source region to provide an accurate 
background subtraction. 

We then flag each sky image based on the position of 
the object in the image frame and the aspect keyword value. We only use images
whose aspect values equal {\tt DIRECT}. The other images
(flagged as $-1$) do not have correct aspect corrections and may lead to 
inaccurate photometric results. The detailed image flagging strategy is
tabulated in Table~\ref{t-uvotflag}. 
We only accept UVOT images flagged with 0, 1 and 2. 

Due to relatively small photon counts or 
positioning uncertainties, a fraction of object images are not 
symmetric or well aligned with the default source region circle 
with a radius of $3^{\prime\prime}$. 
It is necessary to co-align the quasar image with the center of 
the source region circle because the aperture correction process 
assumes a symmetric photon loss outside
the pre-defined aperture. To co-align the locations, we extract a region
of $11\times11$ pixels (about $5^{\prime\prime}\times5^{\prime\prime}$) 
centered on the SDSS quasar coordinates. If
the total photon count inside this region is less than 20, we
regard this image as ``faint" and assign a flag of ``1".
Even if a single image is regarded as ``faint", it 
may contribute to the total object counts when 
stacked with other images (e.g., Lehmer et al.
2007\nocite{leh07}). If an image is not flagged as ``faint", we
attempt to fit the source with a two-dimensional Gaussian
profile the centroid of which is set to be free. The background is determined by
averaging the photons inside the region centered at the SDSS
coordinate and bounded by two squares with sizes of $21\times21$ pixels and
$11\times11$ pixels. If the photon count inside the
$11\times11$ pixel region is less than three times that of the
background level, this image is also treated as ``faint" 
(flagged as ``1"). Otherwise, we fit the $11\times11$ pixel region
with a 2-D Gaussian profile. The new Gaussian centroid is compared with the
original SDSS quasar coordinate. We denote their angular separation as
$\delta$. 

There are three categories of images.
\begin{itemize}
  \item If $\delta\leq0^{\prime\prime}.618$, the two coordinates
        are consistent and the default source region circle is
        used. This image is flagged as ``0".
  \item If $0^{\prime\prime}.618<\delta\leq3^{\prime\prime}$, this image
        is flagged as ``2" and the source region circle is re-centered
        at the Gaussian centroid.
  \item If $\delta>3^{\prime\prime}$, this image is flagged as ``3". This
        large offset could be caused by the unrecorded aspect problem, 
        which requires manual correction or
        non-Gaussian photon distribution. Less than 1\% of sky images are 
        flagged as ``3", which are dropped.
\end{itemize}

In the next step, we define the default source and background
regions for each object, and select the sky image with the longest 
passband central wavelength and the longest exposure,  
because, stellar contamination from host galaxies is larger in 
optical than in UV. If the V band is not available or its exposure is too 
short ($\lesssim100$~seconds), 
we examine the B band. If the B is unavailable, we check the U band, and 
then successively the UVW1, UVM2 and UVW2 bands. For each object, we 
attempt to exclude all stars, galaxies
and bad pixels in the background annulus. If the source region 
is significantly contaminated or most of the background region
must be masked, we remove this object from the UVOT analysis. 

In most cases, the procedures above are sufficient to define
a source and background region for all sky images. However, there
remain a number of observations that require visual inspection. 
It is sometimes necessary to define a different background region to 
exclude the sky images in order to
obtain acceptable photometry. For example, a number of 
images contain stellar ghost rings. Due to differences of
aspect and exposure time, the influence of these factors may
vary for each sky image even for a given object. In addition, although 
the {\tt ASPCORR} keyword may be set to {\tt DIRECT}, the image may 
still suffer from uncorrected aspect problems so that a single 
point source may have multiple images or even smeared images. 
Consequently, it is necessary to visually inspect each sky image 
 to perform the second order region customization
and image selection. After the visual inspection process 
($>50,000$ images), 3183 ($\sim 1\%$) UVOT sky images 
were excluded from our UVOT image set.

We use \emph{uvotsource} (HEADAS version 6.10)
to calculate photometry on \emph{individual} sky images using the 
curves of growth from the \swift\ CALDB
for the aperture correction model. The composite photometry uses the 
photon counts of all available images. For a
given waveband, the mean magnitude is calculated as
\begin{equation}
  \label{e-avemag}
  \langle m\rangle=Z_{\rm pt}-2.5\log{\langle R_{\rm LSS}\rangle}
\end{equation}
in which $Z_{\rm pt}$ is the zero-point
magnitude (the magnitude when the count rate is
1~photon~s$^{-1}$). The values are taken from Table~6 of
\citet{poo08}. $R_{\rm LSS}$ is the source count rate with coincidence-loss, 
aperture, and large-scale sensitivity corrections
applied.
The associated uncertainty in this magnitude measurement is
\begin{equation}
  \label{e-eavemag}
  \delta_{\langle m\rangle}=\sqrt{\Delta^2_{Z_{\rm pt}}+\left(\frac{2.5}{\ln{10}}\langle R_{\rm LSS}\rangle\right)^2\Delta^2_{R_{\rm LSS}}}
\end{equation}
in which
\begin{equation}
  \label{e-avecr}
  \langle R_{\rm LSS}\rangle=\frac{\sum{R_{{\rm LSS},i}\cdot T_i}}{\sum T_i} 
\end{equation}
and 
\begin{equation}
  \label{e-davecr}
  \Delta_{\langle R_{\rm LSS}\rangle}=\frac{\sqrt{\sum{\left(T_i\Delta_{R_{{\rm LSS},i}}\right)^2}}}{\sum{T_i}}
\end{equation}
 where $T_i$ is the exposure time in image $i$. We also calculate the 
flux density at the effective wavelength for each filter for each object 
by multiplying the LSS photon count rate by the conversion factors in Table~10
(conversion from power-law spectra) of \citet{poo08}. 

The above equations can be applied to an individual
sky image as well as a group of sky images. This process produces 
three measurements for each filter: photometry for each individual sky 
image, for all sky images in an OBSID,
and for all OBSIDs of a filter. 

A total of 675 objects in our sample are detected by UVOT. 

%% file: sec3.tex
\section{Quasar SEDs}
\label{swiftq-results}
We classify objects in the database into four types based on 
data availability (see Table~\ref{tab-fourtypes}). In the description 
below, ``useful data" includes both detections and non-detections. 
\begin{itemize}
  \item Type~A contains 637 objects that have useful data from both UVOT 
        and XRT. Among them, 345 are detected by UVOT \emph{and}
        XRT.
  \item Type~B contains the 38 objects that have useful data 
        from UVOT but \emph{not} XRT. These objects cannot be included
        in correlation analysis. 
  \item Type~C contains 168 objects that have only useful 
        data from XRT but not UVOT. We will only create the XRT energy spectra 
        for these objects. 
  \item Type~D contains 191 objects that are bereft of any useful data 
        from UVOT or XRT. They are not included in the final catalog. 
\end{itemize}
\subsection{Supplementary Data}
As the object redshift increases, the UVOT wavebands 
are progressively shifted into the EUV band shortward of
\lya; as a consequence, UVOT data for some quasars cannot be used to 
measure the rest-frame UV power-law. The SDSS and 2MASS data,
observed at longer wavelengths, can be used to
extend the available measurements to UV and optical bands in the
quasars' rest-frames. We supplement our UV photometry with the
five bands (\emph{ugriz}) from the SDSS DR5 quasar catalog and three bands 
(\emph{J, H, Ks}) of the 2MASS \citep{coh03,skr06} survey, when available. 
In the best cases, we have 14 photometric data points in UV/Optical
wavebands: six from \emph{Swift}/UVOT, five from SDSS, and three from 2MASS. 
\subsection{Initial SED Plots}
\label{initialsedplots}
The monochromatic luminosity corresponding to each UVOT filter can 
be easily computed after shifting the flux density to the quasar's 
rest-frame. The frequency band widths are calculated by converting
rest-frame FWHMs of corresponding wavebands.  

The SDSS quasar catalog provides magnitudes that we must convert 
to flux densities. Instead of the \emph{asinh} magnitude used in the 
general SDSS photometric measurements \citep{lup99}, we 
convert the SDSS flux densities from corresponding band magnitude 
using the Pogson definition \citep{pog57} which has a much 
simpler analytical expression $m_1=-2.5\log{f_1/f_0}$, where $f_0$ is the 
zero-magnitude flux. The Pogson magnitude system deviates from 
the asinh system for faint objects. According to the \emph{asinh softening 
parameters (b coefficients)} table in the SDSS DR5 photometric 
calibration document\footnote{\url{http://www.sdss.org/dr5/algorithms/fluxcal.html\#asinh\_table}}, 
the difference between Pogson and asinh magnitudes is less than
1\% for objects brighter than $g=22.60$ and $u=22.12$. 
The majority of our objects are much brighter than the limit. 
The difference between the Pogson 
and asinh systems is therefore negligible. There is one object
(SDSSJ122740.85$+$440604.7) whose B band magnitude is fainter than 22.60 
and two objects (SDSSJ020316.37$-$074832.1, and SDSSJ133613.62$+$025703.8) 
whose U band magnitudes are fainter than 22.12. They are all at high redshift 
 ($z>2.5$) with extremely low X-ray photon counts ($N_{\rm Xph}<10$), so 
they do not play significant roles statistically in the correlation 
analysis result. For wavebands in SDSS, we adopt the 
central wavelength and FWHM from \citet{fuk96}.
For wavebands in 2MASS, we adopt the isophotal central wavelengths
and bandwidths from \citet{coh03}.

For X-ray sources with less than 100 photon counts, we use a different 
binning strategy in \xspec\ from the one presented in 
Table~\ref{t-xrtbin} to avoid large error bars. We use the 
command {\tt setplot rebin} to re-bin the spectrum 
until each bin has a detection at least as large as \onesig\
and no more than 100 bins may be so combined. Error types are set to
{\tt quad} which sums in quadrature the errors on the original bins.
The {\tt rebin} command only affects the plot appearance but not the
fitting results (see Fig.~\ref{fig-xrtexample} for examples), but 
produces a clear representation of the average flux levels and 
associated uncertainties. Because of the relatively large uncertainties in 
the X-ray energy spectrum, we calculate the lower and upper error
bars of each flux point, instead of applying standard error propagation.

The {\itshape initial} quasar SEDs are generated based on all available
UV/optical photometric data points and X-ray energy spectra. Before 
shifting into the rest-frame, we apply Galactic reddening 
corrections to all available wave band flux densities using the standard 
$E(B-V)$ dependent extinction curve \citep{fit99}. Values of 
$E(B-V)$ are calculated following \citet{sch98}. Some examples of 
quasar SEDs are presented in Fig.~\ref{fig-isedexample}.

These initial SEDs cannot be used for SED fitting for
two main reasons.
\begin{itemize}
  \item Broad emission lines can contribute significantly to 
  broad band filter measurements, which may 
  lead to incorrect SED shapes. For example, the band covering
  \lya\ can alter the UV slope by up to 0.2.
  \item The dates of the observation from SDSS, 2MASS, and \swift\ for 
   a given object differ, often by several years in the rest-frame. The data
   may need to be shifted to mitigate the effect of variability. 
\end{itemize}
\subsection{Emission Line Correction}
Emission line corrections can be performed by subtracting 
the broad line contribution based on their average equivalent
width (EW) \citep[e.g.,][in preparation]{elv11}. Because
we have SDSS UV/optical spectra for all quasars,  
we use a more sophisticated method in which we convolve the
response function $R(\lambda)$ of each filter with the observed spectrum 
$f_{\rm tot}(\lambda)$ and the power-law only spectrum $f_{\rm pl}(\lambda)$ 
to calculate the emission line correction factor.
\begin{equation}
\mbox{EC}=\log{\dfrac{F_{\rm tot}}{F_{\rm pl}}}
\end{equation}
in which 
\begin{equation*}
F_{\rm tot}=\int f_{\rm tot}(\lambda) R(\lambda) d\lambda
\end{equation*}
\begin{equation*}
F_{\rm pl}=\int f_{\rm pl}(\lambda)  R(\lambda) d\lambda
\end{equation*}
Because the quasar UV power-law usually extends from \lya\
to $\sim5600$~\AA\ \citep[e.g.,][]{van01}, we only need to perform 
emission corrections to filters covered within this wavelength 
range. However, low-redshift quasar spectra do not cover \lya\
and high-redshift quasar spectra do not cover \lambdauvo. 
In these cases, the emission line corrections are performed
based on the composite spectrum in \citet{van01}. 
Although the shape of the composite spectrum
may not be exactly the same as the real spectra, the mean 
corrections are sufficient for power-law slope estimation. 

To estimate the errors introduced
by performing emission line corrections using the composite spectra,
we examine the emission line correction trends as a function
of redshift for different filters and find that the 
emission line correction is typically less than $0.1$~dex 
(see Fig.~\ref{fig-elctrend}). The real spectrum corrections are
generally distributed around the composite spectrum, meaning
they are generally in agreement. To view their differences
more clearly, we plot the distributions of differences 
between these to corrections, 
$\Delta\mbox{EC}=\mbox{EC}_{\rm real}-\mbox{EC}_{\rm composite}$,
in Fig.~\ref{fig-elcdiff} (only the SDSS $g$ band and UVOT
B band are displayed, but other filters are similar). 
These distributions indicate that on average, the composite
spectrum emission line correction is consistent with the
real spectrum emission line correction; the dispersion
of these distributions are around 0.05 dex. Therefore, we
apply a systematic uncertainty of 0.05 dex to photometry
corrected using the composite spectra.

After emission line correction, the median photometric slope 
agrees well with the median spectroscopic
slope in the UV band, with a value of $-0.43$. In the upper panel of 
Fig.~\ref{fig-auvphauvsp}, we plot these two slopes for objects 
with SED fits. Most data points are distributed along the line representing 
$\alpha_{\nu,{\rm ph}}=\alpha_{\nu,{\rm sp}}$ with considerable scatter. 
We believe that this scatter is mostly caused by low redshift
quasars (typically $z<0.8$) in which the $2200$~\AA\ ``line-free"
rest-frame continuum point is not covered by SDSS spectra. In these
cases, the uncertainty of the UV spectroscopic slope $\alpha_{\nu,{\rm sp}}$ 
may be larger than for high redshift quasar spectra. The value of
$\alpha_{\nu,{\rm sp}}$ can be more accurately measured for objects
at higher redshifts, which explains why the dispersion of 
$\alpha_{\nu,{\rm ph}}-\alpha_{\nu,{\rm sp}}$ is much smaller 
(\emph{bottom panel}, Fig.~\ref{fig-auvphauvsp}). In the description of
correlation and regression analysis, we consistently use \auv\ to 
represent $\alpha_{\nu,{\rm ph}}$ because the data used to fit the slope
is taken simultaneously with the X-ray data. 

\subsection{Photometric Shift\label{photometricshift}}
The photometric data in SDSS and 2MASS are not observed simultaneously 
with the \emph{Swift} UVOT data, so quasar variability can make flux 
comparisons uncertain. To make best use of these data, we must create 
some \emph{pseudo}-simultaneous data points by shifting the observed 
flux levels to be consistent with the \emph{Swift} UVOT photometry, 
assuming the UV/optical SED shapes remain unchanged. With
additional data points, we can place tighter constraints to 
UV slopes and luminosities while still, in some sense, maintaining the 
simultaneous property of the dataset. 
The photometric shift follows strategies described below 
(see Table~\ref{tab-sedshift} and Fig.~\ref{fig-sedshift}). 
\begin{enumerate}
  \item Because all quasars have five simultaneous SDSS photometric 
        measurements we always interpolate or extrapolate the SDSS 
        photometry to obtain fluxes at the UVOT filter effective wavelengths, 
        then shift all SDSS and 2MASS photometry to match UVOT. 
        Because there is no overlap between the 2MASS and UVOT bands, 
        it is impossible to shift 2MASS photometry separately so we use
        the same amount of shift as the SDSS data. The 2MASS photometry is
        only used for a few high redshift quasars so the time difference
        between 2MASS and SDSS data will not affect the UV spectral slope for
        a majority of objects in our sample.
  \item We prefer to match in the UVOT U band. The \swift\ U
        band is very close to the SDSS \emph{u} band, which significantly
        reduces uncertainties introduced by interpolation or extrapolation. 
        The U band, being bluer than the V and B bands, is 
        also less contaminated by host galaxy light.
  \item If \emph{Swift} U band photometry is unavailable,
        we interpolate the SDSS \emph{u} and \emph{g} bands
        to match the \emph{Swift} B band. If neither the U nor the B band 
        is available, we interpolate the SDSS \emph{r} and
        \emph{g} band to match the \emph{Swift} V band.
  \item If none of U, B, or V bands is available, we extrapolate the 
        SDSS \emph{g} and \emph{u} band to match 
        one of the UVW1, UVM2 or UVW2 bands, 
        using the available UVOT filter with the longest 
        effective wavelength. Matching these filters is only done as a 
        last resort; the significant extrapolation required inevitably 
        introduces considerable
        photometric uncertainties. Furthermore, these three
        bands frequently lie in the Lyman forest where the 
        quasar SED suffers from severe intrinsic
        and intervening absorption \citep[e.g.,][]{rau98}. The continuum in 
        this spectral region cannot be approximated as a single power-law.
  \item When selecting the matching filters, we require the SDSS filters and 
        the UVOT filters to \emph{both} fall in the UV range (between \lya\ 
        and $5600$~\AA) or within the EUV range, simply because the UV 
        power-law cannot be extended to EUV region. 
\end{enumerate}
Examples of photometric shifting results are presented in 
Fig.~\ref{fig-sedshiftexample}.
\subsection{Error Analyses}
For UVOT photometry, we adopt the photometric uncertainties 
produced by \emph{uvotsource} and perform error propagation 
assuming these errors follow a Gaussian distribution. The
uncertainties of the SDSS and 2MASS fluxes are calculated based on 
magnitude uncertainties in the SDSS DR5 quasar catalog.
The frequency bandwidths plotted on the SED are 
converted from the corresponding FWHM of each filter. 

For the XRT data, we use the parameter uncertainties produced
by the \xspec\ {\tt error} command. Because the X-ray 
photon counts follow a Poisson distribution, which is 
unsymmetric, we 
calculate the upper and lower error bars separately.  

\subsection{SED Models}
Because of the deficiency of the EUV data,
the exact quasar bolometric luminosities strongly depends on the model 
used to fit the BBB. Traditionally, emission in this ``gap" is 
represented by a power-law
continuum with a slope \aox\ \citep[e.g.,][]{ric06}. In this work,
we consider a model which provides an upper limit to the flux of the BBB
in the canonical case. This shape is inspired from the
presence of a soft X-ray excess over a flat X-ray
component reported by \citet{arn85} in \emph{EXOSAT} spectra of
Seyfert~1 galaxy \object{Mkn~841} and further studied 
by \citet{wal93} and \citet{gie04}. The latter study
found that this soft X-ray excess can be well fit by a black body
of energy 0.1 -- 0.2 keV. This result motivates us to \emph{constrain} 
this feature using a ``bump" shape. In the following context, this model 
will be called the \emph{exponential decay model} (hereafter EXP model).
In addition, we use another model that directly connects the high energy 
limit of the UV power-law and the lower energy limit of the X-ray
power-law at 0.3~keV.  We call it the \emph{triple power-law model} 
(hereafter TPL model). In either model, total flux is 
the result of a UV and an X-ray component.
    \begin{equation}
      \label{expdecmod}
      f_{\nu,{\rm tot}}=f_{\nu,{\rm UV}}+f_{\nu,{\rm X}}.
    \end{equation}

{\bf EXP model} In this model, the two components are 
        \begin{equation}
          \label{expdecuv}
          f_{\nu,{\rm UV}}=10^{\beta_{\rm UV}}\nu^{\alpha_{\rm UV}}e^{(-h\nu/kT_{\rm B})^{\gamma_{\rm UV}}}
        \end{equation}
        \begin{equation}
          \label{expdecx}
          f_{\nu,{\rm X}}=10^{\beta_{\rm X}}\nu^{\alpha_{\rm X}}e^{(-h\nu/kT_{\rm X})^{\gamma_{\rm X}}}
        \end{equation}
        The UV component is a power-law multiplied by an exponential 
        decay term. The power-law slope $\alpha_{\rm UV}$ and scale factor
        $\beta_{\rm UV}$ are obtained by fitting photometric data points 
        covered in the UV region (from 5600~\AA\ to \lya). Canonically 
        $\alpha_{\rm UV}\approx-0.4$ \citep[e.g.,][]{van01}. 
        The combination of exponential and power-law terms create
        a ``bump" in the EUV region.  The value of $\gamma_{\rm UV}$ 
        controls how quickly the flux decays given a value of 
        $T_{\rm B}$ which is obtained from the SED fitting. We fix 
        $\gamma_{\rm UV}$ to be 1.5 so that at $\nu<\nu({\rm Ly}\alpha)$ the
        contribution from the exponential term is negligible and the total
        curve agrees well with the UV power-law given the best fit of 
        $T_{\rm B}$.  For example, if we adopt $\gamma_{\rm UV}=1$ \citep[e.g., ][]{gru10},  
        a typical value of $T_{\rm B}\sim4$~Ryd from the SED fitting produces a discrepancy 
        between the UV power-law and the total SED curve by $\sim0.1$~dex at
        $\nu({\rm Ly}\alpha)$. 

        The X-ray component is also a power-law multiplied by an 
        exponential decay term. The power-law slope $\alpha_{\rm X}$ and scale
        factor $\beta_{\rm X}$ are obtained by fitting the X-ray energy spectra
        using \xspec. The X-ray decay energy $T_{\rm X}$ is fixed at 
        0.3~keV. We adopt $\gamma_{\rm X}=-8/3$ which 
        assumes neutral hydrogen absorption \citep{lon92}.

        We emphasize that both of the UV and X-ray components are 
        only mathematical expressions. Specifically, the value of $T_{\rm B}$ 
        does not reflect the accretion disk temperature. 
        Our goal is to place a reasonable upper limit to the strength of the BBB
        which contributes a significant fraction of the bolometric luminosity 
        \citep[e.g.,][]{mat87,zhew97,lao97,tel02,shaz05}. 
        Therefore, the specific mathematical form is not important. 
 
        Because the XRT observes photons between 0.3 and 10~keV, these 
        measurements are shifted to a higher rest energy range for high 
        redshift quasars, i.e., we lack the soft X-ray data points for these
        objects. This affects our SED fitting because if we only fit 
        the observed 
        data points, the decay energy of the bump will be shifted to a higher 
        energy band as the redshift becomes higher, which leads to
        additional flux contribution. To solve this problem, we supplement 
        artificial data points by extending the X-ray power-law from the 
        minimum energy of the real data down to 0.3~keV. 
        We use 0.3~keV as the lowest energy for the X-ray power-law 
        based on previous X-ray studies of  quasars. These investigations 
        found that the X-ray power-law can extend from $\sim0.1$~keV to 
        $\gtrsim10$~keV \citep[e.g.,][]{tur89,lao97,george00,pag05,you09}. 
        We then fit all the data points with this SED model using the 
        Levenberg-Marquardt algorithm, leaving $T_{\rm B}$ the only free 
        parameter. The median value of \tbb\ is $5.96$~Ryd. 
 
    {\bf TPL model} In this model, the 
        UV and X-ray power-laws are fit in the same way as the EXP model. 
        Instead of multiplying each power-law with an exponential decay 
        term, we simply connect the UV power-law at \lya\ and the X-ray 
        power-law at 0.3~keV and denote the slope as \auvx. 
        The distribution of \auvx\ is presented in Fig.~\ref{fig-auvxdist}. 

In a canonical situation, i.e., \auv$=-0.4$ and \ax$=2$, 
the EXP model generates a bump in the EUV region which produces an upper 
limit while the TPL model yields a lower limit of the strength of the 
BBB. This simple analysis may not apply to strong X-ray 
absorption quasars. An example is shown in the first SED plot in 
Fig.~\ref{fig-sedmodexample}. In this case, the absorption leads to 
a flat UV slope in addition to low X-ray emission in soft X-ray band. In 
these cases, the EXP model does not produce a ``bump" but a ``dip" in the 
EUV region and the corresponding integrated bolometric luminosity is not
reliable. These quasars are flagged as ``red" and will not be included into
our cleaned sample defined below. Examples of fitted SEDs using these 
two models are presented in Fig.~\ref{fig-sedmodexample}. 

\subsection{Bolometric Luminosity and Black Hole Mass}
To obtain black hole masses, we measure broad emission 
line widths with SDSS spectra.
We employ the same software
package used in previous SDSS spectral analyses (e.g., Wu et
al. 2009\nocite{jwu09}; Vanden Berk et al. 2011, in
preparation). The Galactic reddening corrections to all
the spectra are performed using the extinction curve of
\citet{fit99}. Values of $E(B-V)$ are calculated following
\citet{sch98}. 
We use three components to fit a spectrum: a single
power-law in the UV band (between the \lya\ emission line and
$\sim5500$~\AA), the small blue bump, the UV
iron template from \citet{ves01} and the optical iron template
from \citet{ver04}. 
Each broad emission line is fit by a single or
multiple Gaussian profiles. 
To ensure that our software package produces consistent results with
previous studies, we compare our black hole masses to these in 
\citet{shey08}.

Following \citet{shey08}, we adopt different black hole
estimators depending on quasar redshifts. At $z<0.7$, we use
\hbeta\ plus $l_\lambda(5100\mbox{~\AA})$, at $0.7<z<1.9$, 
we use \mgii\ plus $l_\lambda(3000\mbox{~\AA})$, and at $z>1.9$, 
we use \civ\ plus $l_\lambda(1350\mbox{~\AA})$.
The black hole mass is calculated using the following equation
\citep{she06b,shey08}
\begin{equation}
\label{e-bhmass} 
\log{M_{\rm BH}}=a+b\log{\left(\lambda L_\lambda\right)}+2\log{\mbox{FWHM}}
\end{equation}
in which $M_{\rm BH}$ is in solar mass $M_{\odot}$, 
$\lambda L_\lambda$ is in $10^{44}$~ergs~s$^{-1}$, and 
FWHM is in km~s$^{-1}$. The two coefficients, $a$ and $b$, in 
this equation, are
$(a,b)=(0.66,0.53)$ for quasars with $z<0.7$, $(a,b)=(0.505,0.62)$ for 
quasars with $0.7<z<1.9$, and $(a,b)=(0.672,0.61)$ for quasars 
with $z>1.9$ \citep[see Table~\ref{tab-mbhcal}; ][]{mcl02,mcl04,ves06}. 

For comparison purposes, we calculate the bolometric luminosity using
the wavelength-dependent bolometric correction (BC) factors by 
\citet{ric06} which are obtained using the composite SED for a 
sample of SDSS DR3 quasars \citep{aba05,sch05},
so that $L_{\rm bol,SDSS}=\mbox{BC}\cdot\lambda l_\lambda$. 
These are bolometric luminosities \emph{without} using simultaneous 
observations and does not include the BBB component, which differs 
from the bolometric luminosity we obtained by SED integration. 

We then compare some important correlations and distributions 
with \citet{shey08}. The black hole masses vs. bolometric luminosities of 
923 quasars are plotted in 
Fig.~\ref{fig-mbhlbol}\footnote{The others are not displayed because 
of low SDSS spectral quality.}. The locations of these quasars are 
consistent with the locations of quasars in Fig.~11 of \citet{shey08}, 
indicating that most quasars are accreting at a sub-Eddington
level, i.e., $0.1<$\lle$<1$. The few quasars that fall above \lle$=1$ (they 
are accreting at super-Eddington level) are narrow line 
Seyfert 1 (NLS1) galaxies \citep{shej08,gru10}. 
The median values of 
$\log{\mbox{FWHM/(km~s}^{-1})}$ are 3.62, 3.63 and 3.75 for 
\hbeta, \mgii\ and \civ, respectively, which are consistent with \citet{shey08}.
By comparing the FWHM calculated using both of \hbeta\ and \mgii\ or 
\mgii\ and \civ\ we find a potential source of bias in the calculation of 
black hole masses as the \civ\ FWHMs are systematically larger than 
those of \hbeta\ and \mgii. However, we still use the black hole masses 
estimated by \civ\ because this offset is small compared to the 
dispersion of the black hole mass distribution.

We also calculate the bolometric luminosities by integrating 
the two SED models from 5600~\AA\ to 20~keV.
In general, emission in this region contributes the 
majority of the quasar luminosity for typical Type~1 quasars.

In Fig.~\ref{fig-lboldist}, we show the differences between 
the bolometric luminosities measurements by integrating the 
two SED models, and 
the bolometric luminosities calculated using BC. 
The EXP model and the TPL model produce bolometric luminosities 
average higher and lower than the \lbol\ from the BC, respectively. 
Because the bump emission from the EXP model is more sensitive to 
the UV and X-ray observed spectral shapes than the TPL model,
it has a larger dispersion. 
A small fraction of the EXP models produces less bolometric luminosity 
($\log{L_{\rm bol,EXP}}-\log{L_{\rm bol,SDSS}}<0$) than the TPL models. 
Some of these objects are quasars with strong intrinsic 
absorption (the black shaded area in 
Fig.~\ref{fig-lboldist}).
The reason why the TPL model produces less bolometric luminosity on 
average than the BC method is because the bolometric correction by 
\citet{ric06} includes the infrared wavebands, which 
contribute about $37\%$ of the total flux. After correcting 
for contributions from this component, the entire histograms of 
both the EXP and the TPL models should move positively
by $\sim0.2$~dex. Consequentially, the TPL model on average produces
a bolometric luminosity consistent with that from the BC and the
EXP model on average over produces \lbol\ by $\sim0.3$~dex.


%% file: sec4.tex
\section{Catalog Description}
\label{swiftq-catalog}
Our final catalog contains 843 quasars with $0.0129\leq z\leq4.5766$ and 
$-30.24\lesssim M_i\lesssim-22.01$. There are 675 objects observed by
UVOT, 805 observed by XRT and 637 observed by both UVOT and XRT. 
The X-ray detection rate of the entire catalog is $\approx60\%$. Among 
objects in the catalog, 345 objects are \emph{detected} by both 
UVOT and XRT, so that we are able to determine their SEDs from 
UV to X-ray bands. These objects constitute our parent sample to 
evaluate the flux contribution from the BBB. 
This catalog contains parameters directly measured from UVOT and XRT 
data and quantities derived based on these measurements such as black hole 
masses and bolometric luminosity. Columns in this catalog are 
described in Table~\ref{tab-catalog}. Comments for special individual objects
in this catalog are listed in Appendix~\ref{app}. Because 
the catalog contains 63 columns, we publish the full catalog in 
electronic format. In Table~\ref{tab-catalogexample}, we represent ten columns of some 
objects in our catalog.

%% file: sec5.tex
\section{Correlation Analyses}
\label{swiftq-analyses}
In this section, we investigate the \aoxluv\ relationship using
a selected sample of Type 1 quasars from our catalog. This correlation has 
been described in a series of papers
\citep{avn82,mar84,tan86,and87,wil94,pic94,avn95,vig03,str05,ste06,jus07}; 
as quasars become more luminous, their SEDs from UV to X-ray bands become 
softer, i.e., less X-ray emission with respect to UV emission. 
In the study of 372 type 1 quasars 
in Just et al. (2007; hereafter J07\nocite{jus07}), 
the UV/optical data are drawn
from the SDSS DR3 quasar catalog \citep{sch05}, IUE \citep{bog78,kon89} and 
the COMBO-17 survey \citep{wol03} which covers the E-CDF-S \citep{wol04}. 
The X-ray data are from \emph{ROSAT}, \emph{Chandra}, and \emph{XMM-Newton}
observations.
Because the time between UV/optical and X-ray observations 
could span a time scale of years, quasar variability inevitably 
introduces scatters to the \aox--\luv\ relation. In this section, we 
investigate whether this scatter can be reduced by simultaneous UV and X-ray 
observations.
\subsection{Sample Selection}
\label{sampleselection}
In this work, we use 637 quasars with simultaneous observations by 
both UVOT and XRT. Because this sample has a relatively low X-ray detection
rate ($\sim65\%$, see below) we applied an X-ray exposure cut-off to the XRT 
observations. The goal is to obtain a homogeneous, optically-selected
sample with a relatively high X-ray detection rate while retaining a 
sufficiently large sample size. 
The sample size and X-ray detection rate as a function of 
exposure cut-off is presented in Fig.~\ref{fig5-xexpcutoff}. 
The X-ray detection rate is $\sim65\%$ if we use the entire parent sample. 
These censored data can be handled by the ASURV software package 
\citep{lav92}, but the large fraction of non-detections makes the 
results unreliable. On the other hand, if we increase the exposure cut-off, 
the detection rate can rise 
to $>95\%$, but only a few objects remain in the sample and the 
statistics are very poor. 
In this paper, we use two samples to study the \aoxluv\ relation: the 
\emph{large catalog} sample which contains \emph{all} qualified quasars selected 
from the parent sample with no XRT exposure cut-off; and the \emph{small 
catalog} sample which is derived from
the \emph{large catalog} sample except that we apply an XRT exposure cut-off of 
10~ks. The large sample has a sample size of over 400 quasars; the small sample
has a higher X-ray detection rate ($\sim85\%$, see Table~\ref{tab-property}). 
In the following analysis, we supplement both of these samples with 
AGNs from \citet{gru10}, which contains data for 88 AGNs observed 
simultaneously by \swift\footnote{There are 92 objects in total in the 
\citet{gru10} study, but 4 objects were not observed with UVOT photometry.}. 
We define this 
sample \emph{supplemental sample} or the \emph{G10} sample. The G10 sample 
is soft X-ray selected and is composed of low luminosity Type~1 AGNs at low 
redshifts ($z\lesssim0.4$). Because the G10 sample is selected in a different
way from our sample, this sample is not merged with our samples, but is only 
used for comparison purposes.

The catalog parent sample is mixed with different types of quasars. To obtain a 
\emph{clean catalog sample}, we exclude objects that fall into any of
the four categories listed below. The quasar types are determined by 
exploring previously published studies and by calculating relevant 
quantities, e.g., radio loudness. The numbers of quasars excluded in each type are
tabulated in Table~\ref{tab-classification}. Because the G10 sample was not 
selected in the same way as our catalog sample, we only use them to represent 
simultaneously observed objects at relatively low redshift and low luminosity. 
Therefore, the exposure time cut-off is not applied to the G10 sample. 
The G10 sample is X-ray selected, so all objects are detected by XRT.

{\bf RL quasars} They frequently have substantial
        X-ray flux from the radio jet, which leads to a higher \aox\ than
        those of radio quiet quasars \citep{wor87,bri00}.

        We adopt the ``radio loudness" $(R^*)$ defined by the ratio
        of monochromatic luminosities at rest-frame 5~GHz and 
        2500~\AA\ \citep{sto92,del94,luy07,mil09}, e.g., 
        $R^*=L_{5\mbox{~GHz}}/L_{2500\mbox{~\AA}}$. 

        For most of the quasars in the parent sample, we calculate 
        the $k$-corrected \lradio\ from the 20~cm peak flux from the Very Large
        Array (VLA) FIRST Survey \citep{bec95,whi97} listed in the SDSS DR7
        quasar catalog, assuming an average value of radio spectral index 
        $\alpha_{\rm r}=-0.5$ \citep[e.g.,][]{kel89,kom06,luy07}. 

        For objects not detected by the FIRST survey (but still covered), 
        we estimate radio luminosity upper limits using the 
        sensitivity limit of the FIRST survey, which is 1~mJy. This 
        provides upper limits of $R^*$. 
        If $\log{R^*}<1$, we classify this object as RQ. 
        If $\log{R^*}\geq1$, we check
        other resources of radio surveys, such as the
        NRAO VLA Sky Survey \citep[NVSS; ][]{con98}.
        
        For objects not covered by the FIRST survey, we 
        examine $15^\prime$ NVSS images for strong nearby 
        radio sources. If we cannot confirm the quasar is RQ, 
        we conservatively exclude it from the clean sample. 

        We preferentially adopt \luv\ from SDSS spectral fitting results, 
        unless 2500~\AA\ is not covered, in which situation we use 
        \luv\ calculated from the SED fitting. 

{\bf Lensed quasar} The fluxes from these quasars are amplified, 
        so these objects are removed from the clean sample 
        \citep[e.g.,][]{jus07}.

{\bf Blazars} Blazars, including BL Lac objects and FSRQs, usually have large
        amplitude variations in UV/optical \citep[e.g.,][]{ran10} 
        as well as strong radio and X-ray emission
        \citep[e.g.,][]{bec03,pad07}. As a result,
        the values of \aox\ can vary significantly over time.
        The UV/optical spectra of 
        BL Lac objects usually have a featureless continuum without any 
        emission lines \citep[e.g.,][]{bla78,kol94,plo08,abd10}. 
        Many of them are RL and their emissions are believed to 
        be relativistically beamed from the jet \citep{urr95}. 
        Although some FSRQs exhibit similar broad emission line features as
        Type~1 quasars, their radio emission is still beamed and variable. 
        All objects classified as blazars are also excluded from 
        the clean sample. 

{\bf Reddened quasars} Our sample contains a number of quasars with 
        shallow UV/optical slopes and/or strong soft X-ray decline. 
        The relatively flat UV/optical spectra are attributed to dust reddening 
        \citep{ric01,ric03,hop04}, although observations of individual 
        objects suggest that some slopes could be intrinsically steep 
        \citep[e.g.,][]{hal06}. Because the gas density of
        the AGN BLR is much higher and radiative transfer effects are
        not the same as in the low density regime \citep[e.g.,][]{hao05},
        traditional approaches to correct reddening by the Balmer decrement 
        cannot be applied for quasars. 
        In this work, we only flag these objects and exclude them from the 
        cleaned sample. 
        
        Following \citet{ric03}, we use the relative color to define 
        the ``dust-reddened" quasars. The \emph{relative color} is calculated
        by subtracting the median colors of quasars at the redshift of each
        quasar from its measured colors. As argued by \citet{ric03},
        the quantity $\Delta(g-i)=(g-i)-\mbox{Median}(g-i)_z$ is an excellent
        redshift-independent surrogate for the photometric spectral index. 
        In Fig.~\ref{fig-gmi}, we plot 
        $\Delta(g-i)$ vs. redshift for all the objects in our raw catalog. 
        Objects to the right of the dashed line are flagged as ``dust-reddened" 
        quasars. These quasars, which comprise 8.5\% of the raw catalog,
        are excluded from our clean sample.

{\bf Broad absorption line (BAL) quasars} BAL quasars are excluded from the 
        clean sample because the emission line absorption is found to be 
        associated with the continuum absorption in both of the UV/optical and 
        the X-ray bands \citep[e.g.,][]{mat95,bra00}. This property can 
        significantly 
        affect the spectral indices and flux values obtained by fitting 
        photometric data points, which produces inaccurate values of \aox. 
        Most BAL quasars in the parent sample are identified in the BAL quasar 
        catalogs by \citet{tru06} and \citet{gib09}. We found 50 BAL quasars 
        in the parent sample; seven of them are also identified to be red 
        quasars. 

\subsection{The \aoxluvbf\ Relation}
In this section, we compare the dispersions of the \aoxluv\ relation 
between the J07 sample and our cleaned catalog sample. We show that 
the dispersion can be reduced using simultaneous UV and X-ray observations. 
Because the exact values of the dispersion could be method-dependent, we use 
two methods in our analyses. We will present two sets of results based on 
both of the small and large cleaned catalog sample as well as for the 
combined sample.

Because our sample contains censored data, we use the ASURV software package
\citep{lav92} to perform statistical analysis. 
This package includes the Expectation-Maximization (EM; Dempster et 
al.~1977\nocite{dem77}) 
and the Buckley-James (BJ; Buckley \& James~1979\nocite{buc79}) algorithms, 
which we can use to perform linear regression and dispersion estimation. 
The methods differ in that the EM algorithm estimates the residual assuming a
Gaussian distribution while the BJ algorithm assumes the Kaplan-Meier 
distribution. The correlation and regression results are tabulated in 
Table~\ref{tab-correlationregression}. We discuss the results and their
implications below. 

 The clean catalog and the combined samples both exhibit strong 
  correlations between \aox\ and \luv, although the absolute values of 
  the correlation coefficients are slightly lower than the J07 sample.
  The relatively large fraction of undetected objects in the large clean catalog
  sample smears the correlation, but it is more evident in the small clean 
  catalog sample with a higher X-ray detection rate. 

  The slopes ($-0.16\pm0.02$ for the small 
  sample and $-0.15\pm0.01$ for the large sample) are both steeper than the J07 
  sample. The intercepts are also larger, but they are both consistent 
  within \twosig\ uncertainty. The combined sample has a shallower slope than 
  the clean catalog sample; this change is caused by the G10 sample. 
   The possibility that the \aox--\luv\ relation is non-linear is 
   proposed in the study of \citet{wil94} and \citet{and03}.
        As discussed by \citet{wil94}, the difference in slopes is likely caused
        by the varying host galaxy contribution to the \luv\ measurement at 
        low redshift. This may be the reason for the shallower slope of the 
        G10 sample since no host galaxy contribution correction is 
        applied \footnote{We 
        do not correct for host galaxy light in our sample either, but since 
        the majority of our objects 
        are luminous quasars at higher redshift than G10, the host galaxy 
        contamination is much smaller.}.
        \citet{str05} found that their sample does not offer significant 
        evidence for a 
        non-linear \aoxluv\ relation. Although they obtained a shallower
        slope of $-0.09\pm0.02$ for the low luminosity ($\log{L_{\rm UV}}<30.5$)
        and $-0.13\pm0.02$ for the high luminosity ($\log{L_{\rm UV}}>30.5$) 
        sample, they argue that the difference in slopes is likely an artifact 
        of the addition of five outlier AGNs, which are probably X-ray-absorbed
        Seyfert galaxies at $z<0.22$.  From Fig.~\ref{fig-aoxluv}, we do not
        see suspected low luminosity ``outliers", but we still cannot 
        exclude the possibility that the shallower slope is caused by host 
        galaxy contamination. In addition, we also note that 
        the J07 sample contains a number of the most luminous quasars with 
        $\log{L_{\rm UV}}>31.5$, but excluding these high 
        luminosity quasars does not reduce the slope significantly 
        (see regression results of J07T \footnote{The low luminosity sample 
        $\log{L_{2500\mbox{~\AA}}}\leq31$ in J07.} in 
        Table~\ref{tab-correlationregression}).
        A careful removal of host galaxy contribution to the images is 
        probably necessary for further checks, e.g., by using \emph{GalFit} 
        \citep{pen02,pen10}. 

  From Table~\ref{tab-correlationregression}, it is clear that the 
        standard deviation of the clean catalog sample is smaller than the J07 
        sample by about 13\%--19\%. The large clean sample has the largest 
        dispersion (BJ algorithm), which is comparable with the J07 sample, but 
        this large dispersion is mostly caused by a large fraction of 
        undetected X-ray sources. Using the EM algorithm, the J07 sample 
        still has the largest dispersion among all samples.
        The small catalog sample exhibits a dispersion reduced even more 
        by 18\%--25\%, compared with J07. In general, the combined sample has 
        an even smaller intrinsic dispersion because the G10 sample has a 
        higher degree of simultaneity. 

   The dispersion of the \aoxluv\ relation does not exhibit a luminosity
        dependence (see Figure~7 of Just et al. 2007). From 
        Fig.~\ref{fig-aoxluv}, 
        we do not see the dispersion showing any evident dependence on 
        luminosity (at least for X-ray detected quasars), either.
\subsection{Correlations Between UV/optical and X-ray Spectral Indices}
\label{auvax}
Previous studies found that AGNs with bluer optical/UV spectra have relatively 
steeper X-ray spectra \citep{wal93,grupe08}. Using simultaneously observed 
nearby AGNs, \citet{gru10} found a mild correlation between \auv\ and \ax\ 
(See Figure~10 of their paper). In Fig.~\ref{fig-auvax}, we plot \auv\ from 
UVOT photometric data vs. the X-ray spectral slope \ax\ for objects 
from our clean catalog sample (objects whose \ax\ were 
fixed to $-1$ are excluded). 
The Spearman correlation coefficient for the data is 
$\rho_{\rm s}=-0.136\,(P_0=0.125)$, 
which indicates a very weak correlation but the 
value of $P_0$ indicates low confidence level\footnote{The null hypothesis 
here is that the correlation does not exist, so the
lower $P_0$ is, the more confident we feel on this correlation and vice versa.}. 
As claimed by \citet{gru10}, the correlation between \auv\ and \ax\ 
is primarily driven by BLS1 with \ax$>-1.6$. By excluding NLS1 and \ax$<-1.6$ 
spectra objects from our sample, we find that the correlation coefficient 
$\rho_{\rm s}=-0.22$ with $P_0=0.016$, which still does not indicate
a significant correlation.

To investigate whether the lack of this correlation in our sample is due to 
the wider range of redshifts compared to the \citet{gru10} sample, we select the
 low-redshift counterparts from our sample with $z < 0.4$, marked in blue in 
Fig.~\ref{fig-auvax}. These objects
exhibit a stronger correlation, with $\rho_{\rm s}=-0.468\,(P_0=0.058)$. 
Although this subsample contains only 17 objects, we argue that this result 
is expected and due to changes of measured spectral slope in soft and hard 
X-ray bands. In the G10 sample, most AGNs are at low redshift so the rest-frame
X-ray spectra cover both soft and hard X-ray bands. Because this sample is soft 
X-ray selected, objects in this sample usually exhibit strong soft X-ray 
emission with respect to hard X-ray emission, which is caused by the soft 
X-ray excess. The majority of our sample is composed of quasars with much 
higher redshifts, so the rest-frame X-ray spectra cover relatively less 
soft X-ray band. As a result, the measured photon indices are 
mostly based on hard X-ray data. 

We further investigate whether the correlation differences are due to the 
differences between the luminosity range in our sample and that of G10. 
As seen in Fig.~\ref{fig-aoxluv}, the G10 sample is composed mostly of low
luminosity AGNs with a luminosity upper limit of 
$\log{L_{2500\mbox{~\AA}}}\sim30.5$. We select a subsample of 
quasars from the
clean catalog sample with $\log{L_{2500\mbox{~\AA}}}<30.5$ (green diamonds
in Fig.~\ref{fig-auvax}) 
and found that the correlation coefficient is only $0.043$ with a very low 
confidence level $P_0=0.726$. Therefore, our data do not support the argument
that the differences between \auv-\ax\ correlation found by \citet{gru10} and
in our work are due to luminosity differences, but more objects are needed to 
verify this.
\subsection{Correlation between \aox\ and Spectral indices}
Using the low redshift AGN sample, \citet{gru10} found a correlation 
between \aox\ and \ax, i.e., AGNs with softer X-ray spectra tend 
to be X-ray weak relative to UV band, which is consistent with the 
results of \citet{atlee09}. In Fig.~\ref{fig-aoxax}, we plot \aox\ 
vs. \ax\ for 129 objects in our clean catalog sample. 
We do not see a significant correlation with $\rho_{\rm s}=-0.062\,(P_0=0.484)$. 
Similar to the method described in Section~\ref{auvax}, we select a 
subsample with $z<0.4$, which 
is the low redshift counter part of the G10 sample. The \aox\ 
and \ax\ of this subsample exhibit a strong correlation, although this 
subsample contains only 17 objects. 

To see if the lack of correlation is due to redshift, we plot the high 
redshift ($z>1.5$) objects of the catalog sample in red. These objects
are at a different location from the G10 sample with relatively higher
values of \ax. As we argued in Section~\ref{auvax}, the measured value
of spectral slope may depend on the rest-frame energy range. 
Because the observed
energy range is fixed, we do not see the soft X-ray energy spectra for high
redshift quasars and thus photon indices measured for these objects suffer
less from the soft X-ray excess.  

We also display the low luminosity subsample 
($\log{L_{2500\mbox{~\AA}}}<30.5$) using diamonds 
in Fig.~\ref{fig-aoxax}, which contains only eight objects. The subsample
size is too small and we cannot decide if lack of correlation
is related to the observed luminosity range. 

Compared with the positive correlation found in \citet{gru10}, we plot the 
marginal linear correlation found by \citet{you09} in Fig.~\ref{fig-aoxax}, 
which shows a \emph{negative} correlation. The regression by \citet{you09}
in general agrees with our data trend.
The sample in \citet{you09} contains RQ quasars over a redshift range of 
$z=0.11$--$5.41$, which is very similar to our sample. However, 
both the \citet{atlee09} and the G10 sample contain
soft X-ray selected AGNs at low redshift. Because the fixed observed energy 
range, the spectral indices we measured for high redshift quasars represent
the hard X-ray spectral shapes and are less affected by the soft X-ray excess
which produce lower values of \ax. The marginal correlation found in 
\citet{you09} is not seen in our sample is likely because they fit 
their spectral over the 0.5--10~keV band which covers less portion of the
soft energy band and are less vulnerable to the soft X-ray excess. 
Therefore, the correlation found by \citet{gru10} basically implies that
AGNs with softer SEDs over the UV and X-ray bands tend to have 
stronger soft X-ray excess. This is consistent with the argument that
the soft X-ray excess is a tail of the BBB in the EUV band, which originates 
from the thermal emission from the accretion disk. The lack of correlation
in our sample and marginal correlation found by \citet{you09} implies
that the hard X-ray generation process is relatively independent of the
process producing the BBB photons. 

\citet{gru10} also found a correlation between \aox\ and \auv,
which is not seen in our sample. The Spearman correlation coefficient 
is $\rho_{\rm s}=0.166\,(P_0=0.014)$. 
The low redshift subsample ($z<0.4$) contains only nine objects which 
reside at similar location as the G10 sample and appear to exhibit a 
strong correlation with $\rho_{\rm s}=-0.683\,(P_0=0.042)$. However,
the objects in our sample with higher redshifts cover a much wider 
range of \aox\ values. 

The G10 sample is X-ray selected, so this sample will naturally 
include objects with stronger X-ray with respect to UV emission. 
Our sample is optically selected
so it may contain objects with relatively weak X-ray with
respect to UV/optical emission. It is 
clear to see from Fig.~\ref{fig-aoxluv} that our sample contains 
objects with lower values of \aox. If we combine our sample with the 
G10 sample, we see a mild correlation between \aox\ and \auv, but 
with a large dispersion for high \auv\ objects.

\subsection{Correlation Between \lle\ vs. Spectral Slopes}
It has been reported that the Eddington ratio \lle\ is correlated with 
\auv, \ax, and \aox\ \citep[e.g.,][]{gru10}. 
Intuitively, this reflects that the accretion state affects the 
AGN spectral shape. Using the clean catalog sample, we find a mild correlation
between the Eddington ratio and the UV spectral index (Fig.~\ref{fig-lleauv}).
The Spearman correlation coefficient is $\rho_{\rm s}=0.35\,(P_0<10^{-3})$.
The linear regression result is
\begin{equation}
\log{L_{\rm bol}/L_{\rm Edd}}=(1.194\pm0.302)\alpha_{\rm UV}+(0.102\pm0.094)
\end{equation}
In Fig.~\ref{fig-lleauv}, we over plot the linear regression results for
BLS1s (magenta dash dotted line), NLS1s (blue dash dotted line) and both 
(black dotted line). The BLS1 linear relation is clearly more consistent
with our result because most objects in our sample are Type~1 quasars. 
We also distinguish objects in our sample at different redshifts but
we do not see any systematic offset for quasars at different redshifts. 
However, because the Eddington ratios used in Fig.~\ref{fig-lleauv} are 
calculated using the bolometric luminosity under the EXP model, the values of 
\lle\ are not entirely independent of \auv. Thus, an independent measurement
of \lle\ is required to verify this correlation.

Compared to \auv, the value of bolometric luminosity is much less 
dependent on \ax\ (Fig.~\ref{fig-lleax}). The Spearman correlation 
coefficient is $\rho_{\rm s}=-0.36\,(P_0<10^{-3})$ which indicates a 
mild correlation. We perform linear regression to all data points 
in our sample
\begin{equation}
 \log{L_{\rm bol}/L_{\rm Edd}}=(-1.187\pm0.211)\alpha_{\rm x}+(-1.740\pm0.175)
\end{equation}
The slope is generally consistent with that found by 
\citet{she08} ($-0.9\pm0.3$) and shallower than the slope reported by 
\citet{gru10} ($-1.65\pm0.26$).  In Fig.~\ref{fig-lleax}, we distinguish
objects at different redshifts in different colors. Again, 
quasars at low redshift ($z<0.4$) have relatively lower values of \ax,
because of the soft X-ray excess. After excluding these objects
from our clean sample, the correlation coefficient is 
$\rho_{\rm s}=-0.30\,(P_0=0.001)$, and the linear regression slope 
is $-0.88\pm0.61$. 
The enhanced correlation for low redshift
quasars indicates that the accretion state change causes the change of the 
accretion disk temperature. Specifically, higher accretion rate increases 
the disk temperature, which leads to a higher level of soft X-ray excess. 

\citet{gru10} also found a strong correlation between \lle\ and \aox. 
This correlation was not seen in the sample of \citet{she08}.
In Fig.~\ref{fig-lleaox}, we see a very weak correlation between these
two quantities with $\rho_{\rm s}=-0.16\,(P_0<10^{-3})$. By distinguishing
quasars in different redshifts in colors, we notice that the
strong correlation between \lle\ and \aox\ stronger for low redshift
quasars. Because of methods used in sample selection, 
the G10 sample consists of mostly X-ray bright AGNs, thus relatively
higher value of \aox. Because of this bias, many UV/optical bright
and X-ray normal/faint quasars are excluded. After including these
quasars, the correlation becomes much less significant. Even by 
combining G10 sample with our sample, the correlation remains weak
with significant scatter. 
\subsection{X-ray Slope versus Redshift and Luminosity}
We do not find that the X-ray photon index $\Gamma$ has 
any significant correlations
with redshift, UV or X-ray luminosity (Fig.~\ref{fig-xslope}). 
The Spearman correlation coefficients are 
$\rho_{\rm s}=-0.14\,(P_0=0.003)$ for redshift, 
$\rho_{\rm s}=-0.13\,(P_0=0.13)$ for \luv, and
$\rho_{\rm s}=-0.09\,(P_0=0.27)$ for \lx. Adding the G10 sample does not 
make these correlations stronger. This result confirms previous studies
with smaller or comparable samples \citep{pag04,risaliti05,she05,vig05,kel07,you09}.

%% file: sec6.tex
\section{Conclusions}
\label{swiftq-conclusion}
We have compiled an optically selected quasar catalog with serendipitous 
and simultaneous UV/optical and X-ray observations with the \swift] observatory. 
The catalog is generated by matching the $\sim3.5$~year \swift\ pointings 
from November, 2004 to June, 2008 to the SDSS DR5 quasar catalog. 
For each object, the sky images observed by either UVOT or XRT are
carefully selected to ensure high image quality. We derive
the composite UVOT photometry and XRT energy spectra by stacking all
archival data to generate the deepest sky images. The resultant 
SEDs reflect the time-averaged shape of quasar emission with simultaneous
observations at multi-wavebands. 
The catalog contains 843 objects. 
There are 637 objects ($\sim76\%$) that have 
UVOT and XRT observations, 168 objects ($\sim20\%$) that only
have XRT data and 38 objects that only have UVOT data $(\sim4\%)$. 
Among all the 675 objects with X-ray coverage, 460
($\sim60\%$) are detected, which rises to $85\%$ amongst source with at
least 10~ks of XRT exposure time. We construct SEDs for all objects 
with both XRT and UVOT data. In this work, we focus on 637 objects 
with both X-ray and UV observations. We supplement UVOT photometry with
SDSS and 2MASS data if available. All the photometric points are corrected 
for the effects of emission lines, and fluxes from SDSS and 2MASS are shifted 
to match the flux levels of \swift\ UVOT data. 

We fit SEDs using the EXP and TPL models, 
attempting to constrain the flux contribution from the BBB. 
In most cases, the EXP models create a bump in the EUV region, producing
an upper limit on the BBB emission, while the TPL model connects
the SED points at 1216~\AA\ and 0.3~keV, producing a lower limit on
the BBB emission. After correcting for the contribution from IR emission,
the TPL model produces bolometric luminosities consistent with 
those estimated using BCs from composite quasar SEDs, while the EXP model 
produces bolometric luminosities on average 0.3~dex higher than the TPL model. 

We identify two clean samples (large and small) selected from our catalog,
and supplement each sample with 88 nearby AGNs from \citet{gru10}. 
We re-visit the \aoxluv\ relation presented by \citet{jus07}. 
We use the EM and Buckley-James methods to compare the 
intrinsic scatters of the \aoxluv\ relationship of our and the J07 sample.
These two methods consistently indicate that the dispersion based on 
our sample is reduced compared to J07 by 13\% to 19\% using 
the cleaned catalog sample and $18\%$ to 25\% using the 
combined sample. 

Firmly establishing the \aoxluv\ relation in
AGNs is an important step toward understanding energy
generation mechanisms of AGNs. Our work has verified the correlation
again, and has shown that the correlation is even tighter after
reducing or eliminating scatter due to variability. An additional
source of scatter in the correlation could be due to an intrinsic
\aoxluv\ relation, which may have a different slope for each quasar.
In Fig.~7 of \citet{vas09}, they present simultaneously and non-simultaneously
observed SEDs in the UV and X-ray bands. The dramatic difference 
between two SEDs for three of these objects indicates the
variation of \aox\ with time for a given AGN. This variation
then reflects the change of accretion state of the central 
engine and a different slope for an intrinsic and for a global relation would 
produce scatter around the global relation. In order to 
see the intrinsic \aox\ variation with \luv\ for an given AGN/quasar, 
it is necessary to gather long-term simultanous observational data.
In our fugure work, we will perform time-resolved UV/X-ray data anlayses 
to selected targets from our catalog which was observed with high 
cadence, which will determine the contribution to the scatter of the 
global \aoxluv\ relation from the intrinsic variations of X-ray with
respective to UV emissions.

We also investigate correlations between spectral shapes in 
different wavebands and \lle, and compare the results found in 
G10. Our low redshift ($z<0.4$) counterparts
to the G10 sample verify significant correlations
exist between \auv\ and \ax, \aox\ and \ax, \aox\ and \auv,
\lle\ and \auv\, \lle\ and \ax, \lle\ and \aox,
which physically implies that the accretion status plays a fundamental
role in shaping the quasar SED between UV and soft X-ray band. 
This supports the argument that the BBB is produced by the disk emission
and the soft X-ray excess is a result of thermal emission from accretion 
disk. However, for high redshift quasars, 
the measurement of X-ray spectral slope covers less soft energy 
band and is less affected by the soft X-ray excess. As a result, the 
correlations between spectral shapes are much weaker after including 
high redshift quasars. 
This implies that the hard X-ray emission is relatively independent
of the thermal emission on the accretion disk. 


%% file: app.tex
\section{Comments on Individual Objects\label{app}}
The following sources possess 
special spectroscopic or photometric features or have unusual
classifications.

\emph{SDSSJ021702.66$-$082052.3}. This is a flat spectrum radio 
        quasar (FSRQ) cataloged by \citet{mas09}. The optical 
        spectrum exhibits a featureless continuum with very weak \mgiifull\ 
        and \hb\ emission lines and a few narrow lines such as \oiiidoublet. 

\emph{SDSSJ074110.70$+$311200.2}. This is an FSRQ \citep{liu02,hea07}.

\emph{SDSSJ074625.86$+$254902.1}. This is a high redshift FSRQ 
        \citep{mas09} 
        with $z=2.979$ and an MeV blazar discovered by \swift\
        \citep{sam06,tue08,jol09} and thus in the
        BAT-selected AGN catalog by \citet{win09}. It is also
        observed by \emph{Suzaku} \citep{wat09}. The UV spectrum contains
        broad emission lines such as \lya, \civfull\ and \ciiiuvfull, which is
       unusual for a blazar.

\emph{SDSSJ081331.28$+$254503.0}. This is a confirmed lensed quasar 
          \citep{len08,con10,ina10}.

\emph{SDSSJ083148.87$+$042939.0}. This is an FSRQ with peak radio 
         flux of $\sim1$~Jy. Its optical spectrum exhibits a featureless 
         continuum with an extremely weak \ha\ emission line.

\emph{SDSSJ083740.24$+$245423.1}. This object is a blazar and classified 
        as an FSRQ by \citet{hea07}. The UV spectrum exhibits strong
        \mgii\ and \ciiiuv\ emission lines. 

\emph{SDSSJ090821.01$+$045059.4}.
         The \mgii\ emission line at $\sim$2798~\AA\ is barely detected,
         \hbeta\ is extremely weak and the two nearby \oiii\ lines 
        are quite prominent, indicating that the BLR is obscured and only the 
         NLR is seen. This object has been identified as a strong radio 
         source \citep[e.g.,][]{gri95} but the exact classification is not 
         yet determined. 

\emph{SDSSJ092703.01$+$390220.8}. This is an FSRQ with a 
         strong radio jet \citep{liu02,hea07}. The UV / optical spectrum 
         also contains prominent
         \mgiifull, \hbeta\ and \hgamma\ emission lines. 

\emph{SDSSJ094215.12$+$090015.8}. This is quasar with double-peaked
         \halpha. The FWHM of the broad component is extremely wide and reaches
         $\sim40,600$~\kms, which is the broadest known \citep{wan05}.
         The broad \halpha\ suggests that the emission region is 
        close to the black hole $r~\sim80$--$100\,r_{\rm g}$ \citep{str06}.

\emph{SDSSJ101405.89$+$000620.3}. This quasar is classified as a 
        Seyfert~1.8 by \citet{don05}. The optical spectrum exhibits a strong and
        double-peaked \halpha\ line but a weak \hbeta\ line. The excess 
        emission over the power-law around these two lines is likely to be 
        contributed by host galaxy light.

\emph{SDSSJ101541.14$+$594445.2}. This is a RL quasar exhibiting 
         two pairs of radio lobes in an X-shape \citep{che07}.  
        The origin of the X-shape wings in this radio source is unclear.   

\emph{SDSSJ101810.98$+$354239.4}. This is an FSRQ \citep{hea07}. 

\emph{SDSSJ102738.53$+$605016.5}.
         The optical spectrum of this object exhibits extremely broad and
        double-peaked \hbeta, and is cataloged as a double-peaked 
        emission line quasar by \citet{wux04}. They measured the FWHM of \hbeta\
        of $\sim16,200$~\kms\ and a black hole mass of 
        $\log{(M_{\rm BH}/M_\odot)}=9.649$.

\emph{SDSSJ103303.70$+$411606.2}. This is an FSRQ \citep{hea07}.

\emph{SDSSJ121826.51$+$294846.5}. This object, also known as Mkn~766 and
        NGC~4253, is a local Seyfert~1.5 galaxy ($z=0.013$) and is resolved 
        in the UVOT image. Because of the strong host galaxy contamination, 
       it is difficult to isolate the AGN component. It is not included in 
       our UVOT processing list. The X-ray spectrum is complicated; 
      it cannot be fit by any model we described in Table~\ref{tab-models}. 
        This object was previously studied in detail in the \swift\ AGN 
        catalog by \citet{turner06,turner07,gru10}.

\emph{SDSSJ135516.54$+$561244.7}. This is a typical NLS1 
       \citep{zho06b,grupe99,gru10}.

\emph{SDSSJ141927.49$+$044513.8}.
         This object has a featureless UV spectrum with a few narrow absorption
        lines. It is a BL Lac object in the catalogs of \citet{col05},
        \citet{plo08} and \citet{mas09}.
        The narrow absorption lines are most likely intervening.

\emph{SDSSJ142921.87$+$540611.2}. This object is classified as 
        an FSRQ by \citet{hea07}, and \citet{mas09}, and listed as a 
        lensed quasar candidate by \citet{kin99}. \citet{bro03} rejected 
        the lensing hypothesis based upon surface brightness and spectral 
          indices criteria. 

\emph{SDSSJ154929.43$+$023701.1}. This object is classified as 
        an FSRQ \citep{hea07} with high polarization \citep{sca97}. The optical
        spectrum from SDSS, however, contains strong \mgiifull, \hbeta, 
        and even
        \hgamma\ lines. Even some weak forbidden lines are prominent, e.g., 
        the two [Ne~{\sc v}] lines around 3400~\AA. This is another
        case in which a blazar has a regular broad line quasar spectrum. 

\emph{SDSSJ161742.53$+$322234.3}. This is the strong ratio quasar
          3C~332. The \hbeta\ line is relatively weak, but the \halpha\ 
          line exhibits a very prominent double peak structure with a FWHM of
          $19,600$~\kms\ measured by \citet{str06} and $23,200$~\kms\ measured
          by \citet{wux04}. The black hole mass is large,
          $\log{(M_{\rm BH}/M_\odot)}=9.334$ \citep{wux04}. The \halpha\ line
          also displays long-term profile variability which can be explained by 
          a low, smooth, secular change in disk illumination \citep{gez07}.

\emph{SDSSJ162901.30$+$400759.9}. This object is classified as a blazar
        by \citet{mas09}, an FSRQ by \citet{fal04}, and a NLS1 galaxy by 
        \citet{bad95,grupe04,kom06}.
        The optical spectrum exhibits strong \halpha, \hbeta, and a complex of
        low ionization \feii\ emission. 

\emph{SDSSJ170231.06$+$324719.6}. This is a typical NLS1 
         \citep{grupe04,zho06b,gru10}.

\emph{SDSSJ213638.58$+$004154.1}. This $z=1.9414$ FSRQ has a strong 
        radio jet \citep{liu02,hea07,mas09}. The UV spectrum contains strong 
        \mgiifull, \civfull, and \ciiiuvfull\ emission lines.

%% file: fig.tex
\begin{figure}
\centering
\includegraphics[width=10cm,angle=90]{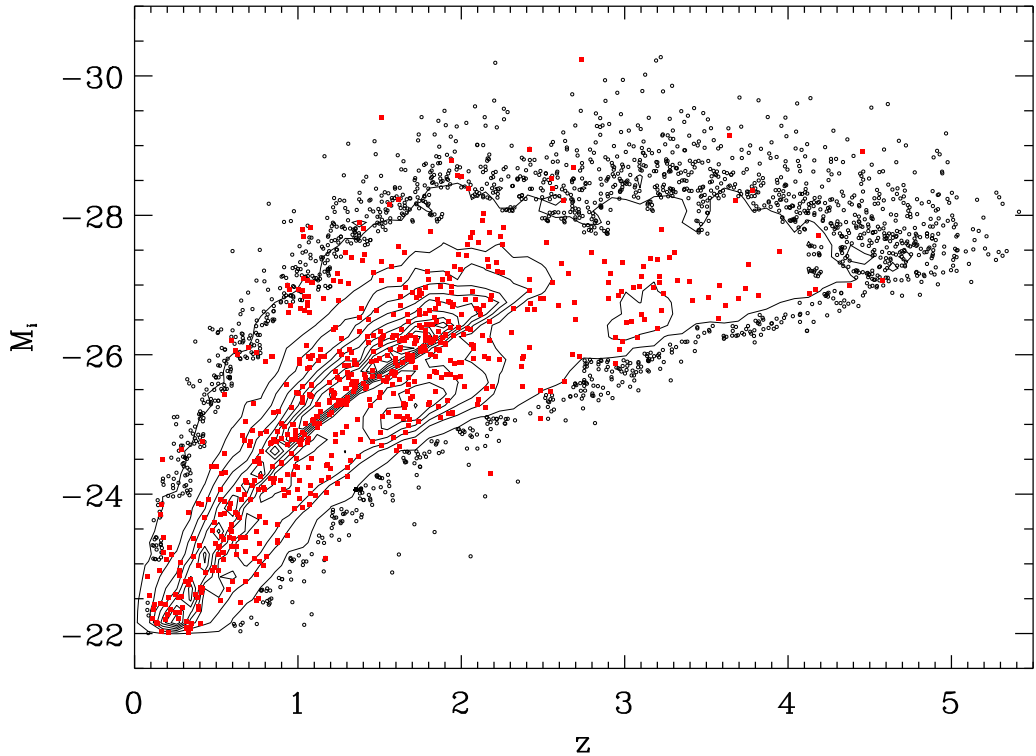}
\caption[Absolute magnitude-redshift diagram of raw catalog 
        quasars]{Distribution of our sample in the redshift vs. SDSS $i$
        band absolute magnitude diagram. The SDSS DR5 quasar catalog 
        objects are represented by open circles. Their distribution is
         represented by a set of linear contours when the density of open
         circles in this two-dimensional space exceeds a certain threshold and
         the plot symbols begin to overlap. Objects in the \emph{Swift} quasar catalog
         are represented by red filled squares. The lower luminosity limit 
         occurs because the SDSS DR5 quasar catalog includes only quasars 
         more luminous than $M_i=22.0$. 
         \label{fig-miz}}
\end{figure}
\clearpage

\begin{landscape}
\begin{figure}
  \centering
  \begin{tabular}{cc}
      \vspace{3mm}
      \includegraphics[width=9cm,angle=0]{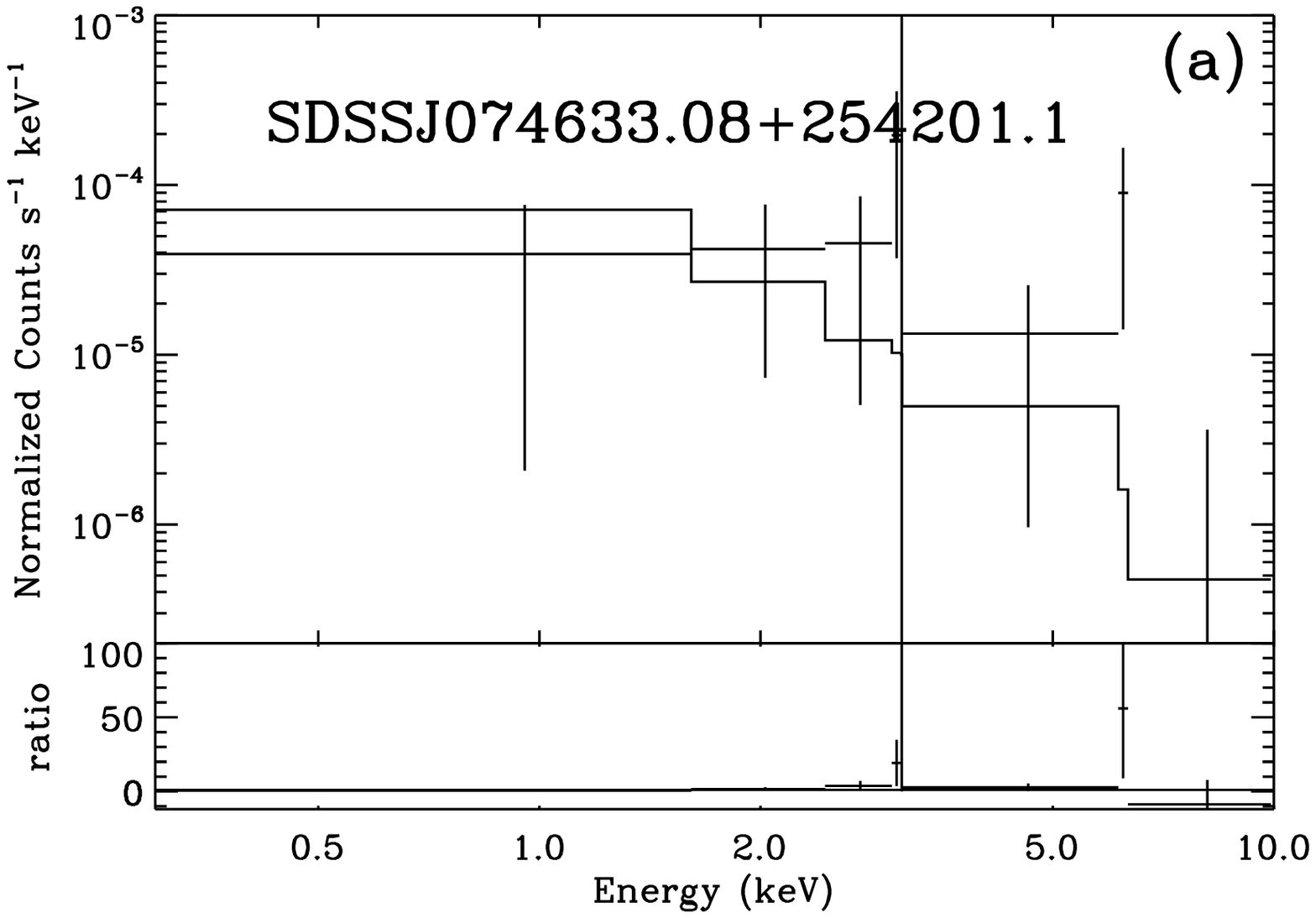}&
      \includegraphics[width=9cm,angle=0]{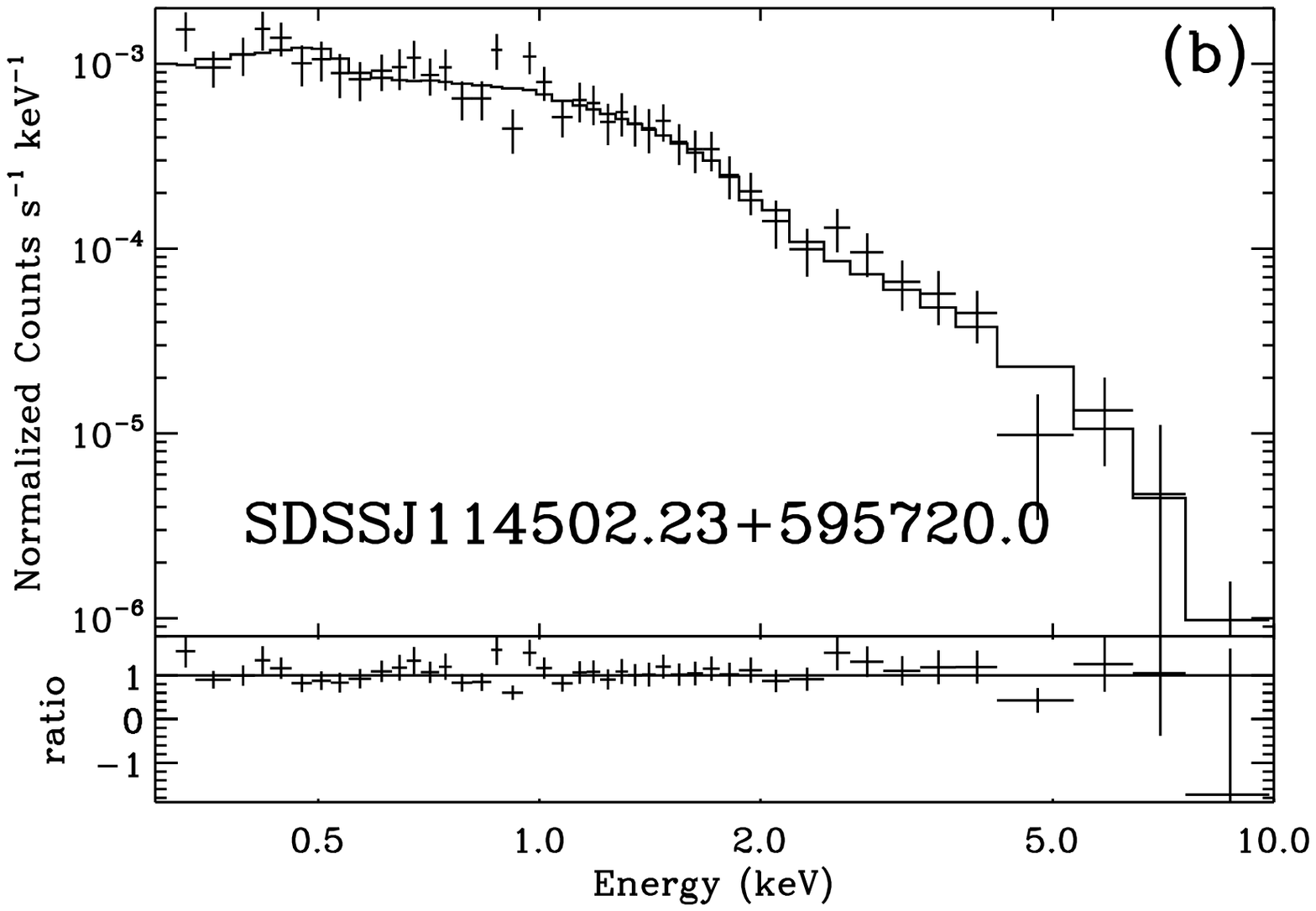}\\
    $N_{\rm Xph}=60$, $\alpha_{\rm x}=-1$ & $N_{\rm Xph}=971$, $\alpha_{\rm x}=-1.268_{-0.063}^{+ 0.065}$\\
      \vspace{3mm}
      \includegraphics[width=9cm,angle=0]{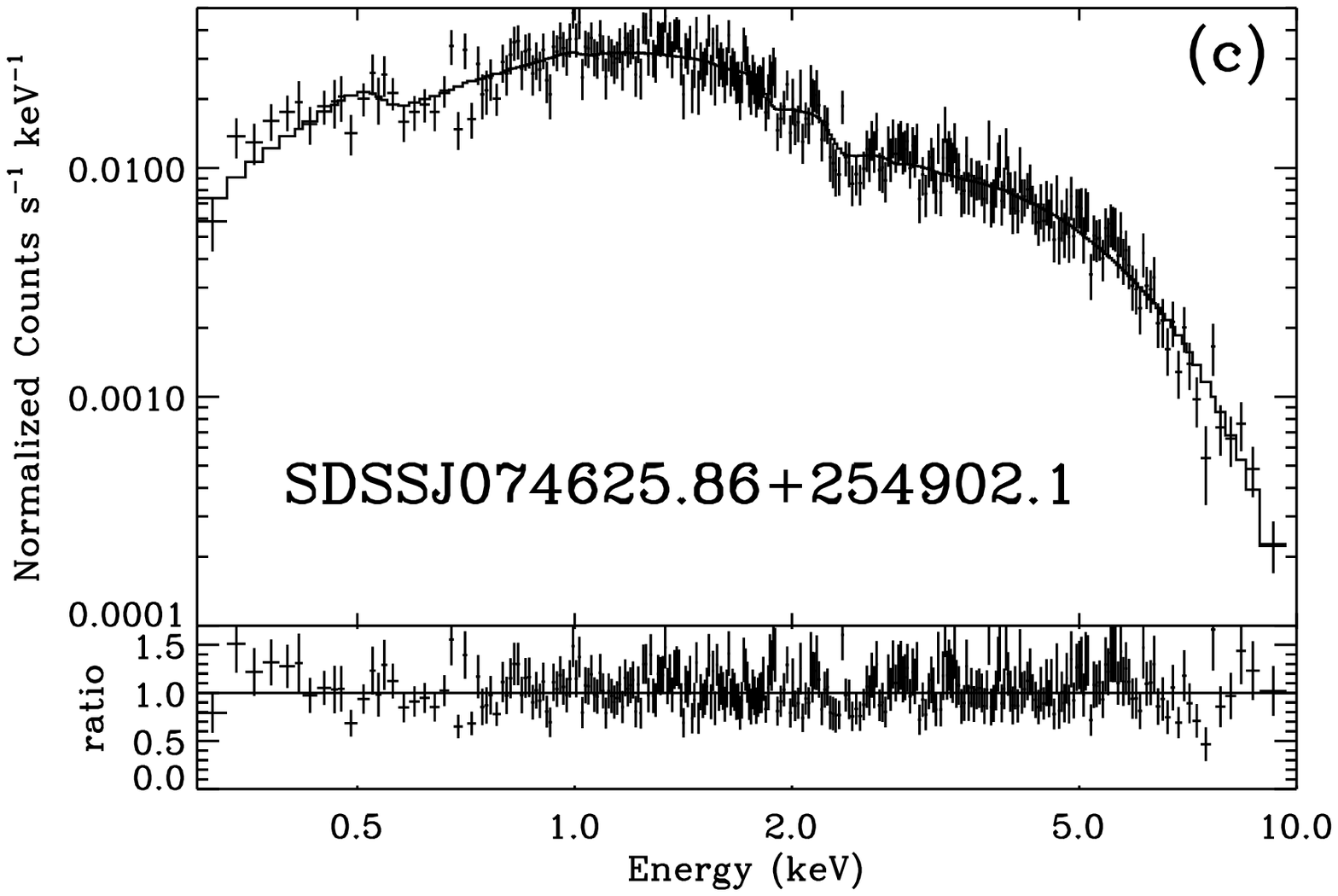}&
      \includegraphics[width=9cm,angle=0]{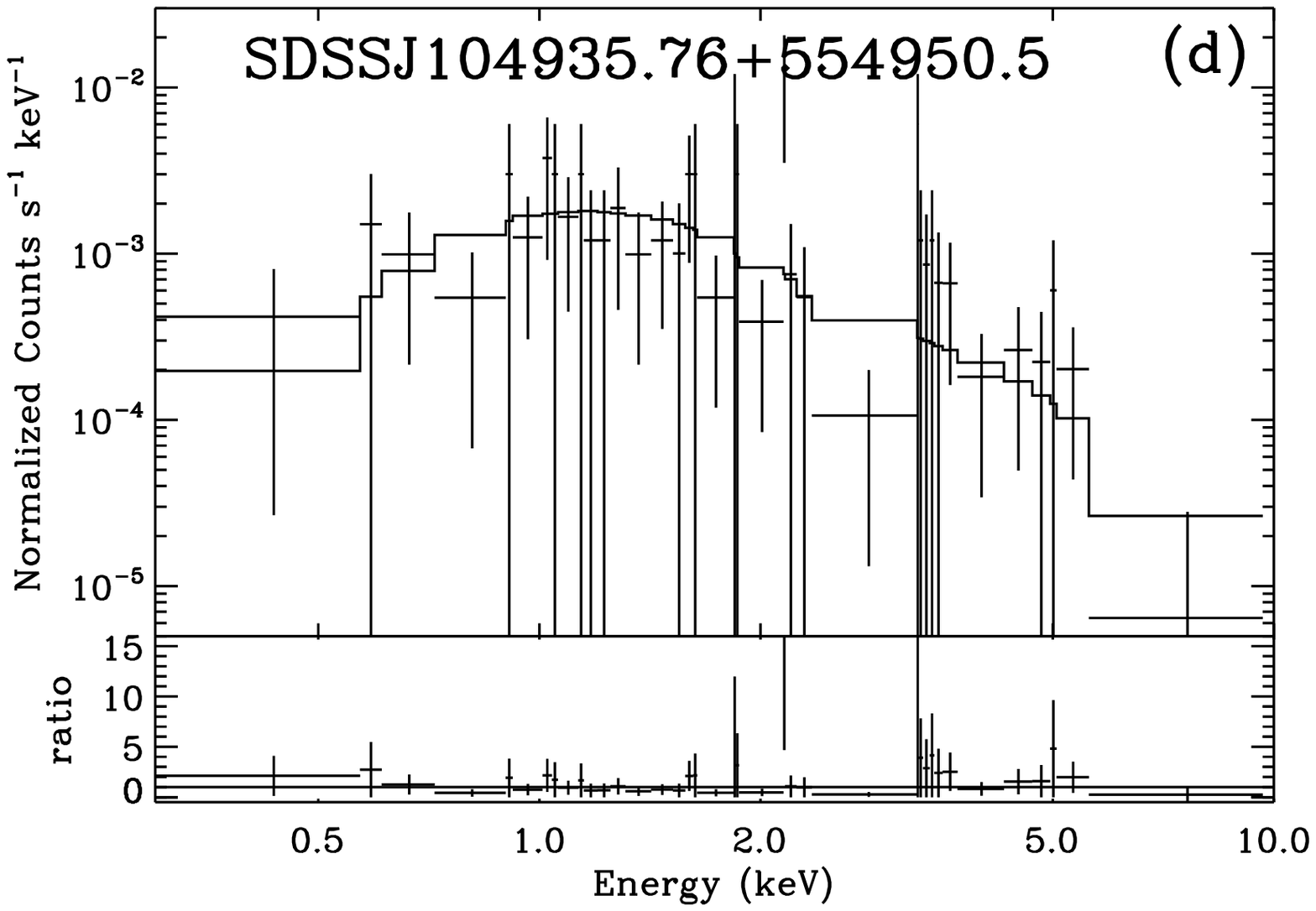}\\
    $N_{\rm Xph}=8258$,$\alpha_{\rm x}=-0.216_{-0.022}^{+ 0.022}$, $N_{\rm H,i}=0.3447^{+1.449}_{-1.374}$&
    $N_{\rm Xph}=58$, $\alpha_{\rm x}=-1$,$N_{\rm H,i}=1.417^{+0.634}_{-0.485}$\\
  \end{tabular}
    \caption[XRT spectra fitting examples]{Four examples of fits to the 
    observed XRT spectra. Intrinsic column densities  ($N_{\rm H,i}$)
             are in units of $10^{22}$~\psqcm. Spectra shown in (b) and 
            (c) are binned as listed in Table~\ref{t-xrtbin}. Spectra shown 
            in (a) and (d) are \emph{rebinned} using the \xspec\ command {\tt 
            setplot rebin}
            (see Section~\ref{initialsedplots})\label{fig-xrtexample}.}
\end{figure}
\end{landscape}
\clearpage

\begin{landscape}
\begin{figure}
  \centering
  \begin{tabular}{cc}
      \vspace{5mm}
      \includegraphics[width=6cm,angle=90]{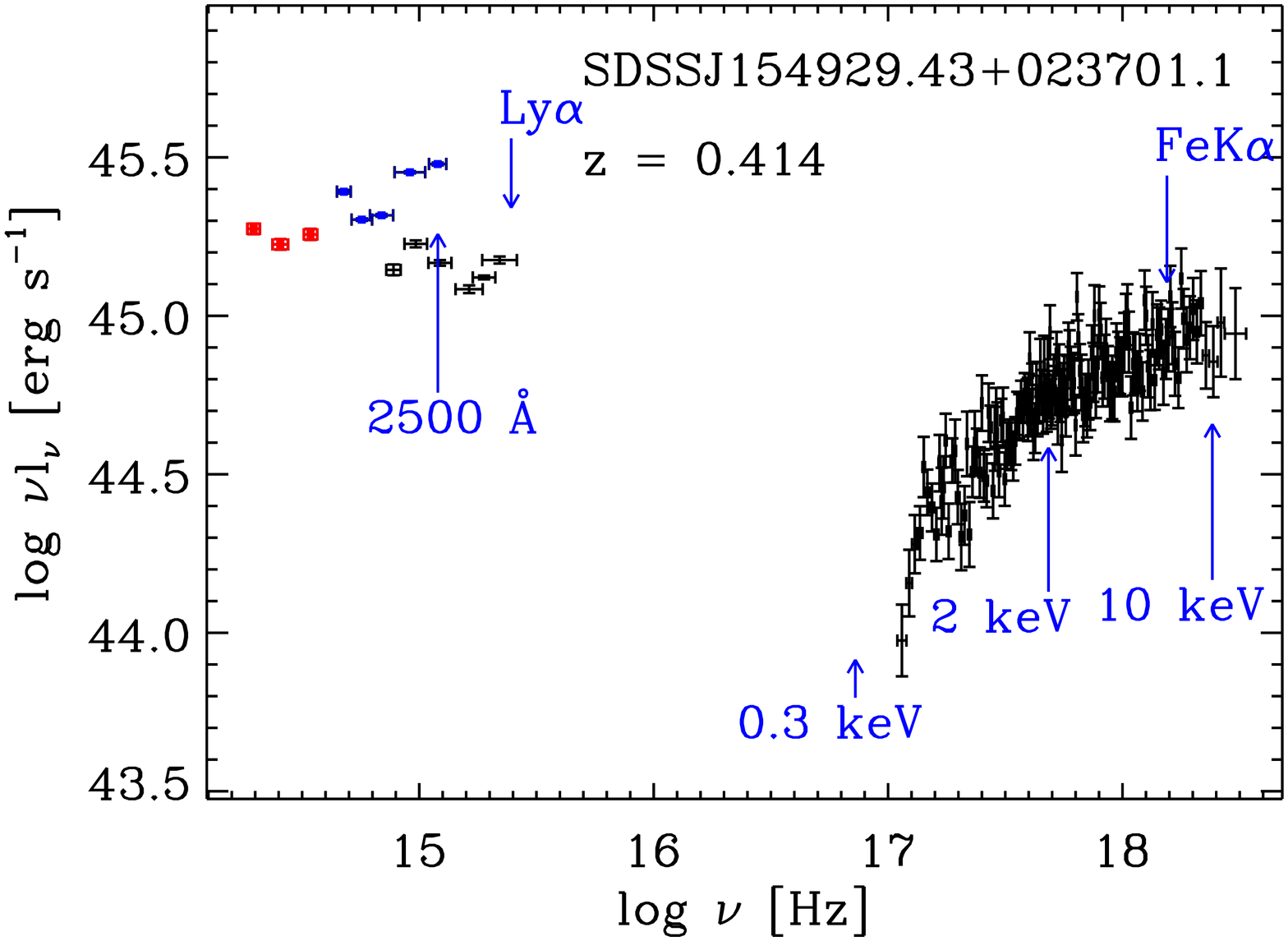} \space{\ \ \ }
      \includegraphics[width=6cm,angle=90]{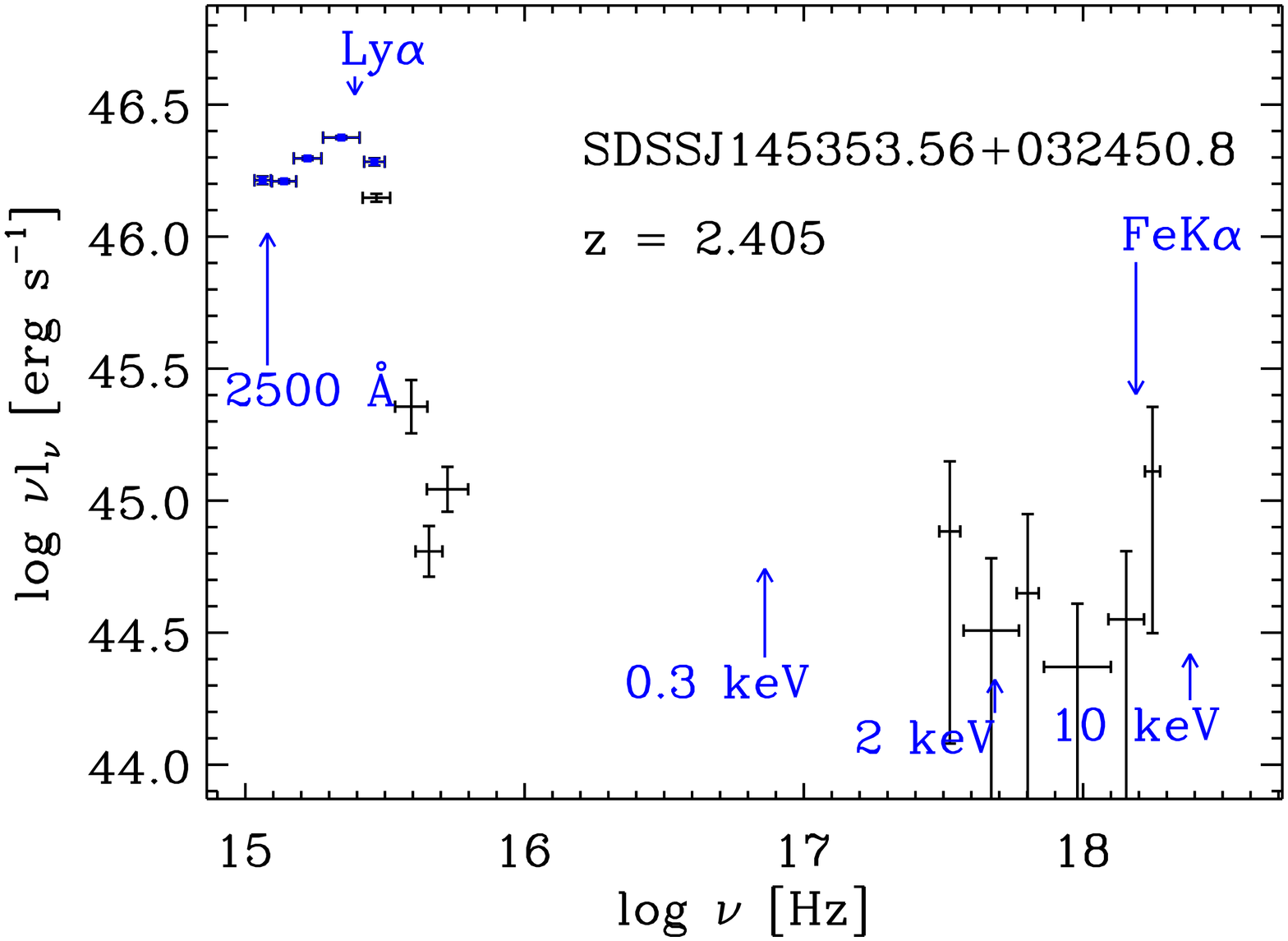}\\
      \vspace{5mm}
      \includegraphics[width=6cm,angle=90]{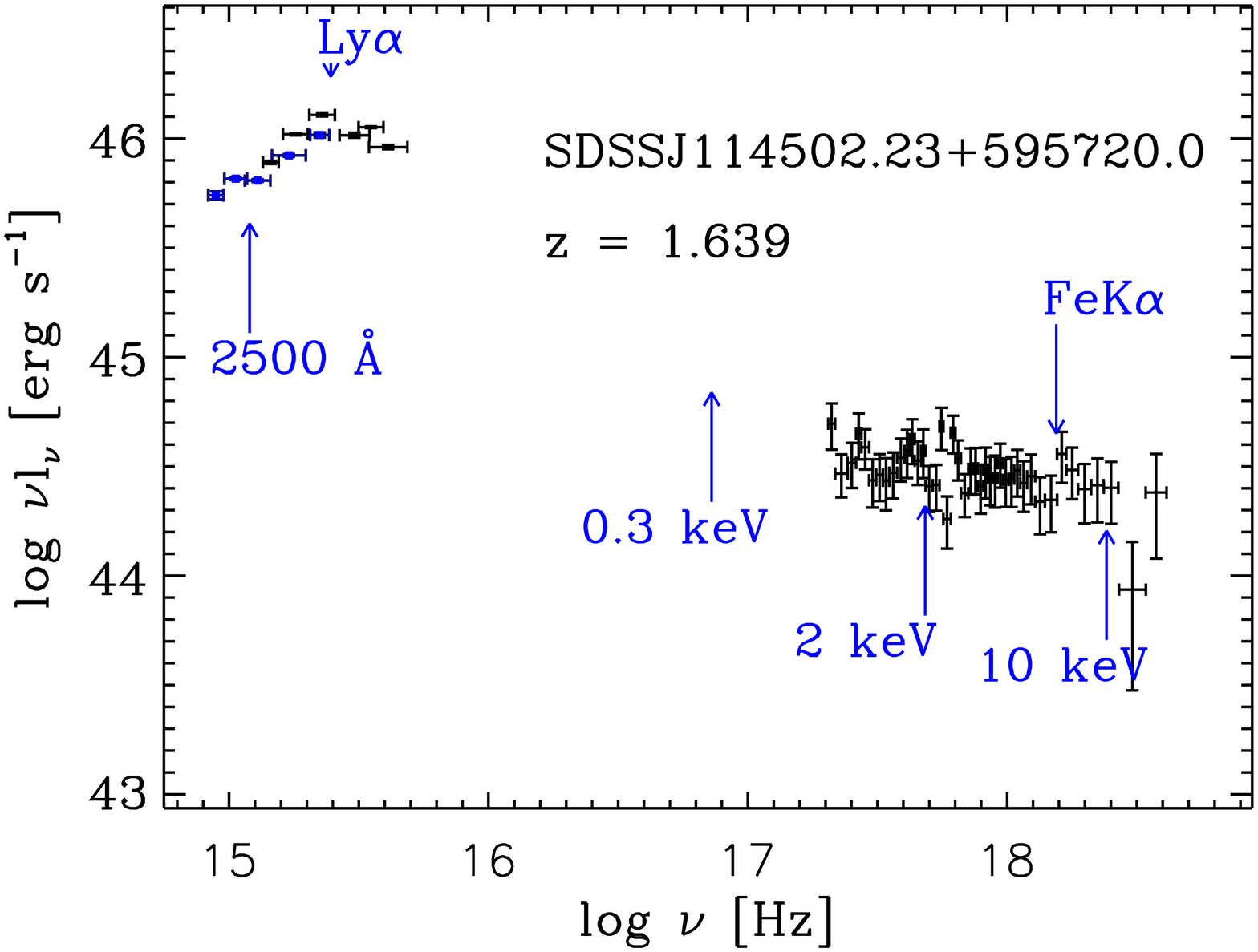} \space{\ \ \ }
      \includegraphics[width=6cm,angle=90]{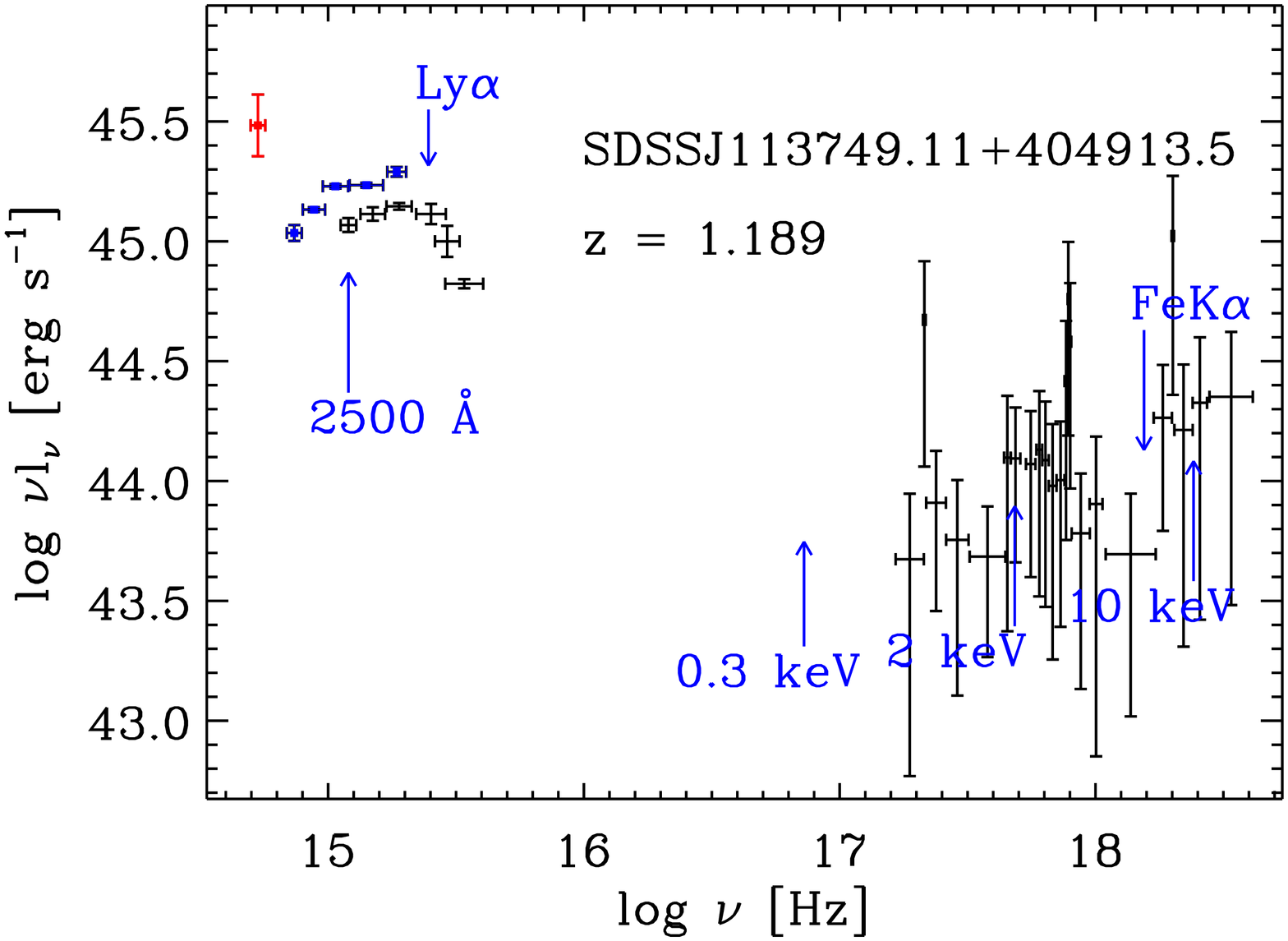}
  \end{tabular}
    \caption[Examples of initial SEDs]{\label{fig-isedexample}Examples of initial 
SEDs, showing data from \swift\ UVOT and XRT (black), SDSS (blue) and 2MASS (red). 
There are clear flux offsets between SDSS and UVOT measurements.}
\end{figure}
\end{landscape}
\clearpage

\begin{figure}
  \centering
    \begin{tabular}{c}
    \vspace{5mm}
      \includegraphics[width=7cm,angle=90]{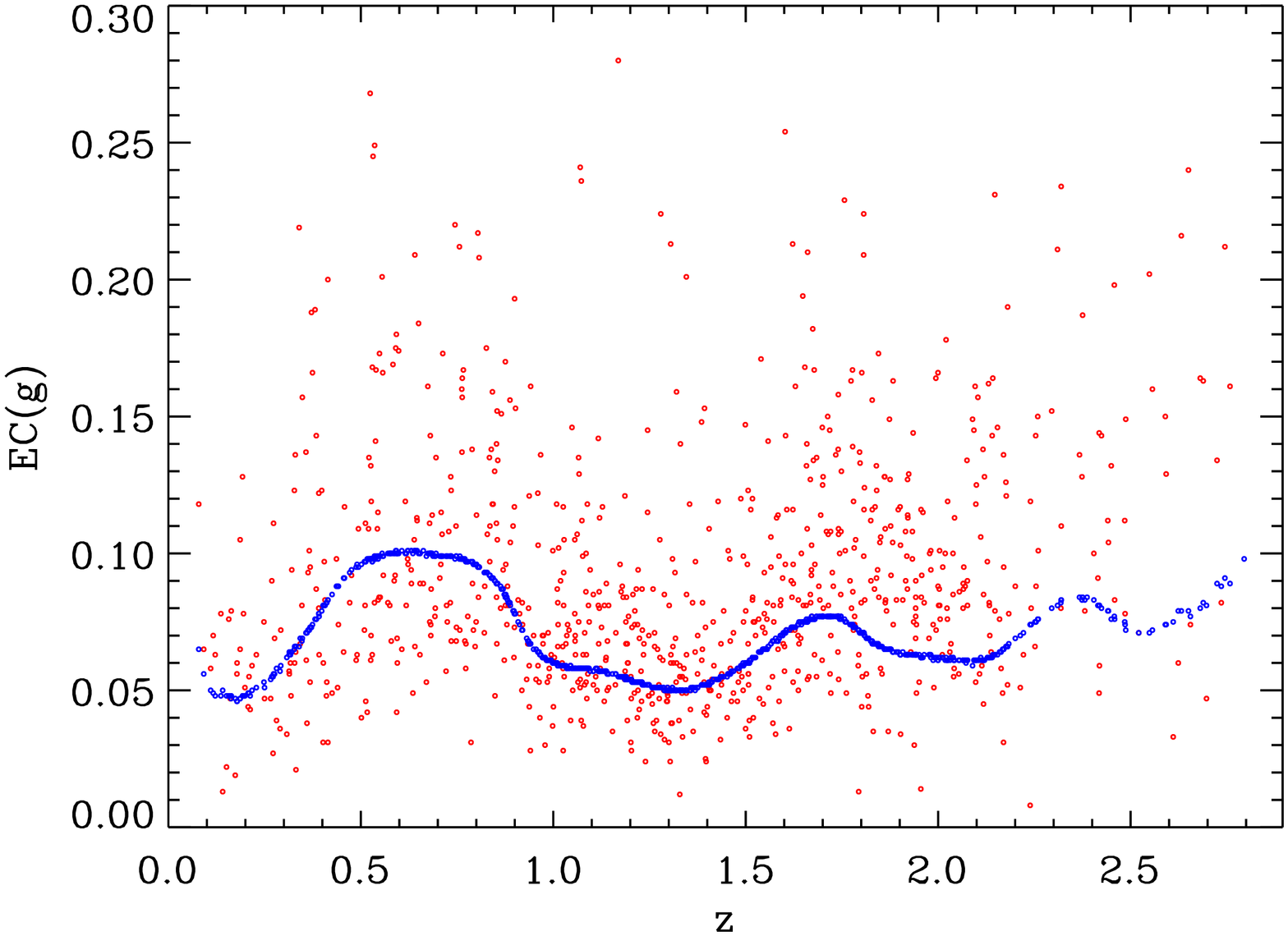}\\
    \vspace{5mm}
      \includegraphics[width=7cm,angle=90]{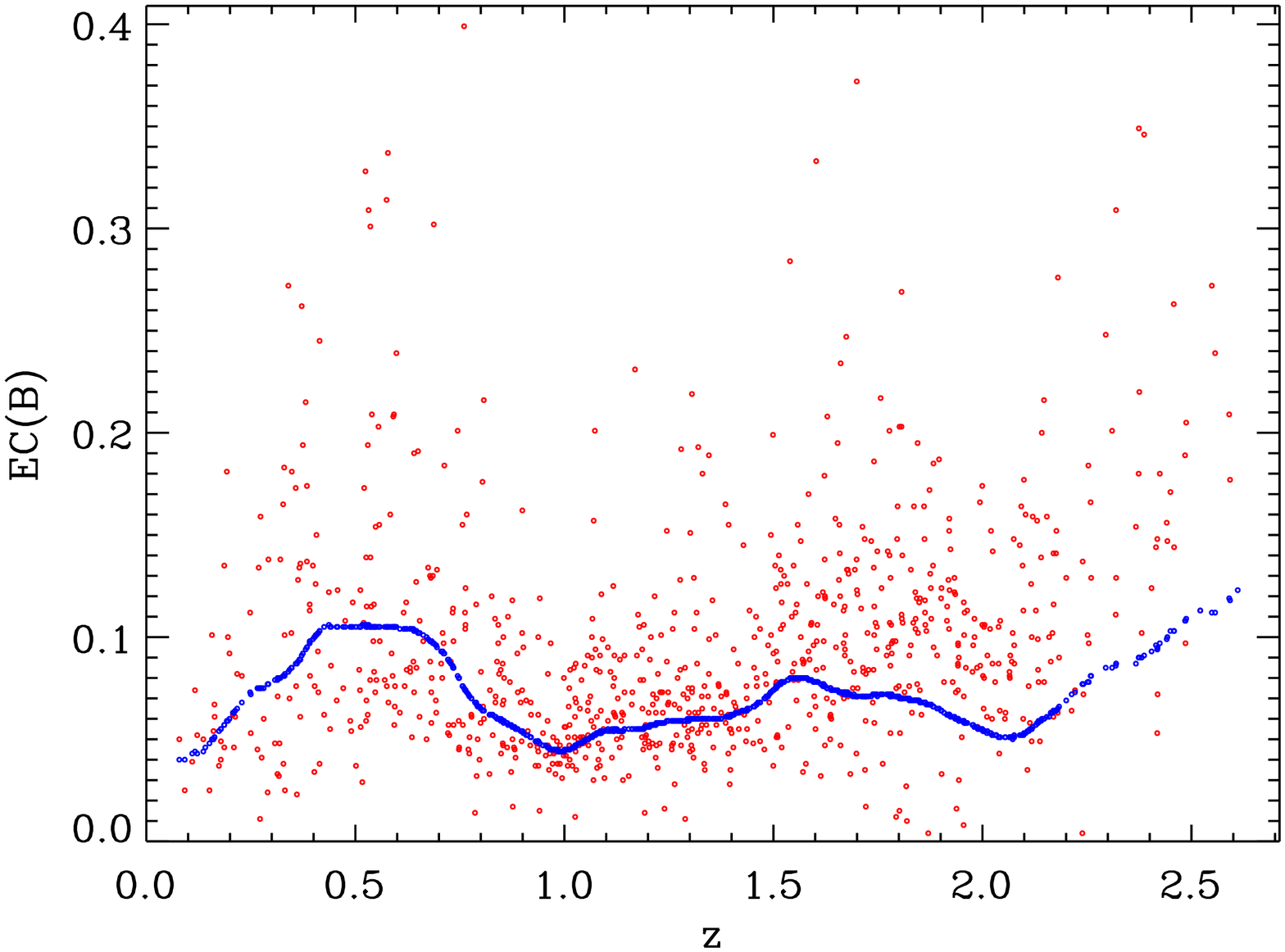}
    \end{tabular}
    \caption[Emission line corrections as a function of redshift]{Emission 
    line corrections (EC) of the SDSS \emph{g} band (\emph{Upper} panel)
    and the UVOT B band (\emph{lower} panel) as a function of redshift for different filters.
    Red circles are EC performed on real spectra. 
    Blue circles are EC performed on the composite spectrum by \citet{van01}
    shifted to the real spectrum redshift.
    The EC represented by blue circles varies with redshift as 
    emission lines are shifted within the coverage of a filter. 
    These plots indicate that the ECs based on real spectra are generally
    in agreement with ECs based on the composite spectrum. \label{fig-elctrend}}.
\end{figure}
\clearpage

\begin{figure}
  \centering
    \begin{tabular}{c}
      \vspace{5mm}
      \includegraphics[width=7cm,angle=90]{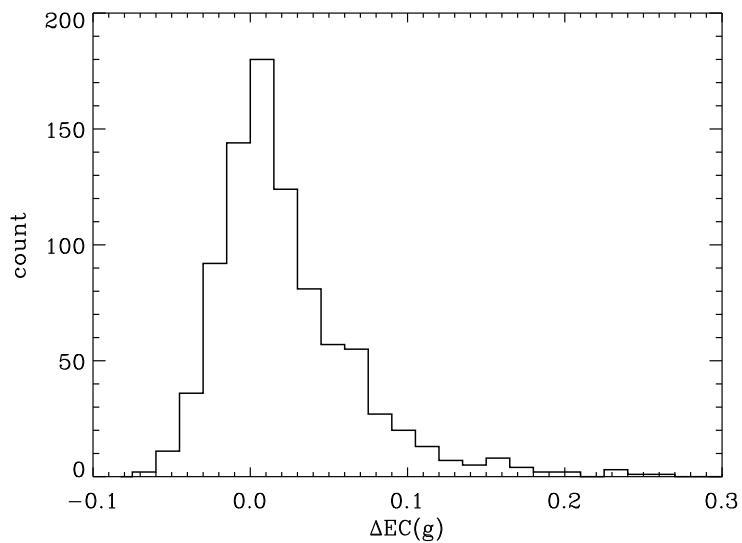}\\
      \vspace{5mm}
      \includegraphics[width=7cm,angle=90]{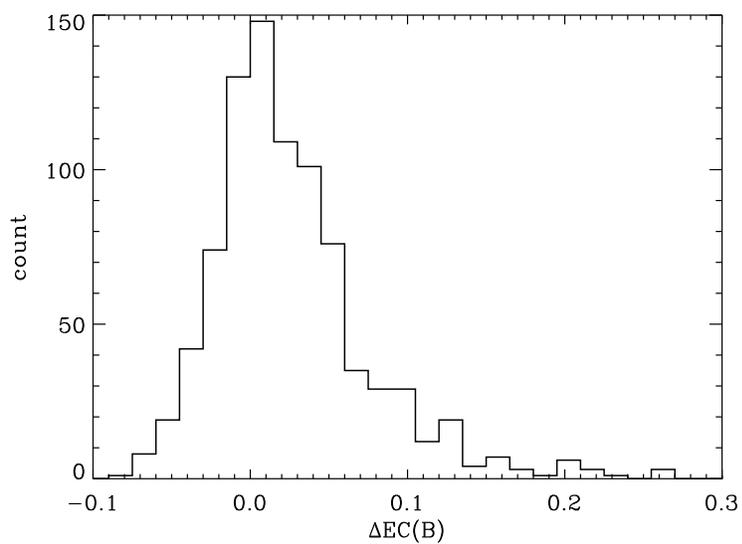}
    \end{tabular}
    \caption[Distributions of emission line correction differences]{Distributions of emission line correction differences 
    $\Delta\mbox{EC}=\mbox{EC}_{\rm real}-\mbox{EC}_{\rm composite}$ in the
    SDSS \emph{g} band and the UVOT B band
    based on the results shown in Figure~\ref{fig-elctrend}. 
    These figures illustrate that we can in general obtain consistent 
    ECs based on real and composite spectra at different bands. The
    dispersion of their difference is typically 0.05. 
    \label{fig-elcdiff}}
\end{figure}
\clearpage

\begin{figure}
  \centering
   \vspace{2cm}
      \begin{tabular}{c}
      \vspace{5mm}
      \includegraphics[width=7cm,angle=90]{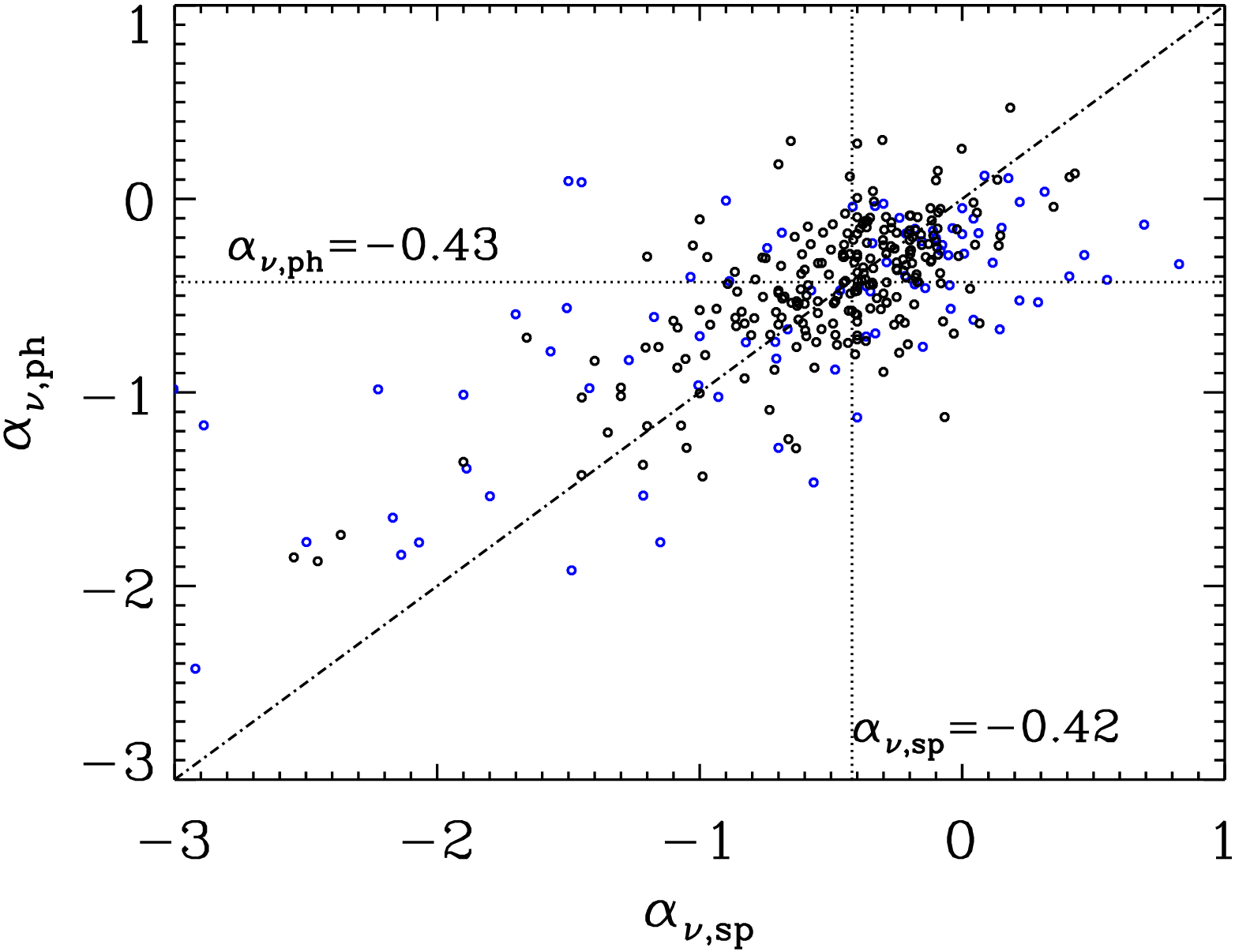}\\
      \vspace{5mm}
      \includegraphics[width=7cm,angle=90]{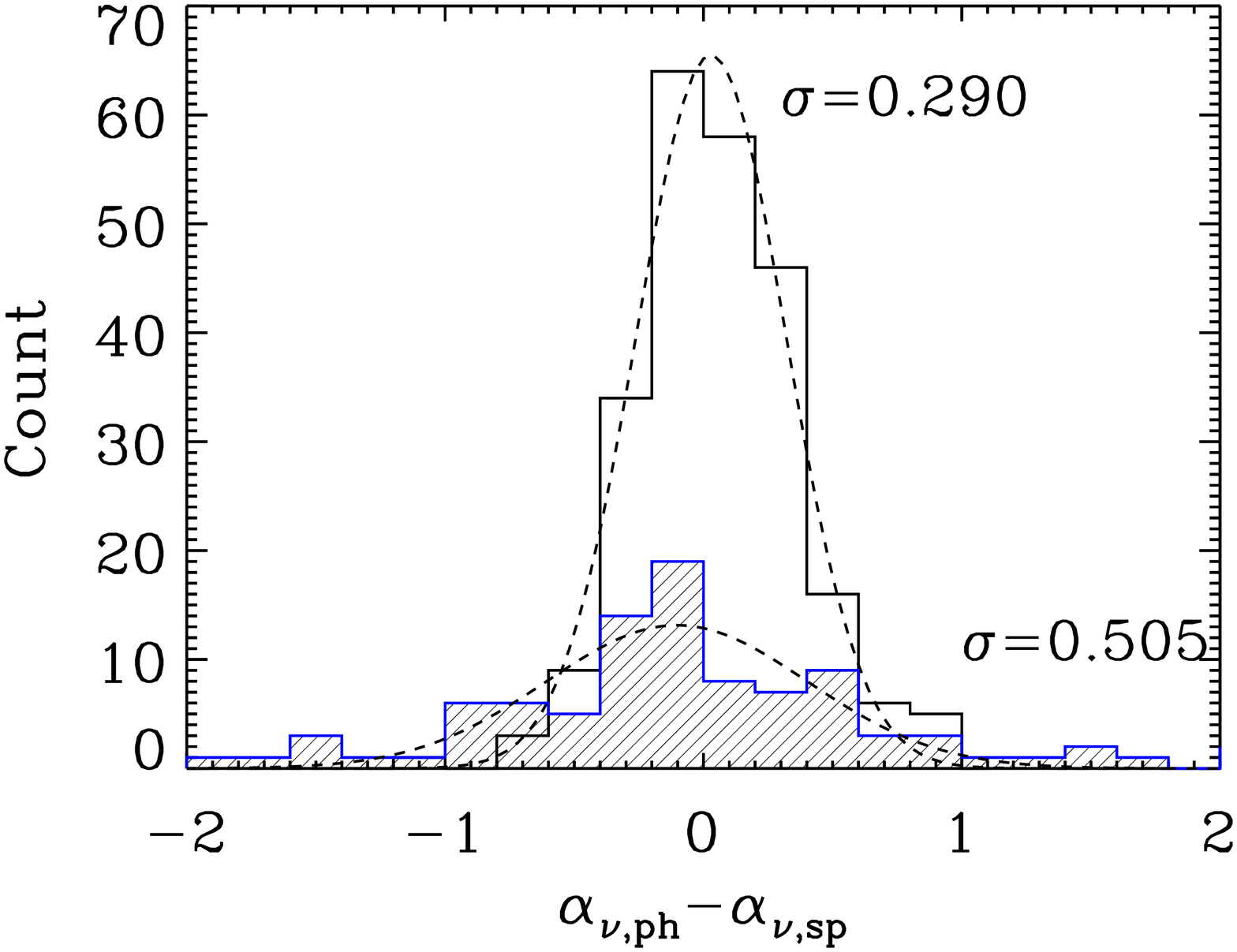}
     \end{tabular}
   \caption[Photometric UV slope vs. spectroscopic UV slope]{
    \emph{Upper: }photometric UV slope $\alpha_{\nu,{\rm ph}}$ vs. 
         spectroscopic UV slope $\alpha_{\nu,{\rm sp}}$. Low-redshift objects
         ($z<0.8$) are in blue and objects at higher redshifts ($z>0.8$) 
         are in black. The dash dotted line is 
         $\alpha_{\nu,{\rm ph}}=\alpha_{\nu,{\rm sp}}$. The dotted lines
         represent the median values of $\alpha_{\nu,{\rm ph}}=0.43$ and 
         $\alpha_{\nu,{\rm sp}}=0.42$, respectively. 
   \emph{Lower: } Distribution of differences between $\alpha_{\nu,{\rm ph}}$
         and $\alpha_{\nu,{\rm sp}}$ for low-redshift ($z<0.8$, in blue) 
        and high-redshift ($z>0.8$, in black) objects. We also fit
        these two histograms with Gaussian profiles. The low redshift
        sample has a dispersion ($\sigma\approx0.5$) much larger than the
        higher redshift sample dispersion ($\sigma\approx0.3$).
        \label{fig-auvphauvsp}}.
\end{figure} 
\clearpage

\begin{figure}
\centering
\includegraphics[width=13cm]{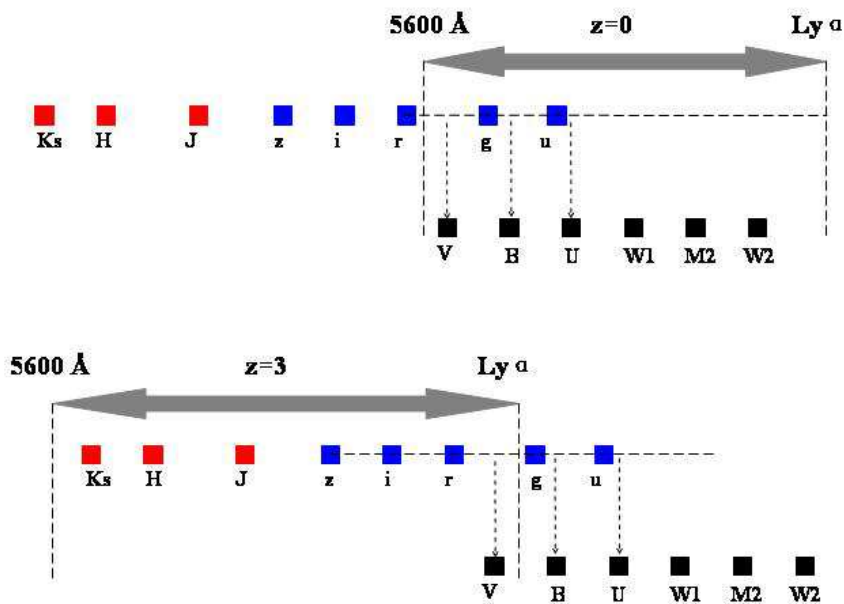}
\caption[Photometric shift strategy]{Illustration of the photometric 
         shift strategy at $z=0$ and $z=3$.
         In either case, SDSS photometry is interpolated or extrapolated
         to the wavelengths of the UVOT filters. The SDSS and 2MASS data are then
         shifted by the difference between real and interpolated luminosities.
         Because the 2MASS do not overlap with the UVOT wavebands, we 
         shift the 2MASS photometry with the same value as SDSS photometry
         (see Section~\ref{photometricshift}).
         \label{fig-sedshift}}
\end{figure}
\clearpage

\begin{landscape}
\begin{figure}
  \centering
  \begin{tabular}{cc}
      \vspace{5mm}
      \includegraphics[width=6cm,angle=90]{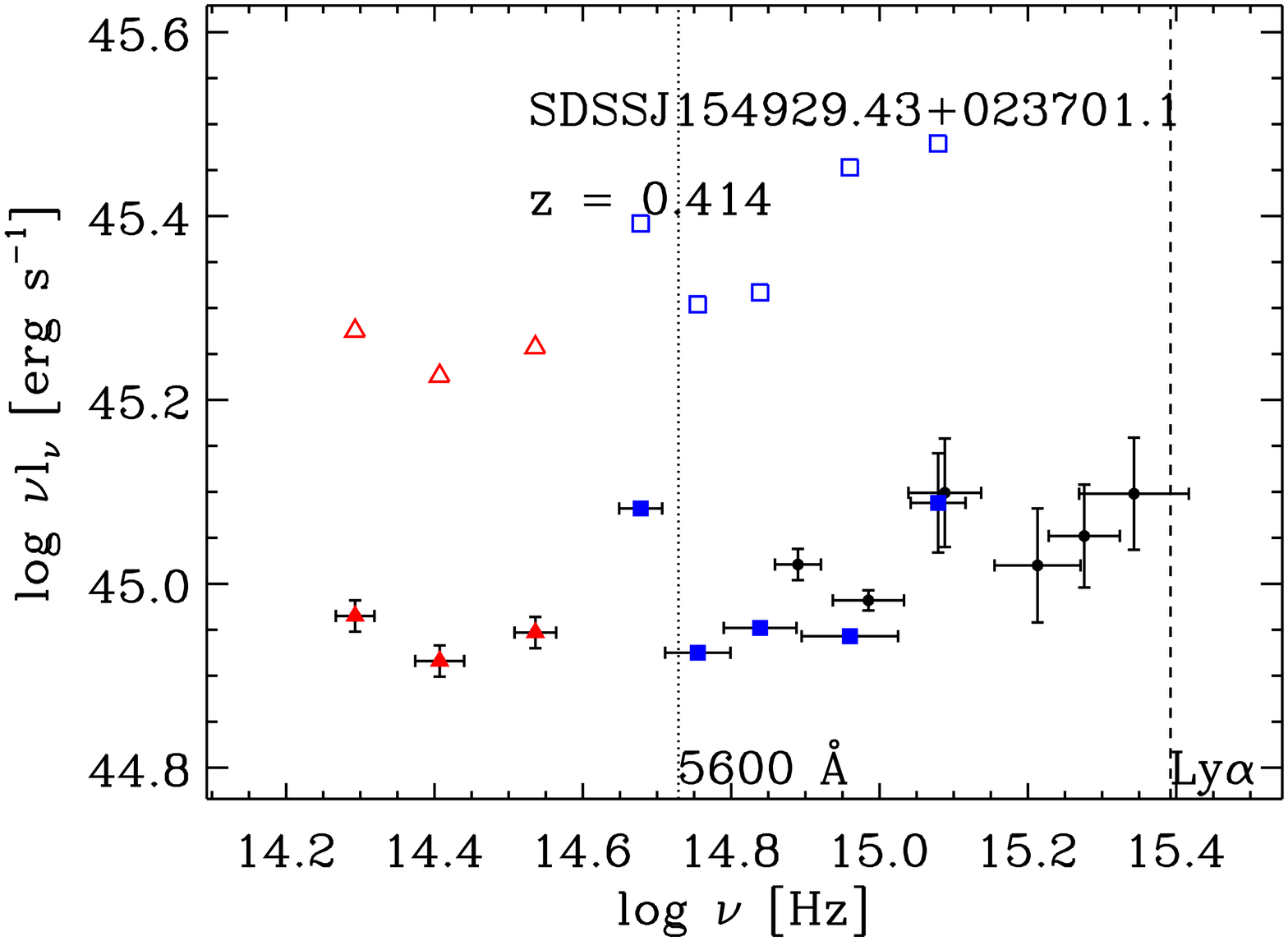} \space{\ \ \ }
      \includegraphics[width=6cm,angle=90]{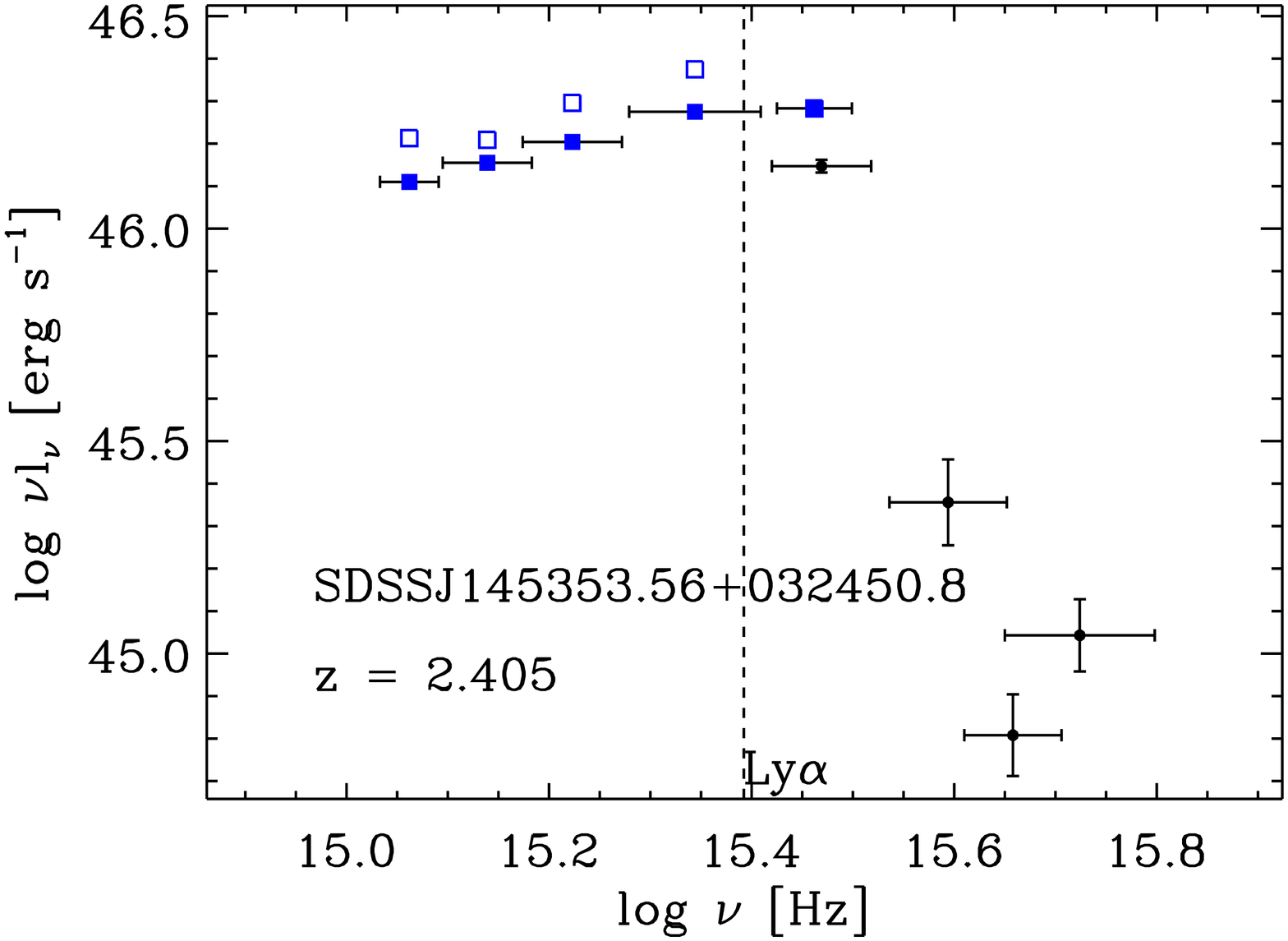}\\
      \vspace{5mm}
      \includegraphics[width=6cm,angle=90]{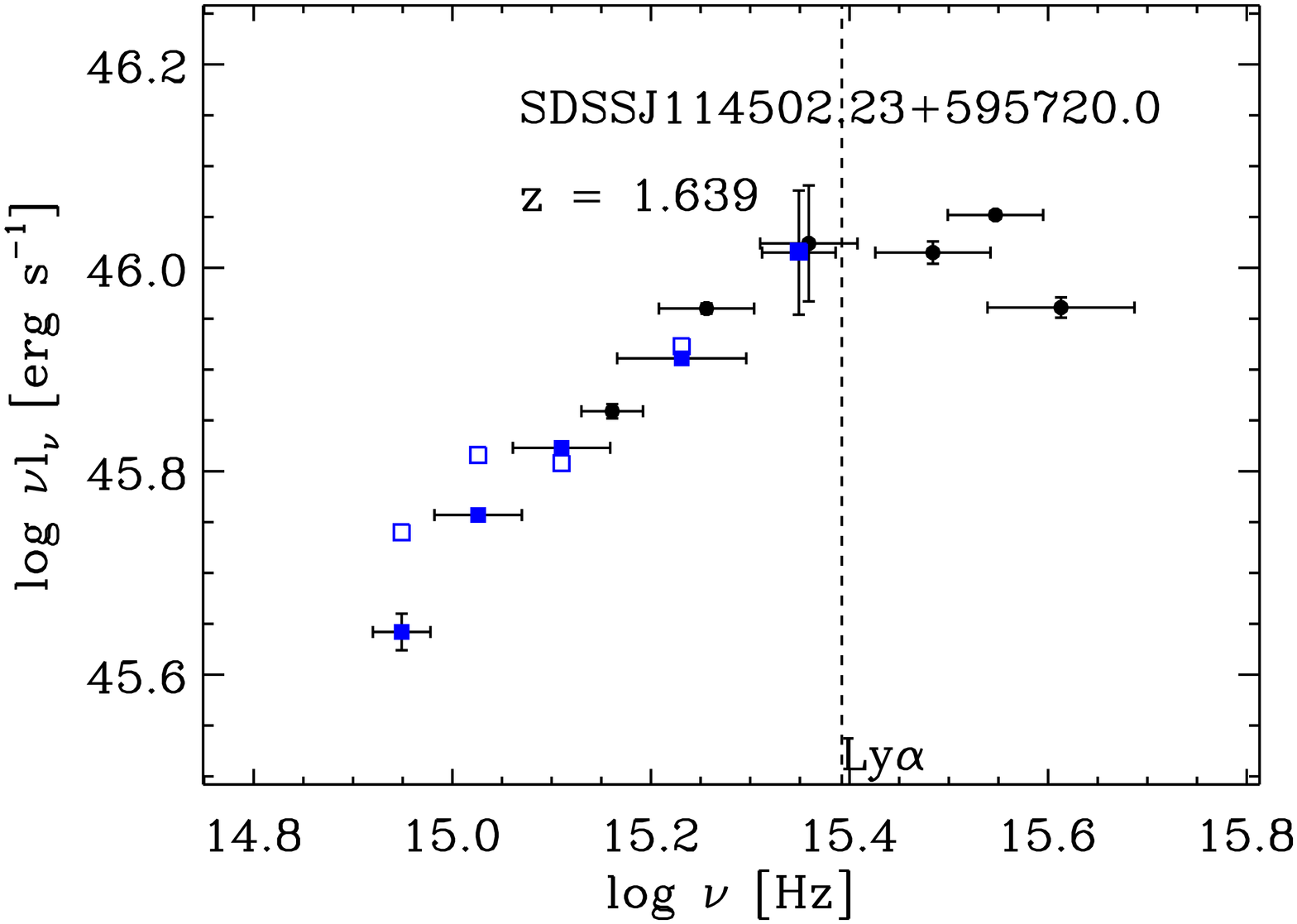} \space{\ \ \ }
      \includegraphics[width=6cm,angle=90]{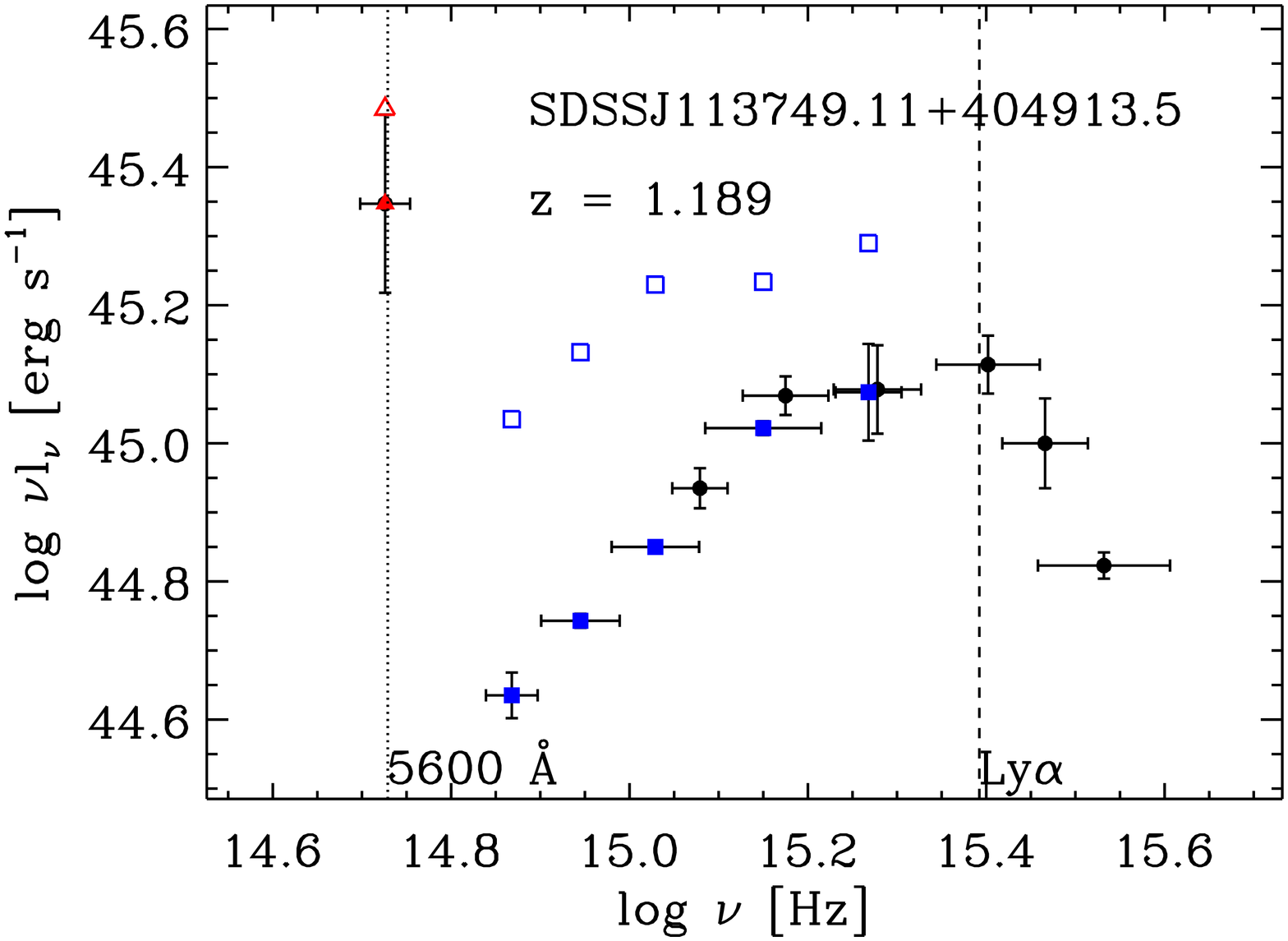}
  \end{tabular}
    \caption[Examples of corrected SEDs]{Emission line corrected and 
    photometric shifted SEDs for four quasars. The X-ray data 
    points ($\log{\nu}>17$) are not plotted because the corrections are only 
    applied to UV/optical data. In the band with
    $\log{\nu(\mbox{5600~\AA})}<\log{\nu}<\log{\nu(\mbox{Ly~}\alpha)}$,
    data points from UVOT are represented by black dots with error bars;
    SDSS data points are represented by blue squares and 2MASS data points
    are represented by red triangles. Open shapes represent photometry 
    before correction, while filled shapes show photometry after correction.  
    Photometric points outside the \lya--\lambdauvo\ region are not
    corrected for line emission.
    \label{fig-sedshiftexample}}.
\end{figure}
\end{landscape}
\clearpage

\begin{figure}
\centering
\vspace{5mm}
\includegraphics[width=8cm,angle=90]{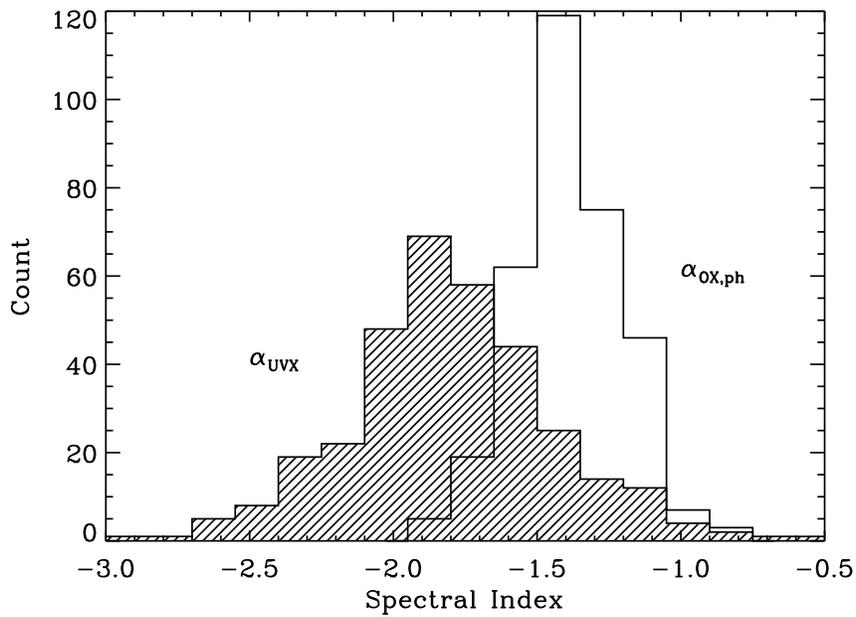}
\caption[Distributions of \auvx\ vs. \aoxph]{Distribution of 
\auvx\ (shaded) and \aox\ (unshaded) using photometric data. The median value
 of \auvx\ is considerably steeper than that for \aox\ ($-1.8$ vs. $-1.39$).
\label{fig-auvxdist}}
\end{figure}
\clearpage

\begin{landscape}
\begin{figure}
  \centering
    \begin{tabular}{cc}
      \vspace{5mm}
      \includegraphics[width=6cm,angle=90]{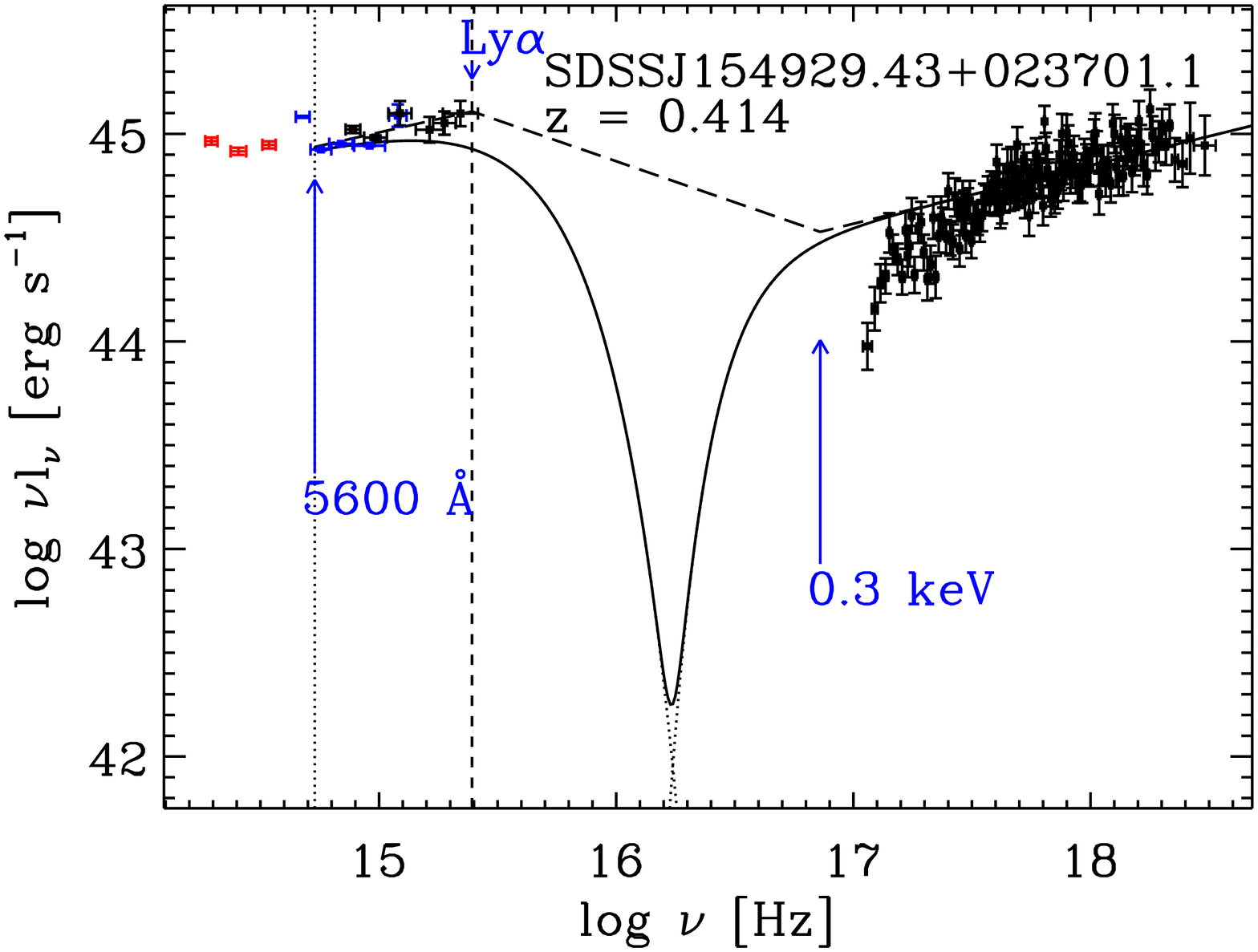} \space{\ \ \ }
      \includegraphics[width=6cm,angle=90]{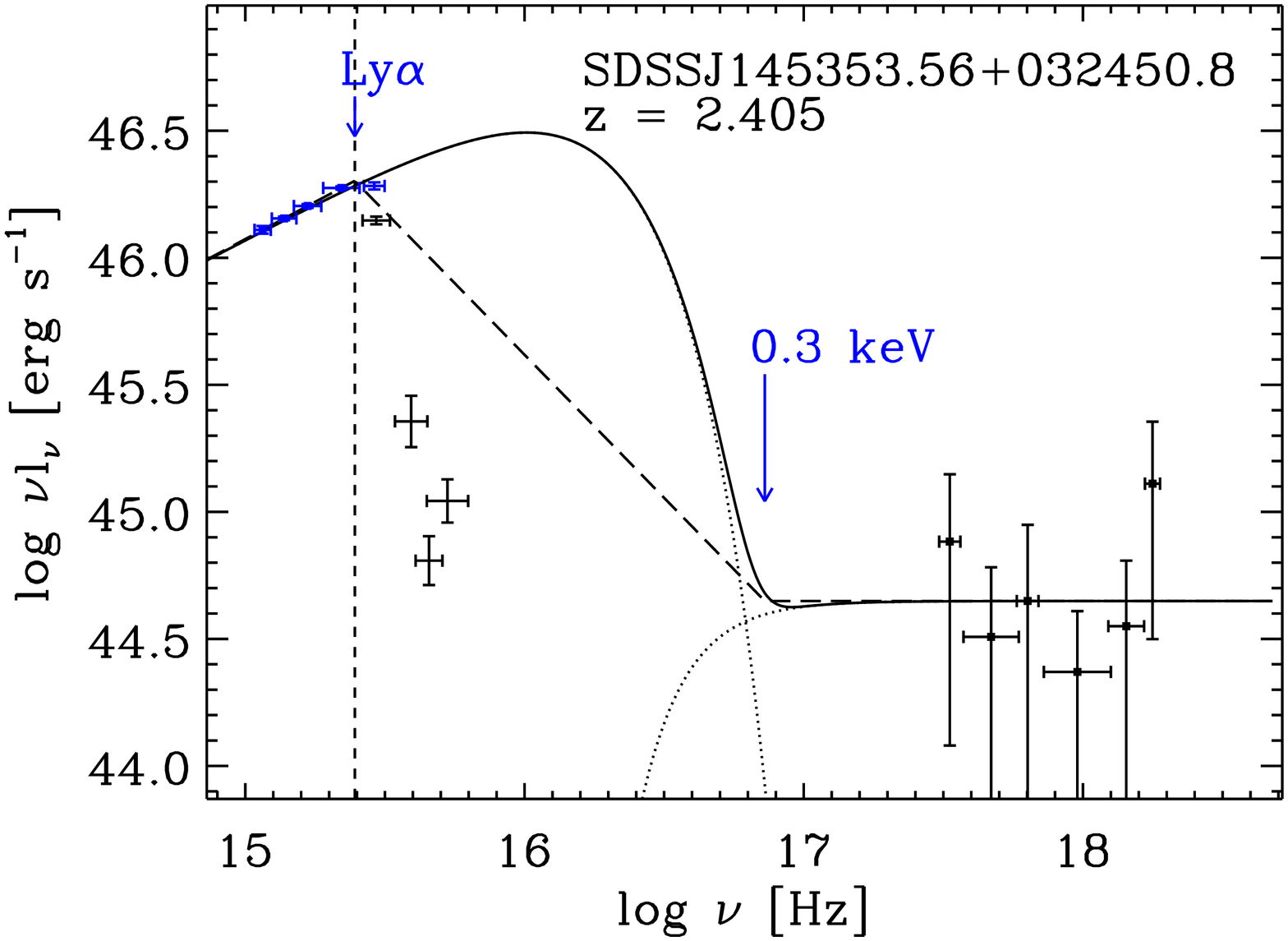}\\
      \vspace{5mm}
      \includegraphics[width=6cm,angle=90]{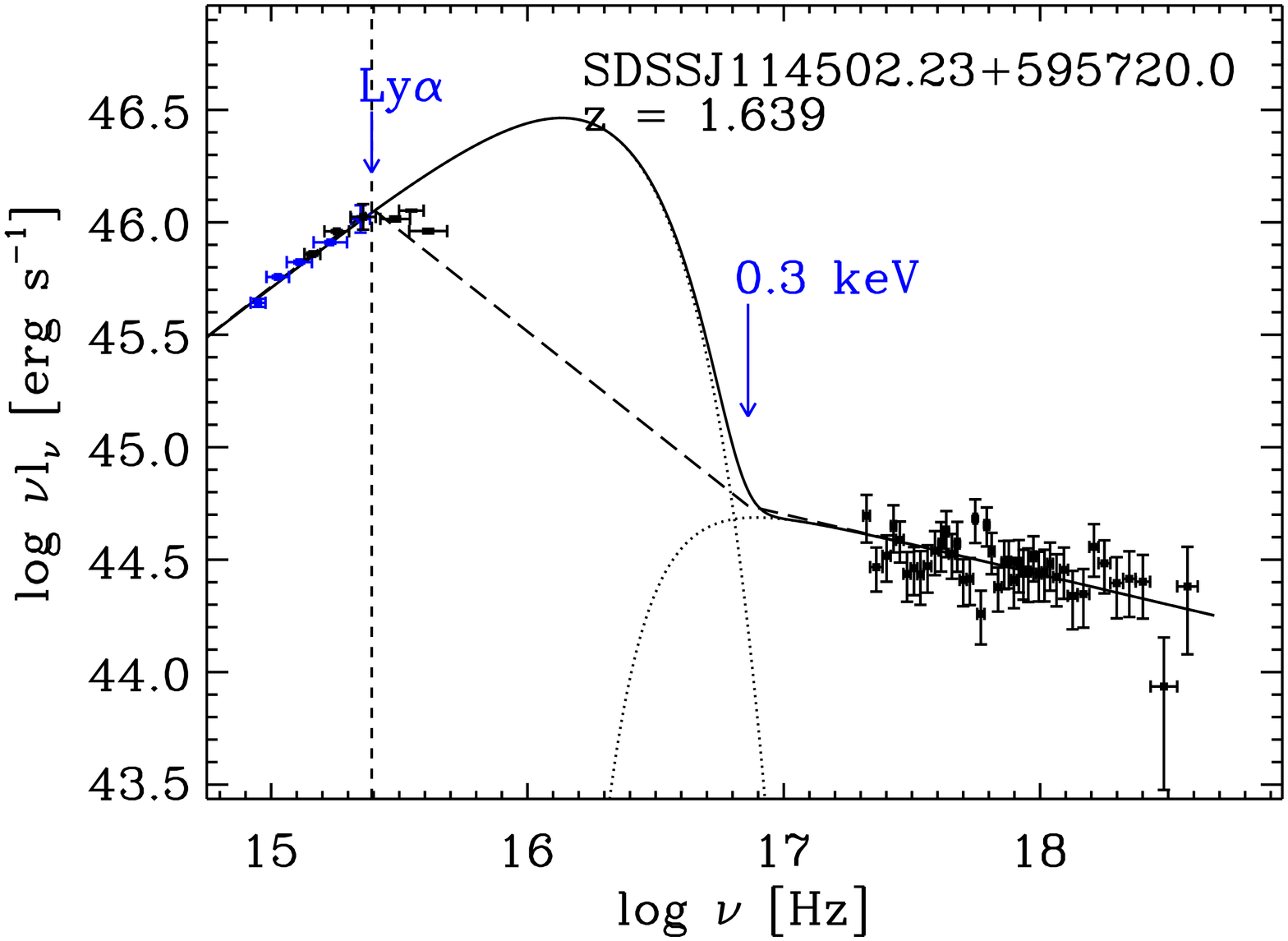} \space{\ \ \ }
      \includegraphics[width=6cm,angle=90]{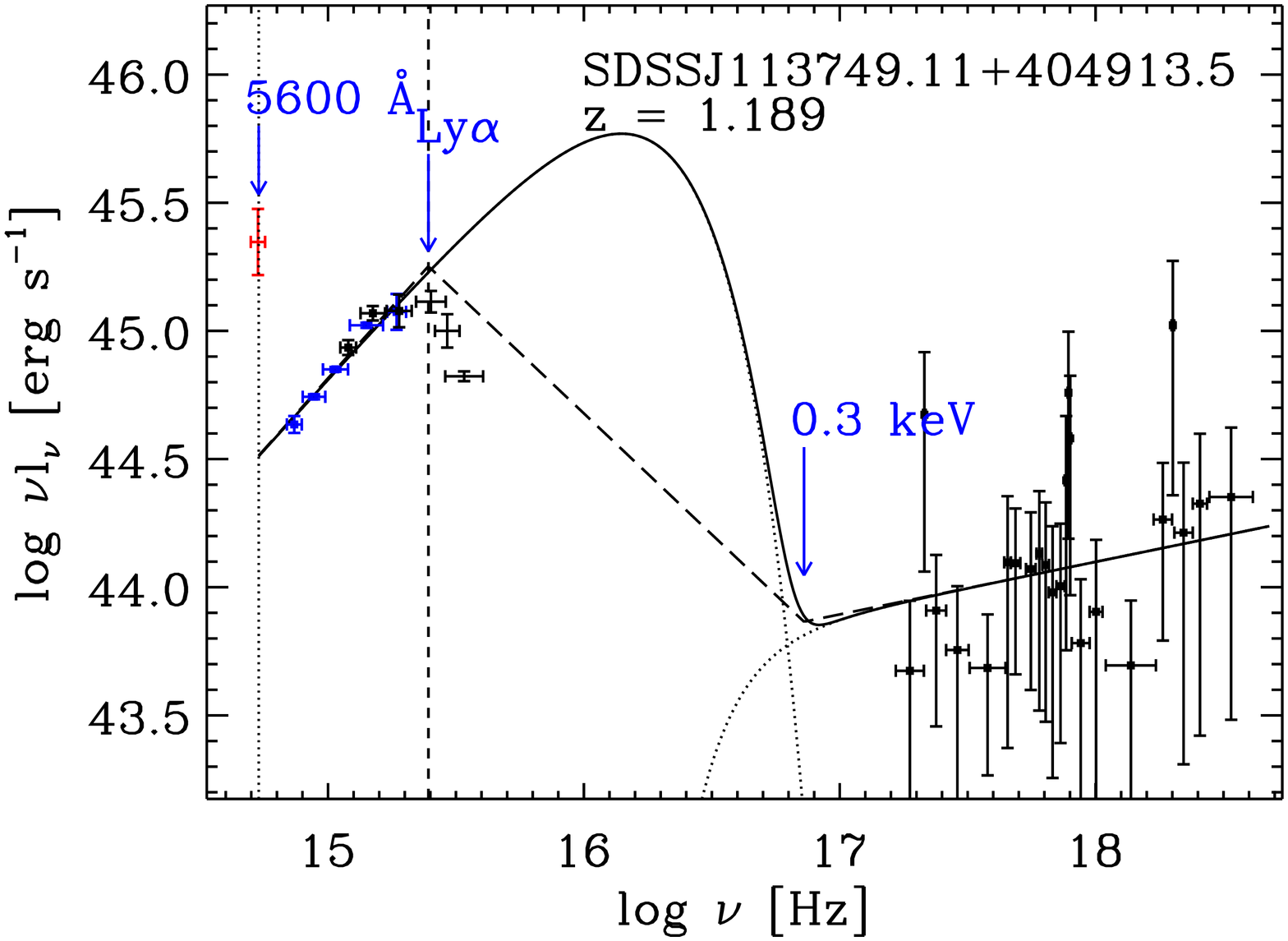}
    \end{tabular}
    \caption[Examples of SEDs fitting in EXP and TPL models]{Examples of SED 
    fits with the EXP (solid curves) and TPL (dashed lines) models. The UV/optical fitting region is 
    bounded with a vertical dotted line at 5600~\AA\ and a vertical dashed 
    line at \lya. 
    UV/optical data points outside this region are not used for SED fitting. 
    \swift\ UVOT data are plotted in black; SDSS data are plotted
    in blue; 2MASS data are plotted in red. Dotted 
    curves are UV and X-ray components for the EXP model.  
    SDSSJ154929.43+023701.1 is a case with strong UV and X-ray
    absorption, in which the EXP model no longer provides an upper limit to
    BBB emission. 
             \label{fig-sedmodexample}}.
\end{figure}
\end{landscape}
\clearpage

\begin{figure}
\centering
\includegraphics[width=10cm,angle=90]{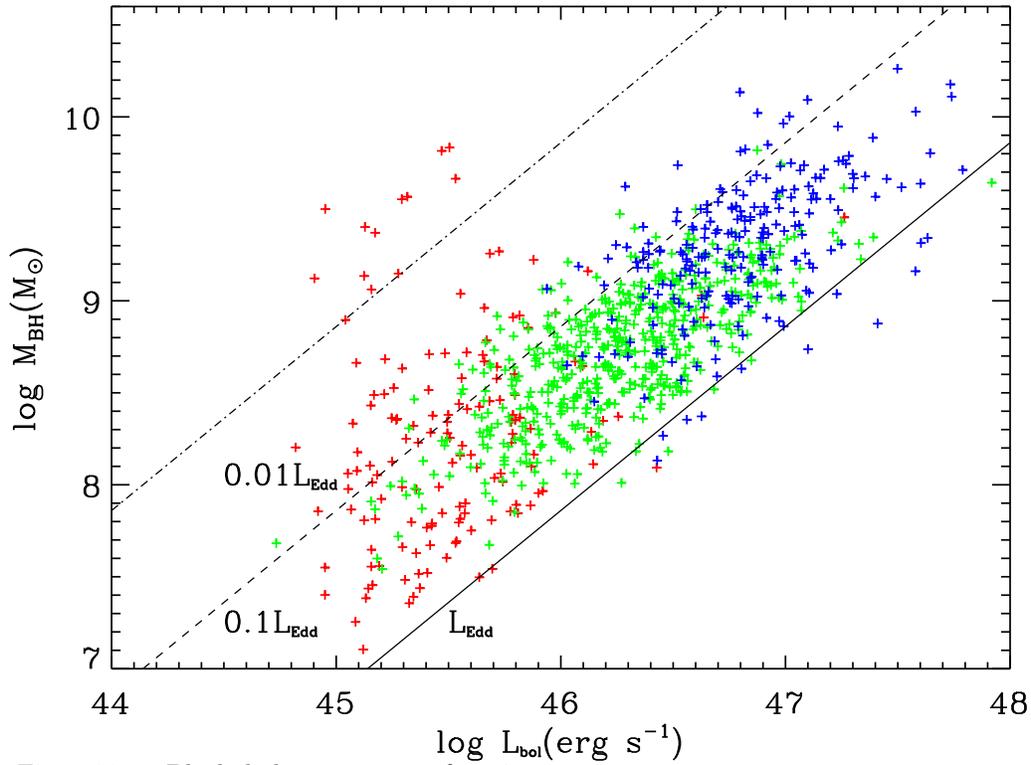}
\caption[Black hole mass as a function of bolometric luminosity]{Black hole mass as a function of bolometric luminosity for
923 quasars in our catalog. Quasars are color-coded based on their
redshift ranges following the same convention as \citet{shey08}: 
red for $z<0.7$, green for $0.7<z<1.9$, and blue for $z>1.9$. 
We also plot solid, dashed and dash-dot
lines when the Eddington ratio is 1, 0.1 and 0.01, respectively.
\label{fig-mbhlbol}}
\end{figure}
\clearpage

\begin{figure}
\centering
\includegraphics[width=10cm,angle=90]{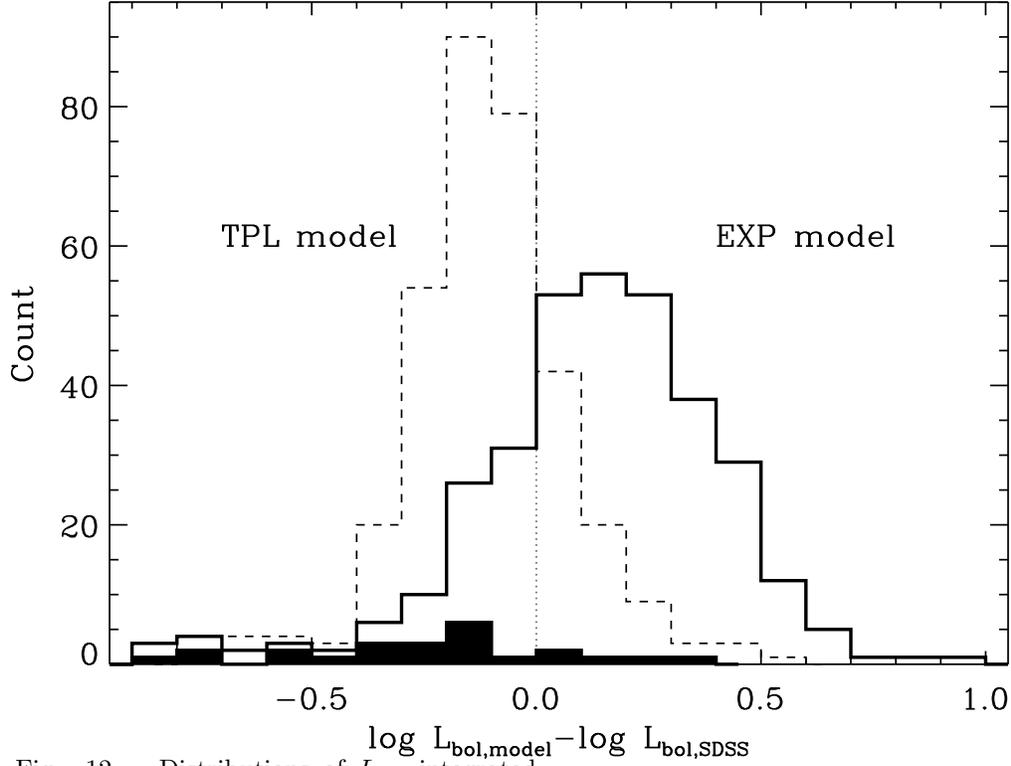}
\caption[Distributions of bolometric luminosities from EXP and TPL 
         models]{Distributions of \lbol\ integrated from
         the EXP (thick solid line) and TPL (thin dashed line) models 
         with respect to \lbol\ calculated using the BC correction.
         The black shaded region under the EXP model histogram 
         represents objects we flagged as ``red" which
         suffer from strong intrinsic absorption (see 
         the ``Reddened quasars" in Section~\ref{sampleselection}). 
         \label{fig-lboldist}}
\end{figure}
\clearpage

\begin{figure}
\centering
\includegraphics[width=15cm,angle=90]{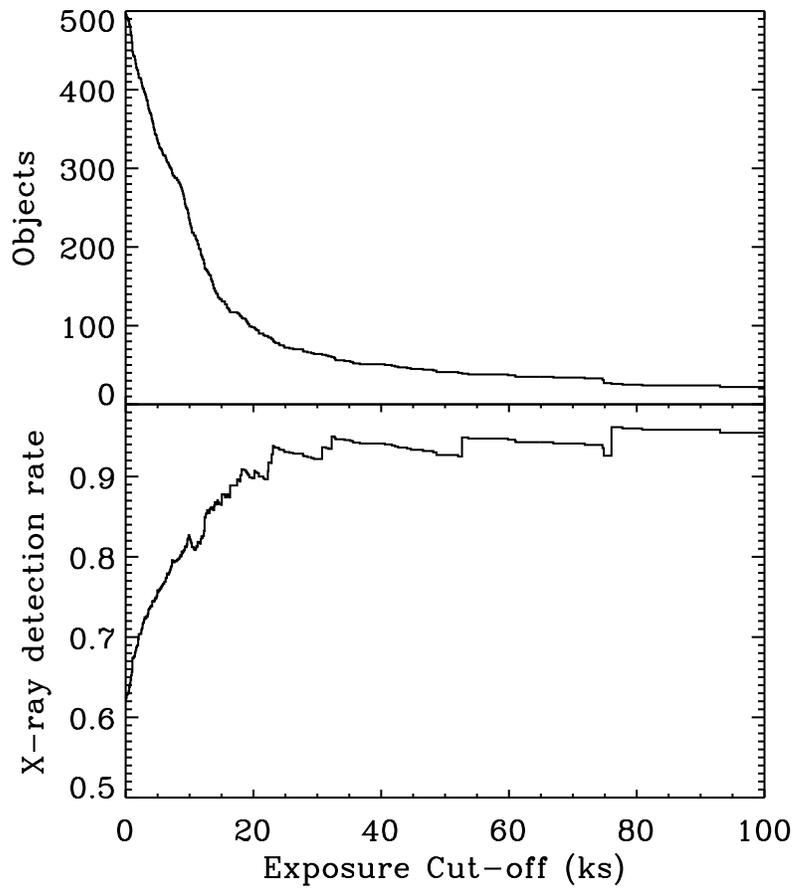}
\caption[Sample size and X-ray detection rate vs. exposure time clean catalog 
        sample]{\label{fig5-xexpcutoff}
        The clean catalog sample size (\emph{Upper} panel) and X-ray detection 
        rate (\emph{Lower panel}) as a function of XRT exposure cut-off in kilo-seconds.}
\end{figure}
\clearpage

\begin{figure}
\centering
\includegraphics[width=10cm,angle=90]{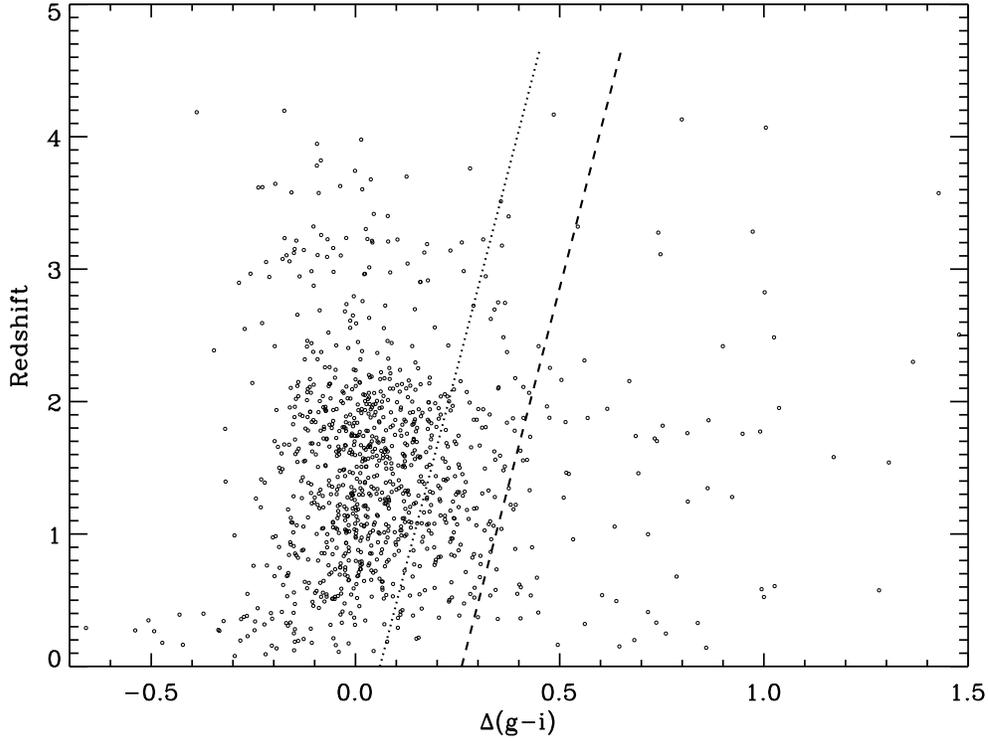}
\caption[Relative color vs. redshift diagram]{\label{fig-gmi} The relative 
color $\Delta(g-i)$ vs. redshift diagram of all the objects in the raw 
catalog. The dotted line shows the effect
   of SMC-type reddening as a function of redshift with 
   $E(B-V)=0.04$ (see Richards et al.~2003\nocite{ric03} for the 
   choice of 0.04 as the value of 
   $E(B-V)$). The dashed line is the dotted line shifted by 0.2 to match the 
   dust-reddening quasar definition of \citet{ric03}. Quasars to the
   right of the dashed line are considered to be dust-reddened.}
\end{figure}
\clearpage

\begin{figure}
\centering
\includegraphics[width=15cm,angle=90]{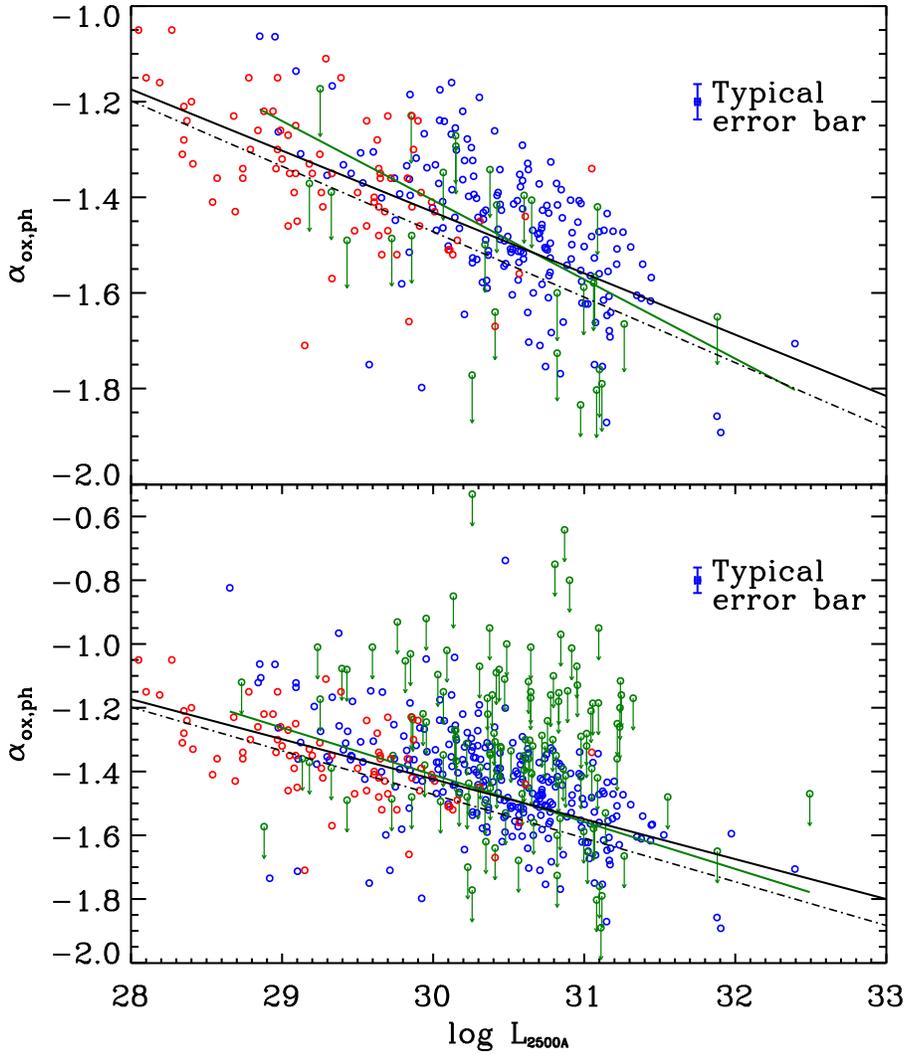}
\caption[\aoxluv\ relation for \swift\ AGN sample]{\label{fig-aoxluv}The 
          \aoxluv\ relation for the clean sample with a 10~ks XRT exposure 
          cut-off (\emph{Upper} panel) and the total clean sample 
          (\emph{Lower} panel), including the large clean catalog sample 
          (green+blue, blue points are X-ray detected), 
          and the G10 sample (red). Blue points are X-ray detected, 
          while green points
          with arrows are upper limits of undetected objects. 
          The solid black line is the best linear fit to the combined sample 
          using the EM method and the solid green line 
          is the best linear fit to the cleaned catalog sample only. 
          The dot dashed 
          line is the best fit of \citet{jus07}. The \aox\ in these plots
          are obtained by fitting photometric data points 
          rather than from spectra. 
          The typical error bar is displayed at the upper right corner.}
\end{figure}
\clearpage

\begin{figure}
\centering
\includegraphics[width=10cm,angle=90]{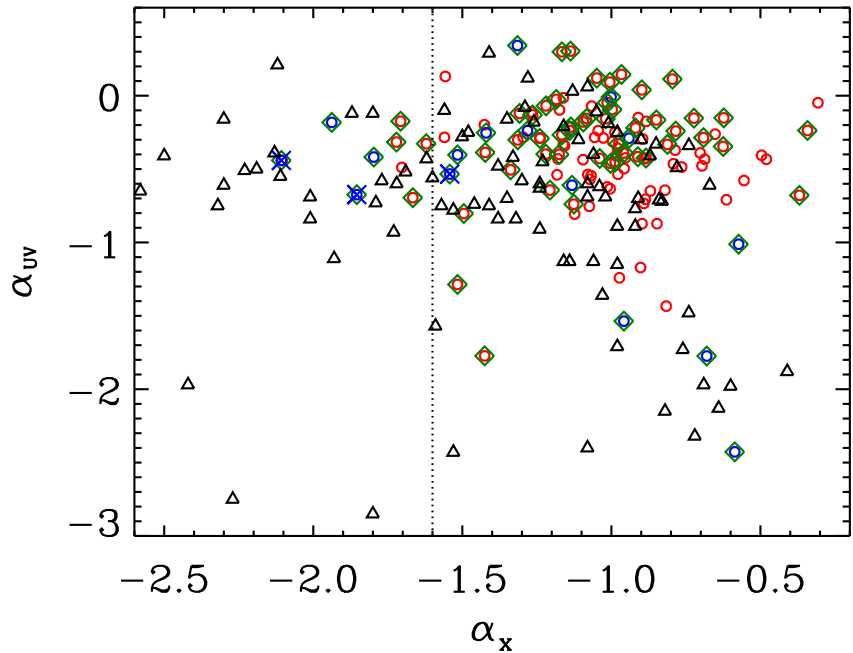}
\caption[\auv--\ax\ relation for \swift\ AGN sample]{\label{fig-auvax}The 
          relation between \auv\ and \ax. Objects are selected from the clean 
          catalog sample (open circles), excluding objects with fixed values 
          of \ax. The 17 low-redshift 
          ($z<0.4$) objects are colored in blue and the remaining objects 
          with $z>0.4$ are in red. NLS1s are marked with a cross on top of 
          circles. We plot green open diamonds on top of low luminosity 
          quasars with $\log{L_{2500\mbox{~\AA}}}<30.5$. 
          For comparison, we also plot the AGN sample from \citet{gru10} 
          (open triangles). The vertical dotted line marks the position 
          where $\alpha_{\rm x}=-1.6$. The Spearman correlation coefficient for all objects 
          from the clean catalog is $\rho_{\rm s}=0.14\,(P_0=0.125)$ which does 
          not indicate significant correlation. The correlation coefficient 
          for low redshift objects is 
          $\rho_{\rm s}=0.47\,(P_0=0.058)$, for objects with $\alpha_{\rm x}>-1.6$ it is 
          $\rho_{\rm s}=0.22\,(P_0=0.017)$ and for low luminosity quasars it is
          $\rho_{\rm s}=-0.04\,(P_0=0.726)$.}
\end{figure}
\clearpage

\begin{figure}
\centering
\includegraphics[width=10cm,angle=90]{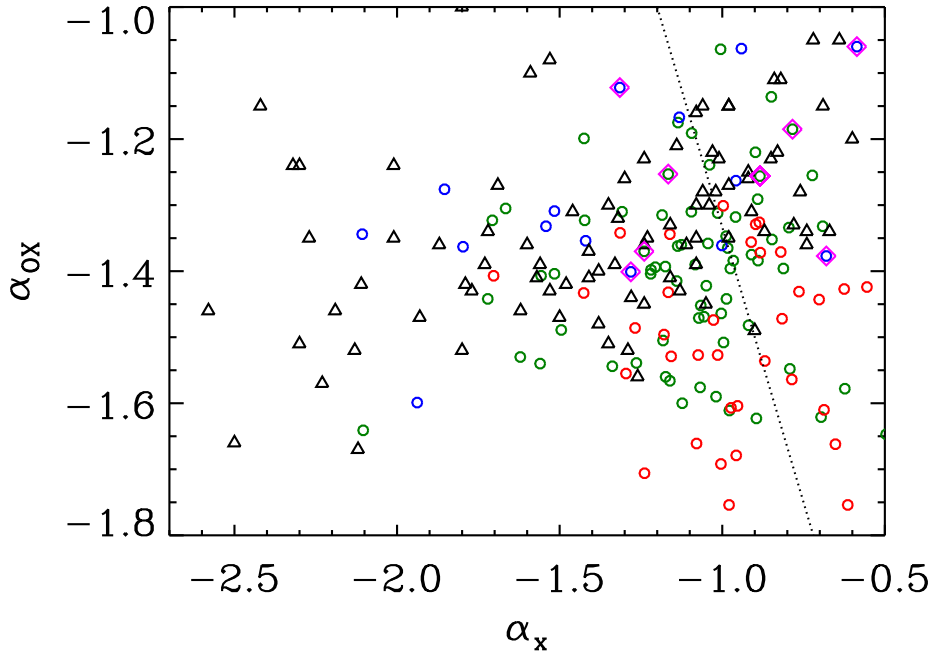}
\caption[$\Gamma$--\aox\ relation for \swift\ AGN sample]{\label{fig-aoxax}The 
          relation between \aox\ and \ax\ 
          for 129 objects (open circles), selected from the clean catalog 
          sample, excluding objects whose X-ray spectral indices are fixed 
          during the fitting process or without \luv\ measured from UVOT photometry. 
          Data points are color-coded depending on redshift with $z<0.4$ in blue, 
          $0.4<z<1.5$ in green and $z>1.5$ in red. 
          Objects with \luv$<30.5$ are flagged with larger diamonds in magenta.
          The correlation coefficient of objects in our work is 
          $\rho_{\rm s}=-0.06\,(P_0=0.484)$, which does not exhibit a 
          significant correlation (though with low confidence 
          level). The dotted line is the weak correlation found by \citet{you09}. 
          For comparison, we over plot the G10 sample in open triangles.}
\end{figure}
\clearpage

\begin{figure}
\centering
\includegraphics[width=10cm,angle=90]{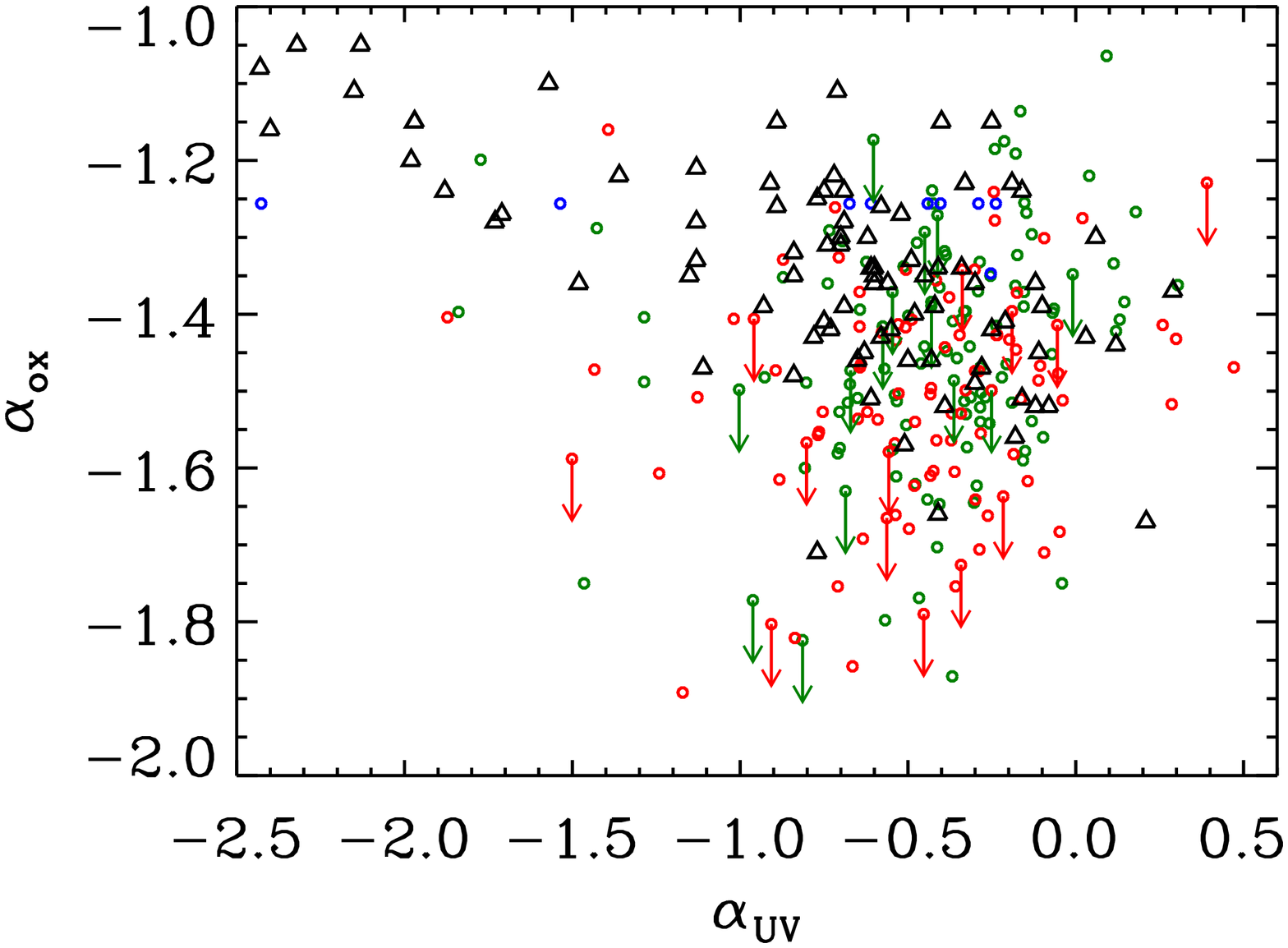}
\caption[\auv--\aox\ relation for \swift\ AGN sample]{\label{fig-auvaox}The 
          relation between the UV/optical spectral index \auv\ and \aox\
          for 217 objects, selected from the clean catalog sample, excluding 
          objects whose \lx\ is not available or without \luv\ measured from 
          UVOT photometry. Data points are color-coded in the same way as 
          Fig.~\ref{fig-aoxax}. The correlation coefficient of objects in our work 
          is $\rho_{\rm s}=-0.17$, which does not exhibit a significant 
          correlation. Arrows represent quasars not detected by XRT. 
          The low redshift ($z<0.4$) quasars in the clean catalog sample
          are color-coded in blue. }
\end{figure}
\clearpage

\begin{figure}
\centering
\includegraphics[width=10cm,angle=90]{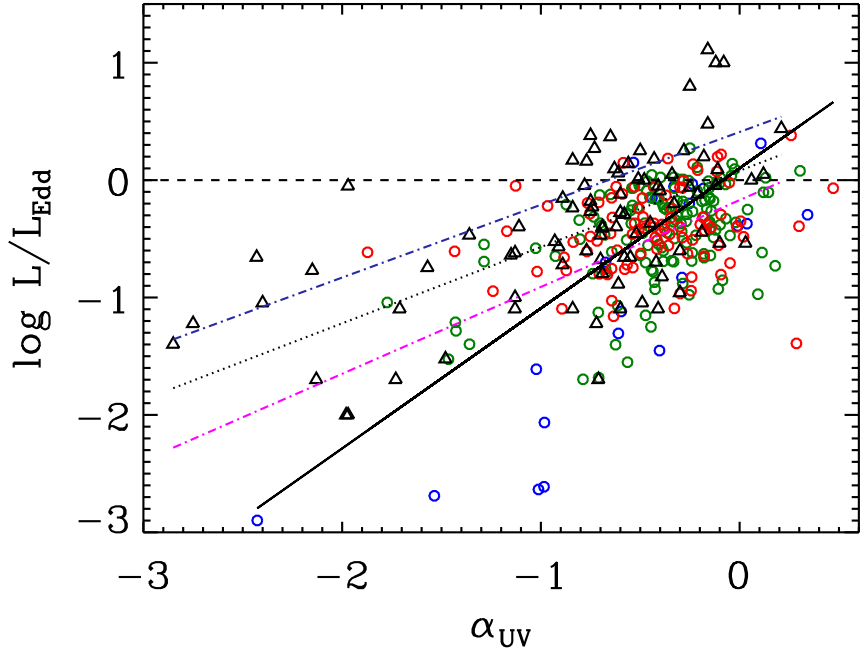}
\caption[\lle--\auv\ relation for \swift\ AGN sample]{\label{fig-lleauv}The 
         relation between the Eddington ratio \lle\ and the UV/optical 
         spectral index \auv\ for 247 objects selected from the 
         clean catalog sample, excluding objects with no bolometric luminosity 
         measurements. Data points are color-coded in the same way as Fig.~\ref{fig-aoxax}.
         The correlation coefficient of objects in our work 
         is $\rho_{\rm s}=0.35\,(P_0<10^{-3})$. The 
         solid straight line is the best linear fit to our data by the 
         BCES method. For comparison, we also plot the linear regression results
         of the BLS1, NLS1 AGNs and the combine of them obtained by \citet{gru10}
         in magenta dash-dotted, blue dash-dotted and dotted lines, respectively.}
\end{figure}
\clearpage

\begin{figure}
\centering
\includegraphics[width=10cm,angle=90]{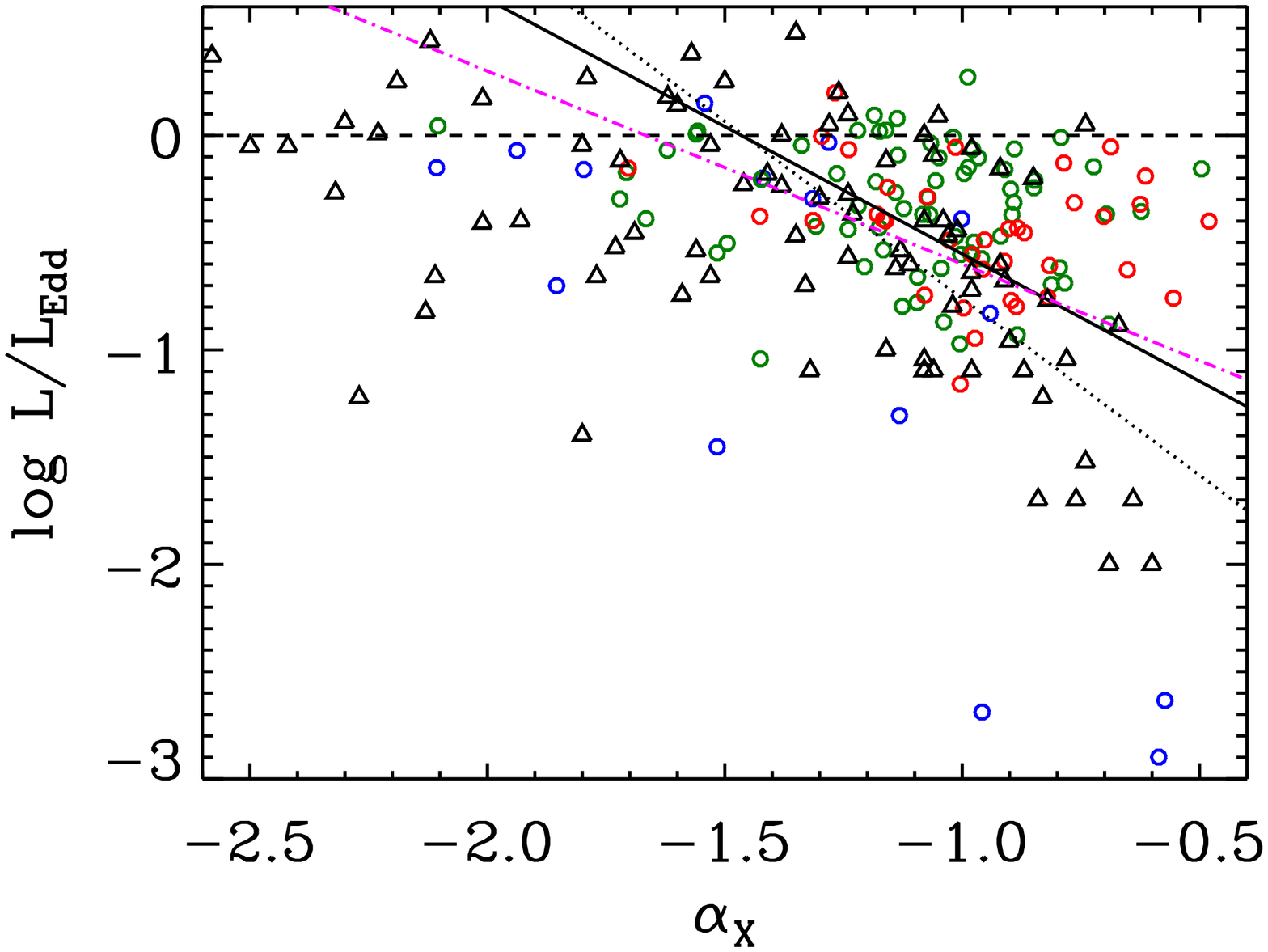}
\caption[\lle--\ax\ relation for \swift\ AGN sample]{\label{fig-lleax}The 
         relation between \lle\ and \ax\ for 129 objects selected from the 
         clean catalog sample, excluding objects with fixed values of 
         photon indices. 
         Data points are color-coded in the same way as Fig.~\ref{fig-aoxax}. 
         The AGNs from \citet{gru10} are represented with black open triangles. 
         The correlation coefficient of objects 
         in our work is $\rho_{\rm s}=-0.26\,(P_0<10^{-3})$. The 
         solid black, dotted and dash-dotted lines represent the linear 
         regression results to our sample, the G10 sample and the 
         \citet{she08} sample. 
        }
\end{figure}
\clearpage

\begin{figure}
\centering
\includegraphics[width=10cm,angle=90]{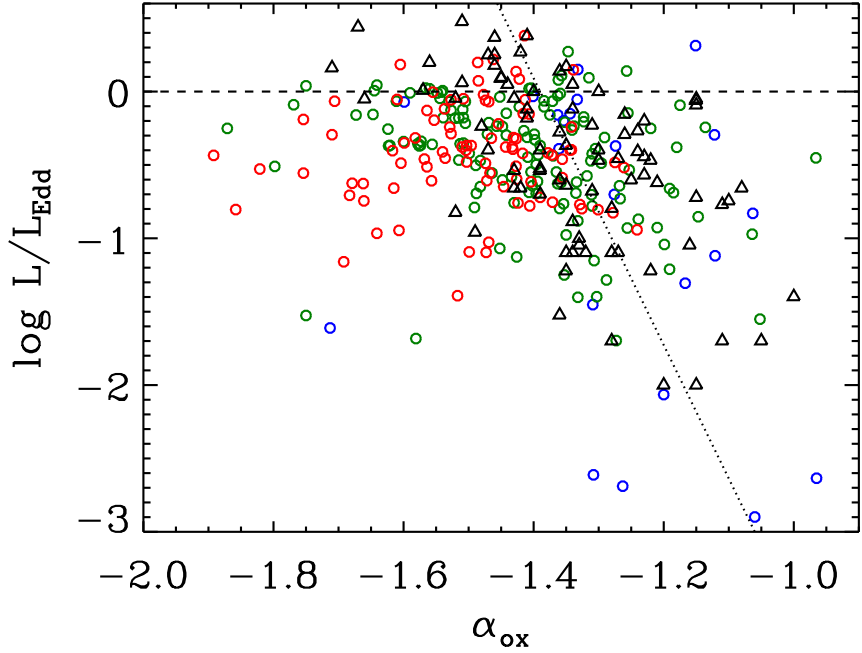}
\caption[\lle--\aox\ relation for \swift\ AGN sample]{\label{fig-lleaox}The 
         relation between the Eddington ratio \lle\ and \aox\ for 247 objects 
         selected from the clean catalog sample, excluding objects
          whose \aox\ values are unavailable. Data points are color-coded in the
         same way as Fig.~\ref{fig-aoxax}. The AGNs from \citet{gru10} are 
          represented with black open triangles. The dotted line is the linear
         regression result by \citet{gru10}.}
\end{figure}
\clearpage

\begin{landscape}
\begin{figure}
  \centering
    \begin{tabular}{c}
      \vspace{0mm}
      \includegraphics[width=8cm,angle=90]{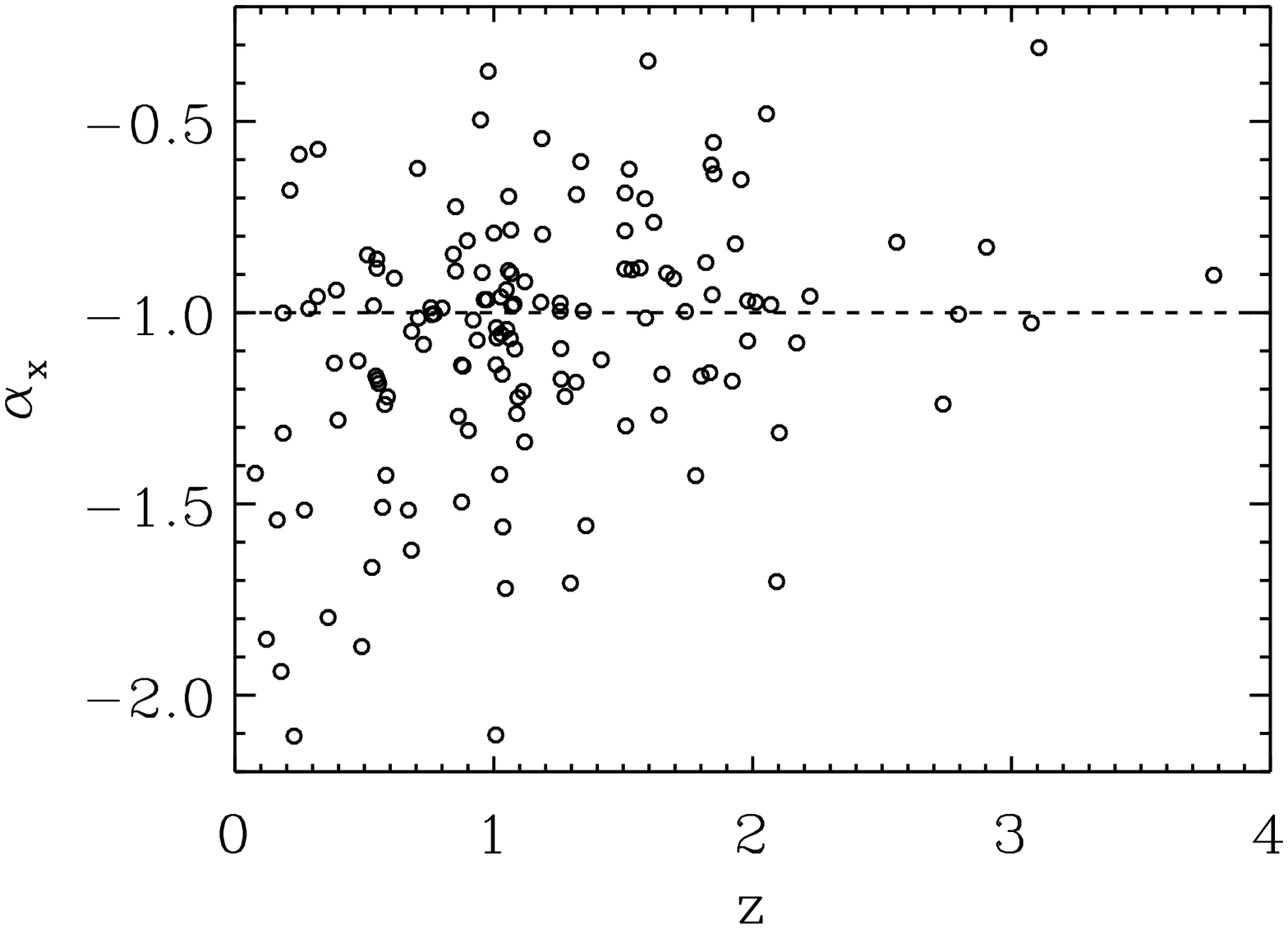}\\
      \vspace{0mm}
      \includegraphics[width=8cm,angle=90]{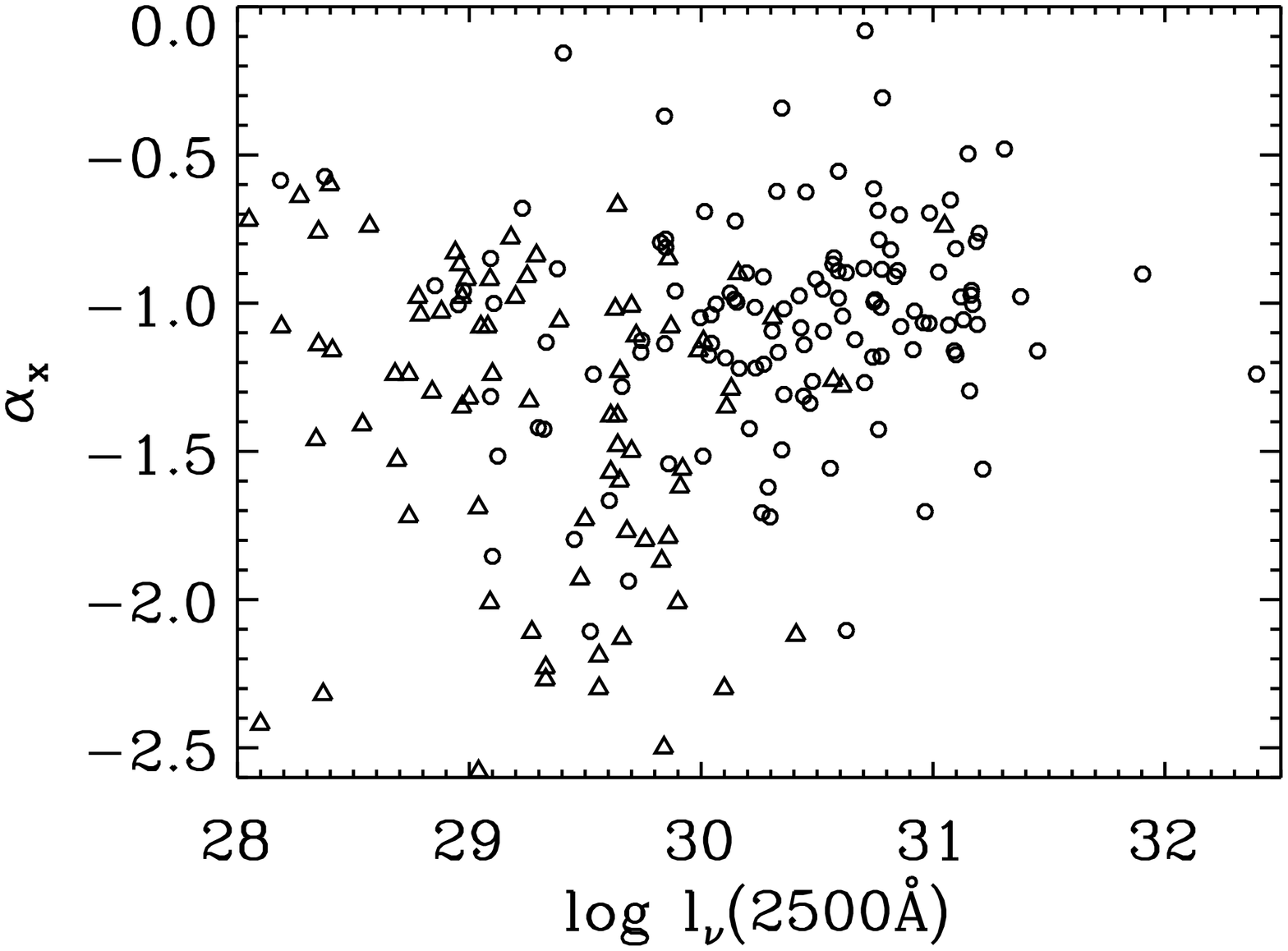}
      \includegraphics[width=8cm,angle=90]{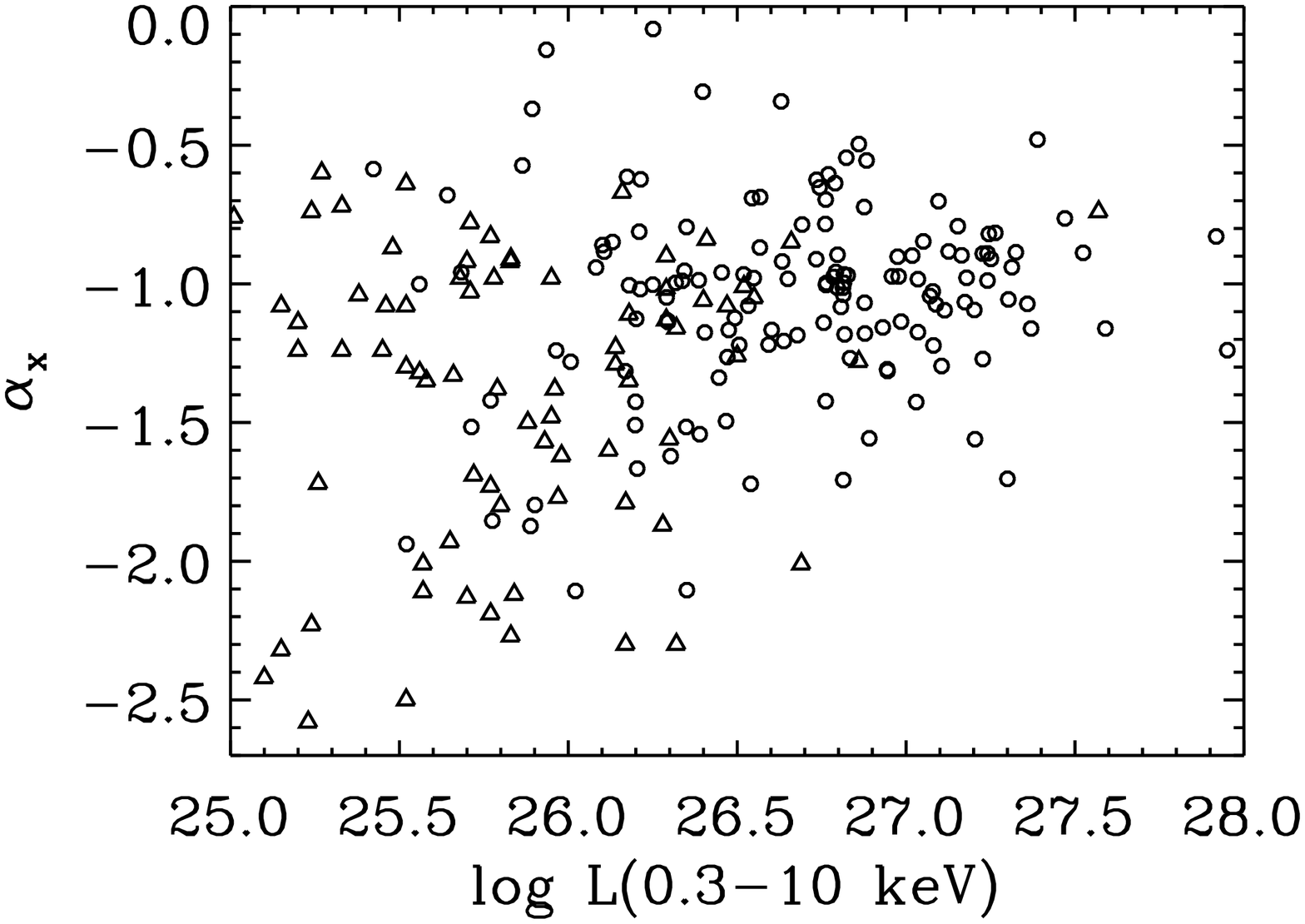}
    \end{tabular}
    \caption[X-ray slope trends]{X-ray slope $\Gamma$ vs. redshift (upper 
     panel), \luv\ (lower left), and \lx\ (lower right). In each panel, open 
     circles represent data points in the clean catalog sample. Objects whose 
     X-ray slopes are fixed during the fitting process are excluded. Open 
     triangles represent data points in the G10 sample. \label{fig-xslope}}.
\end{figure}
\end{landscape}
\clearpage

%% file: table.tex
\begin{deluxetable}{cccl}
\tablecolumns{4} \tabletypesize{\scriptsize}
\tablecaption{XRT Data Binning Strategy.\label{t-xrtbin}}
\tablewidth{0pc}
\tablehead{
\colhead{$N_{\rm Xph}$}       &
\colhead{Photon\# per Bin}    &
\colhead{Statistics}          &
\colhead{Flag}}
\startdata
\smallskip
$N_{\rm Xph}\geq200$     & 20                 & $\chi^2$ & \emph{g}(good)\\
\smallskip
$100\leq N_{\rm Xph}<200$ & $N_{\rm Xph}/10$  & $\chi^2$ & \emph{g}(good)\\
\smallskip
$10\leq N_{\rm Xph}<100$ & 1   & Cash & \emph{a}(acceptable)\\
\smallskip
$N_{\rm Xph}<10$ & Group Min 1\tablenotemark{1}& -\tablenotemark{2} & \emph{w}(weak)\\
\smallskip
$N_{\rm Xph}\sim0$ & $\cdots$ & $\cdots$ & \emph{o}(Out of FOV)
\enddata
\tablenotetext{1}{In these cases, we group spectral bins with a mininum of 1 photon per bin.}
\tablenotetext{2}{If the total number of X-ray photons is less than 10,
                  we do not fit the X-ray spectrum but only calculate flux or flux limit.}
\end{deluxetable}
\clearpage

\begin{deluxetable}{ccccc}
\tablecolumns{5}
\tablewidth{0pc}
\tablecaption{Models used to fit XRT spectra. \label{tab-models}}
\tablehead{
\colhead{Model}    &
\colhead{\ax}      &
\colhead{\nhi}     &
\colhead{Objects}  &
\colhead{Percentage}}
\startdata
A    & Fixed to $-1$ & Fixed to $0$&177 & 44\%\\
B    & Free          & Fixed to $0$&195 & 49\%\\
C    & Free          & Free        &15  &  4\%\\
D    & Fixed to $-1$ & Free        &12  &  3\%
\enddata
\end{deluxetable}
\clearpage

\begin{deluxetable}{cccc}
\tablecolumns{4} \tabletypesize{\scriptsize}
\tablecaption{UVOT Sky Image Flagging Description.\label{t-uvotflag}}
\tablewidth{0pc}
\tablehead{
\colhead{}           &
\colhead{}           &
\colhead{Source}     &
\colhead{Background}\\
\colhead{Flag}       &
\colhead{Description}&
\colhead{Region}     &
\colhead{Region}\\
}
\startdata
$-2$ & Bad Aspect    & None    & None\\
$-1$ & Out of Image  & None    & None\\
$1$  & Faint         & Default & Default\\
$0$  & $\delta\leq0^{\prime\prime}.618$ & Default\tablenotemark{a} & Default\tablenotemark{a}\\
$2$  & $0^{\prime\prime}.618<\delta\leq3^{\prime\prime}$ & Customized\tablenotemark{b} & Default\tablenotemark{a}\\
$3$  & $\delta>3^{\prime\prime}$ & None & None\\
$4$  & Near edge ($d_{\rm E}\leq27^{\prime\prime}.5$) & None & None\\
$5$  & Not in FOV & None & None\\
$10+$ & $27^{\prime\prime}.5<d_{\rm E}\leq35^{\prime\prime}$ & TBD\tablenotemark{c} & TBD\tablenotemark{c}\\
$100+$ & $d_{\rm V}\leq100^{\prime\prime}$ & TBD\tablenotemark{c} & TBD\tablenotemark{c}
\enddata
\tablecomments{$\delta$ is the separation between the SDSS coordinate and the
  Gaussian centroid resulting from fitting the object image (see 
  Section~\ref{swiftq-uvotdata}); $d_{\rm E}$ is
  the distance between the source position and the sky image edge; $d_{\rm V}$
  is the distance between the source position and the nearest vertex of a 
  sky image. }
\tablenotetext{a}{See Section~\ref{swiftq-uvotdata} for descriptions of the 
                  default source and background region files.}
\tablenotetext{b}{Source region circles are centered at the new Gaussian 
                  centroid.}
\tablenotetext{c}{These images are passed to the visual inspection process (Section~\ref{swiftq-uvotdata})}. 
\end{deluxetable}
\clearpage

\begin{deluxetable}{cccccr}
\tablecolumns{6} \tabletypesize{\scriptsize}
\tablecaption{Four types of objects in the raw catalog.\label{tab-fourtypes}}
\tablewidth{0pc}
\tablehead{
\colhead{Type} &
\colhead{UVOT}    &
\colhead{XRT}  &
\colhead{In Catalog} &
\colhead{SED} & 
\colhead{Number}
}
\startdata
A & Y & Y & Y & Y & 637\\
B & Y & N & Y & N & 38\\
C & N & Y & Y & N & 168\\
\smallskip
D & N & N & N & N & 191\\
Total & 675 & 805 & 843 & 637 & 1034
\enddata
\end{deluxetable}	
\clearpage

\begin{deluxetable}{rccc}
\tablecolumns{6} \tabletypesize{\scriptsize}
\tablecaption{Photometric shift strategies. \label{tab-sedshift}}
\tablewidth{0pc}
\tablehead{
\colhead{} &
\colhead{UV Bands} &
\colhead{EUV Bands} & 
\colhead{} \\
\colhead{Priority\tablenotemark{1}} &
\colhead{Condition} &
\colhead{Condition} & 
\colhead{Method\tablenotemark{2}} 
}
\startdata
1 & \emph{g}+\emph{u}+\emph{U}&                            & E \\ 
2 &                           & \emph{g}+\emph{u}+\emph{U} & E \\
3 & \emph{g}+\emph{B}+\emph{u}&                            & I \\
4 &                           & \emph{g}+\emph{B}+\emph{u} & I \\
5 & \emph{r}+\emph{g}+\emph{B}&                            & E \\
6 & \emph{r}+\emph{V}+\emph{g}&                            & I \\
7 &                           & \emph{r}+\emph{V}+\emph{g} & I\\
8 & \emph{V}+\emph{u}+\emph{g}&                            & E\\
9 & \emph{i}+\emph{r}+\emph{V}& \emph{i}+\emph{r}+\emph{V} & E\\
10& \emph{g}+\emph{u}+\emph{UVW1}&                         & E\\
11&                         &\emph{g}+\emph{u}+\emph{UVW1} & E\\
12& \emph{g}+\emph{u}+\emph{UVM2}&                         & E\\
13& \emph{g}+\emph{u}+\emph{UVW2}&                         & E\\
14&                         & \emph{g}+\emph{u}+\emph{UVW2}& E\\
15&                         & \emph{g}+\emph{u}+\emph{UVM2}& E
\enddata
\tablenotetext{1}{The highest priority is 1.}
\tablenotetext{2}{E for extrapolation; I for interpolation}
\end{deluxetable}
\clearpage

\begin{deluxetable}{ccccccl}
\tablecolumns{7} \tabletypesize{\scriptsize}
\tablecaption{Parameters for black hole mass calculation.\label{tab-mbhcal}}
\rotate
\tablewidth{0pc}
\tablehead{
\colhead{Redshift} &
\colhead{FWHM}    &
\colhead{$\lambda_{\rm c}$\tablenotemark{a}} &
\colhead{$a$}  &
\colhead{$b$} & 
\colhead{BC} & 
\colhead{Reference}}
\startdata
\smallskip
$z<0.7$    & FWHM(\hbeta) & 5100~\AA & 0.660 & 0.53 & 9.26  &\citet{mcl04}\\
\smallskip
$0.7<z<1.9$& FWHM(\mgii)  & 3000~\AA & 0.505 & 0.62 & 5.15 & \citet{mcl02,mcl04}\\
\smallskip
$z>1.9$    & FWHM(\civ)   & 1350~\AA & 0.672 & 0.61 & 3.81 & \citet{ves06}
\enddata
\tablenotetext{a}{Wavelength of continuum monochromatic luminosity.}
\end{deluxetable}
\clearpage

\begin{deluxetable}{rlcl}
\tablecolumns{4}
\tablewidth{0pc}
\tabletypesize{\scriptsize}
\tablecaption{Catalog description.\label{tab-catalog}}
\tablehead{
\colhead{Column}&
\colhead{Format}&
\colhead{Symbol}&
\colhead{Description}}
\startdata
1    &A18   &SDSSID& SDSS DR5 Designation hhmmss.ss$+$ddmmss.s(J2000)\\
2    &F10.6 &RA    & SDSS right ascension in decimal degrees (J2000)\\
3    &F10.6 &DEC   & SDSS declination in decimal degrees (J2000)\\
4    &F6.4  &$z$   & Redshift from SDSS DR5 quasar catalog\\
5    &F6.2  &$M_i$ & Absolute magnitude at $i$ band from SDSS DR5 quasar catalog\\
6    &F6.3  &$u$   & BEST PSF $u$ magnitude (not corrected for Galactic extinction)\\
7    &A1    &Quality Flag& Data quality (\emph{a/b/c/d})\\
8    &I1    &Catalog Flag& $1=$in final catalog, $0=$not in final catalog\\
9    &I1    &SED    & $1=$has SED plots, $0=$no SED plot\\
10   &I1    &$N_{\rm 2MASS}$ & Number of 2MASS photometric points\\
11   &I1    &$N_{\rm UVOT}$  & Number of UVOT photometric data points\\
12   &A1    &XRT Flag & Quality of XRT data (\emph{g/a/w/o})\\
13   &A6    &QSO Type & Classification of quasar\\
14   &A1    &XRT Model& Model used to fit XRT spectrum (\emph{a/b/c/d})\\
15   &I2    &Red Flag & $1=$color is ``red", $0=$color is not red\\
16   &I1    &SDSS Fit & $1=$SDSS spectrum is fit, $0=$SDSS spectrum is not fit\\
17   &F7.3  &$\Gamma$&Photon index between 0.3 and 10 keV\\
18   &F7.3  &$\delta^-(\Gamma)$&\onesig\ lower error bar of photon index\\
19   &F7.3  &$\delta^+(\Gamma)$&\onesig\ upper error bar of photon index\\
20   &F7.3  &HR                &Hardness ratio\tablenotemark{3}\\
21   &F7.3  &\logfxabs&Observed flux between 0.3 and 10~keV\\
22   &F7.3  &\logfxunabs&Unabsorbed flux between 0.3 and 10~keV\\
23   &F7.3  &\loglx        &Monochromatic luminosity at 2~keV in \luvunit\ in logarithmic scale\\
24   &F7.3  &\eloglx       &\onesig\ uncertainty of monochromatic luminosity at 2~keV\\
25   &F7.3  &\loglxint     &Integrated luminosity between 0.3 and 10~keV in \ergs\\
26   &I2    &XRT Detect    &$1=$detection, $0=$non-detection, $-1=$not observed\\
27   &E9.2  &CR            &Source count rate in $10^{-3}$~\cps\\
28   &F8.1  &$T_{\rm XRT}$           &Total XRT exposure time in seconds\\
29   &F5.3  &\nhg          &Galactic column density in $10^{20}$~\psqcm\\
30   &F7.3  &\nhi          &Intrinsic column density in $10^{22}$~\psqcm\\
31   &F7.3  &\nhim         &\onesig\ lower error bar in $10^{22}$~\psqcm\ of \nhi\\
32   &F7.3  &\nhip         &\onesig\ upper error bar in $10^{22}$~\psqcm\ of \nhi \\
33   &I2    &V Flag &$1=$has V band photometry, $-1=$no V band photometry\\
34   &F8.1  &$T_{\rm V}$ &Total exposure time (seconds) in V band\\
35   &F7.3  &$\log{f(\mbox{V})}$        &Flux density at \swift\ V band\tablenotemark{1}\\
36   &F7.3  &$\delta\log{f(\mbox{V})}$ &\onesig\ error bar of flux density at \swift\ V band\\
37   &I2    &B Flag &$1=$has B band photometry, $-1=$no B band photometry\\
38   &F7.1  &$T_{\rm B}$ &Total exposure time (seconds) in B band\\
39   &F7.3  &$\log{f(\mbox{B})}$        &Flux density at \swift\ B band\tablenotemark{1}\\
40   &F7.3  &$\delta\log{f(\mbox{B})}$ &\onesig\ error bar of flux density at \swift\ B band\\
41   &I2    &U Flag &$1=$has U band photometry, $-1=$no U band photometry\\
42   &F7.1  &$T_{\rm U}$ &Total exposure time (seconds) in U band\\
43   &F7.3  &$\log{f(\mbox{U})}$        &Flux density at \swift\ U band\tablenotemark{1}\\
44   &F7.3  &$\delta\log{f(\mbox{U})}$ &\onesig\ error bar of flux density at \swift\ U band\\
45   &I2    &UVW1 Flag &$1=$has UVW1 band photometry, $-1=$no UVW1 band photometry\\
46   &F8.1  &$T_{\rm UVW1}$ &Total exposure time (seconds) in UVW1 band\\
47   &F7.3  &$\log{f(\mbox{UVW1})}$     &Flux density at \swift\ UVW1 band\tablenotemark{1}\\
48   &F7.3  &$\delta\log{f(\mbox{UVW1})}$ &\onesig\ error bar of flux density at \swift\ UVW1 band\\
49   &I2    &UVM2 Flag &$1=$has UVM2 band photometry, $-1=$no UVM2 band photometry\\
50   &F8.1  &$T_{\rm UVM2}$ &Total exposure time (seconds) in UVM2 band\\
51   &F7.3  &$\log{f(\mbox{UVM2})}$     &Flux density at \swift\ UVM2 band\tablenotemark{1}\\
52   &F7.3  &$\delta\log{f(\mbox{UVM2})}$ &\onesig\ error bar of flux density at \swift\ UVM2 band\\
53   &I2    &UVW2 Flag & $1=$has UVW2 band photometry, $-1=$no UVW2 band photometry\\
54   &F8.1  &$T_{\rm UVW2}$ &Total exposure time (seconds) in UVW2 band\\
55   &F7.3  &$\log{f(\mbox{UVW2})}$        &Flux density at \swift\ UVW2 band\tablenotemark{1}\\
56   &F7.3  &$\delta\log{f(\mbox{UVW2})}$ &\onesig\ error bar of flux density at \swift\ UVW2 band\\
57   &F7.3  &\auvph&UV spectral index by fitting photometric data\\
58   &F7.3  &\eauvph&\onesig\ error bar of \auvph\\
59   &F7.3  &\logluvph&\logluv\ by fitting UV photoemtric data in \luvunit\\
60   &F7.3  &\elogluvph&\onesig\ error bar of \logluvph\\
61   &F7.3  &\aoxph&\aox\ calculated using \logluvph\ and \loglx\\
62   &F7.3  &\eaoxph&\onesig\ error bar of \aoxph\\
63   &F7.3  &$\log$\mbh&Black hole mass\tablenotemark{2}\\
64   &F7.3  &\loglbolexp&Bolometric luminosity by the EXP model in \ergs\\
65   &F7.3  &\loglboltpl&Bolometric luminosity by the TPL model in \ergs
\enddata
\tablecomments{Data entry is usually set to $-99.9$ if unavailable.}
\tablenotetext{1}{Fluxes are in $10^{-17}$~\fwunit, and are corrected for Galactic reddening.}
\tablenotetext{2}{Calculated using emission line FWHM and corresponding continuum flux
                  calculated by power-law fitting of photometric data points.}
\tablenotetext{3}{The hardness ratio here is defined as 
                  $(N_{\rm H}-N_{\rm S})/(N_{\rm H}+N_{\rm S})$, in which $N_{\rm H}$
                  is the X-ray photon count between 1 and 10 keV and $N_{\rm S}$ is 
                  the X-ray photon count between 0.3 and 1 keV.}
\end{deluxetable}
\clearpage

\enlargethispage{2in}
\thispagestyle{empty}
\newgeometry{textwidth=\textwidth,textheight=3\textheight}
\begin{deluxetable}{llcccllrlrll}
\tablecolumns{12}
\tablewidth{0pc}
\tabletypesize{\scriptsize}
\tablecaption{Data of selected catalog columns for object examples.\label{tab-catalogexample}}
\rotate
\tablehead{
\colhead{}    & 
\colhead{}    & 
\colhead{Quality}    & 
\colhead{XRT} & 
\colhead{XRT} & 
\colhead{}   &  
\colhead{$\log{L_{\rm 2~keV}}$}   &  
\colhead{XRT} & 
\colhead{\nhi}  &  
\colhead{} & 
\colhead{\logluv} & 
\colhead{}\\ 
\colhead{SDSSID}&
\colhead{$z$}   &
\colhead{Flag} &
\colhead{Flag}  & 
\colhead{Model} & 
\colhead{$\Gamma$}  & 
\colhead{[\luvunit]} & 
\colhead{Detect} & 
\colhead{[$10^{22}\mbox{~cm}^{-2}$]} & 
\colhead{\auvph} & 
\colhead{[\luvunit]} & 
\colhead{\aox}  
}
\startdata
$000639.20+142156.1$ & $1.3920$ & A & a & d & $ 2.000$ & $ 27.187\pm  0.096$ & $ 1$ & $  2.517_{ -1.054}^{+  1.424}$ & $-0.927\pm 0.013$ & $31.049\pm 0.002$ & $-1.482\pm 0.037$\\
$000654.40+141442.7$ & $1.6359$ & C & w & $\cdots$ & $\cdots$ & $ 28.411$ & $ 0$ & $  0.000$ & $\cdots$ & $\cdots$ & $\cdots$\\
$001141.40-004722.6$ & $1.6483$ & A & w & $\cdots$ & $\cdots$ & $ 27.050$ & $ 0$ & $  0.000$ & $ 0.083\pm 0.142$ & $30.086\pm 0.032$ & $-1.160$\\
$001217.08-005437.6$ & $3.6030$ & C & w & $\cdots$ & $\cdots$ & $ 28.490$ & $ 0$ & $  0.000$ & $\cdots$ & $\cdots$ & $\cdots$\\
$001746.50-093546.1$ & $0.5790$ & A & a & b & $ 2.240_{-0.220}^{+ 0.238}$ & $ 25.964\pm  0.062$ & $ 1$ & $  0.000$ & $-0.291\pm 0.060$ & $29.535\pm 0.013$ & $-1.370\pm 0.024$\\
$001904.83+003436.5$ & $2.1178$ & A & w & $\cdots$ & $\cdots$ & $ 30.320$ & $ 0$ & $  0.000$ & $-0.290\pm 0.073$ & $30.976\pm 0.011$ & $-0.250$\\
$001917.31+002735.4$ & $2.4183$ & C & a & a & $ 2.000$ & $ 27.202\pm  0.152$ & $ 1$ & $  0.000$ & $\cdots$ & $\cdots$ & $\cdots$\\
$001927.17+003539.0$ & $1.2889$ & A & a & a & $ 2.000$ & $ 26.588\pm  0.114$ & $ 1$ & $  0.000$ & $-0.673\pm 0.089$ & $30.472\pm 0.013$ & $-1.491\pm 0.044$\\
$001927.87+003359.9$ & $1.6234$ & A & w & $\cdots$ & $\cdots$ & $ 26.936$ & $ 0$ & $  0.000$ & $\cdots$ & $\cdots$ & $\cdots$\\
$001954.60+004114.1$ & $1.9081$ & A & w & $\cdots$ & $\cdots$ & $ 26.988$ & $ 0$ & $  0.000$ & $-0.959\pm 0.297$ & $30.652\pm 0.058$ & $-1.406$\\
$001957.60+003936.2$ & $0.9945$ & A & a & a & $ 2.000$ & $ 26.472\pm  0.092$ & $ 1$ & $  0.000$ & $-0.513\pm 0.074$ & $29.959\pm 0.014$ & $-1.338\pm 0.036$\\
$002303.15+011533.6$ & $0.7285$ & A & g & b & $ 2.083_{-0.154}^{+ 0.162}$ & $ 26.807\pm  0.044$ & $ 1$ & $  0.000$ & $-0.155\pm 0.046$ & $30.430\pm 0.010$ & $-1.390\pm 0.017$\\
$002740.38+010608.6$ & $1.5062$ & A & a & a & $ 2.000$ & $ 26.192\pm  0.181$ & $ 1$ & $  0.000$ & $ 0.471\pm 0.157$ & $30.021\pm 0.031$ & $-1.469\pm 0.070$\\
$002828.34-011014.1$ & $1.1930$ & A & w & $\cdots$ & $\cdots$ & $ 26.595$ & $ 0$ & $  0.000$ & $-0.370\pm 0.106$ & $30.378\pm 0.017$ & $-1.452$\\
$003359.38+000230.0$ & $1.6367$ & C & w & $\cdots$ & $\cdots$ & $ 27.870$ & $ 0$ & $  0.000$ & $\cdots$ & $\cdots$ & $\cdots$\\
$003409.08+000318.4$ & $1.2015$ & A & w & $\cdots$ & $\cdots$ & $ 26.621\pm  0.288$ & $ 1$ & $  0.000$ & $-0.039\pm 0.051$ & $30.208\pm 0.008$ & $-1.377\pm 0.111$\\
$003415.77-000030.8$ & $1.9451$ & A & w & $\cdots$ & $\cdots$ & $ 26.940$ & $ 0$ & $  0.000$ & $\cdots$ & $\cdots$ & $\cdots$\\
$003431.78-000957.4$ & $1.5043$ & A & a & a & $ 2.000$ & $ 26.823\pm  0.138$ & $ 1$ & $  0.000$ & $-0.566\pm 0.156$ & $30.444\pm 0.028$ & $-1.390\pm 0.054$\\
$003435.13-000947.8$ & $1.6711$ & A & w & $\cdots$ & $\cdots$ & $ 26.995$ & $ 0$ & $  0.000$ & $-0.338\pm 0.155$ & $30.450\pm 0.023$ & $-1.326$\\
$003922.44+005951.7$ & $1.9889$ & A & w & $\cdots$ & $\cdots$ & $ 26.879\pm  0.391$ & $ 1$ & $  0.000$ & $-0.463\pm 0.288$ & $30.723\pm 0.040$ & $-1.475\pm 0.150$\\
$003940.23+004241.5$ & $1.7010$ & B & o & $\cdots$ & $\cdots$ & $\cdots$ & $-1$ & $\cdots$ & $ 0.037\pm 0.040$ & $30.117\pm 0.009$ & $\cdots$\\
$005446.22+140019.0$ & $0.5015$ & A & g & a & $ 2.000$ & $ 25.019\pm  0.089$ & $ 1$ & $  0.000$ & $-1.465\pm 0.220$ & $29.577\pm 0.040$ & $-1.750\pm 0.038$\\
$005503.52+140806.5$ & $1.6679$ & A & g & c & $ 1.897_{-0.151}^{+ 0.143}$ & $ 27.164\pm  0.077$ & $ 1$ & $  0.647_{ -0.337}^{+  0.372}$ & $-0.872\pm 0.086$ & $30.626\pm 0.019$ & $-1.329\pm 0.030$\\
$011056.90+001912.0$ & $0.8056$ & A & w & $\cdots$ & $\cdots$ & $ 26.837$ & $ 0$ & $  0.000$ & $-0.412\pm 0.016$ & $30.149\pm 0.003$ & $-1.271$\\
$011119.81+002652.0$ & $1.7475$ & A & a & a & $ 2.000$ & $ 27.005\pm  0.098$ & $ 1$ & $  0.000$ & $-0.234\pm 0.088$ & $30.724\pm 0.013$ & $-1.427\pm 0.038$\\
$011124.42+002647.0$ & $1.0029$ & A & a & a & $ 2.000$ & $ 26.402\pm  0.105$ & $ 1$ & $  0.000$ & $-0.285\pm 0.061$ & $30.365\pm 0.012$ & $-1.521\pm 0.041$\\
$113749.11+404913.5$ & $1.1886$ & A & a & b & $ 1.795_{-0.226}^{+ 0.233}$ & $ 26.350\pm  0.069$ & $ 1$ & $  0.000$ & $ 0.113\pm 0.110$ & $29.825\pm 0.018$ & $-1.334\pm 0.027$\\
$114502.23+595720.0$ & $1.6385$ & A & g & b & $ 2.268_{-0.063}^{+ 0.065}$ & $ 26.834\pm  0.019$ & $ 1$ & $  0.000$ & $-0.111\pm 0.045$ & $30.705\pm 0.009$ & $-1.486\pm 0.008$\\
$145353.56+032450.8$ & $2.4045$ & A & a & a & $ 2.000$ & $ 26.965\pm  0.128$ & $ 1$ & $  0.000$ & $-0.415\pm 0.001$ & $31.041\pm 0.000$ & $-1.564\pm 0.049$\\
$154929.43+023701.1$ & $0.4144$ & A & g & b & $ 1.719_{-0.024}^{+ 0.024}$ & $ 27.075\pm  0.009$ & $ 1$ & $  0.000$ & $-0.741\pm 0.079$ & $29.948\pm 0.015$ & $-1.103\pm 0.007$\\
$165004.94+313354.6$ & $1.6948$ & A & g & c & $ 1.911_{-0.276}^{+ 0.271}$ & $ 26.734\pm  0.127$ & $ 1$ & $  0.772_{ -0.552}^{+  0.688}$ & $-0.416\pm 0.075$ & $30.268\pm 0.016$ & $-1.356\pm 0.049$
\enddata
\tablecomments{Refer to Table~\ref{tab-catalog} for conventions of notations and meanings of flags. Values without error 
bars are usually because they are fixed (e.g., $\Gamma$ and \nhi) or they represent upper limits (e.g., \aox\ and \loglx).
Entries without values are because they cannot be measured. The complete catalog content is published in 
its entirety in the electronic version of the {\it Astrophysical Journal Supplement}. The portion is shown here for guidance
regarding its form and content.}
\end{deluxetable}
\restoregeometry

\begin{deluxetable}{cr}
\tablecolumns{2} \tabletypesize{\scriptsize}
\tablecaption{Quasar types excluded 
            from clean catalog sample. \label{tab-classification}}
\tablewidth{0pc}
\tablehead{
\colhead{Classification} &
\colhead{Objects}}
\startdata
BAL quasar    & 50 \\ 
BL Lac        & 1  \\
Lensed quasar & 2  \\
Extended      & 1  \\
RL            & 97 \\
Sy~1.8        & 1  \\
Dust reddened & 50 
\enddata
\tablecomments{Classifications may overlap.}
\end{deluxetable}
\clearpage

\begin{deluxetable}{cccc}
\tablecolumns{4}
\tabletypesize{\scriptsize}
\tablewidth{0pc}
\tablecaption{\scriptsize Sample selection and properties.\label{tab-property}}
\tablehead{
\colhead{Sample}     & 
\colhead{Catalog}     & 
\colhead{G10}     &
\colhead{Total} }
\startdata
\smallskip
Parent Sample  & 637 & 88\tablenotemark{1} & 725\\
Small Clean Sample & 214 & 88 & 302\\
\smallskip
X-ray Detection Rate &85\%&100\% & 89\%\\
Large Clean Sample & 426 &88 & 514\\
X-ray Detection Rate & 65\%& 100\% & 71\%
\enddata
\tablenotetext{1}{The sample in \citet{gru10} contains 92 objects, but 4 objects do not 
have UVOT observations.}
\end{deluxetable}
\clearpage

\begin{deluxetable}{ccc|ccc|ccc}
\tablecolumns{9}
\tabletypesize{\scriptsize}
\tablewidth{0pc}
\tablecaption{\scriptsize Correlation and regression analysis.\label{tab-correlationregression}}
\rotate
\tablehead{
\colhead{}     & 
\colhead{}     & 
\colhead{}     &
\multicolumn{3}{c}{EM} & 
\multicolumn{3}{c}{BJ} \\
\cline{4-6}\cline{7-9}\\
\colhead{Sample} & 
\colhead{Size} & 
\colhead{$\rho_{\rm s}(P_0)$\tablenotemark{3}}&
\colhead{Slope}   & 
\colhead{Intercept} &
\colhead{Dispersion\tablenotemark{1}}    & 
\colhead{Slope}   & 
\colhead{Intercept}    & 
\colhead{Dispersion\tablenotemark{2}}}
\startdata
Small clean catalog     & 207 & $-0.55$ & $-0.156\pm0.022$ & $3.244\pm0.676$ & $0.122$ & $-0.157\pm0.022$ &3.276&0.115\\
Small clean catalog+G10 & 295 & $-0.68$ & $-0.140\pm0.014$ & $2.773\pm0.420$ & $0.113$ & $-0.142\pm0.013$ &2.830&0.107\\
Big clean catalog       & 426 & $-0.47$ & $-0.148\pm0.013$ & $3.018\pm0.398$ & $0.148$ & $-0.146\pm0.013$& $2.972$ & $0.140$\\
Big clean catalog+G10   & 514 & $-0.57$ & $-0.125\pm0.008$ & $2.338\pm0.246$ & $0.140$ & $-0.126\pm0.008$& $2.355$ & $0.131$\\
J07                     & 372 & $-0.76$ & $-0.140\pm0.007$ & $2.704\pm0.212$ & $0.150$ & $-0.140\pm0.006$ &2.723&0.132\\
J07T\tablenotemark{4}   & 289 & $-0.66$ & $-0.134\pm0.010$ & $2.541\pm0.310$ & $0.154$ & $-0.137\pm0.009$ &2.612&0.132
\enddata
\tablecomments{EM: Expectation-Maximization algorithm; BJ: Buckley-James algorithm.}
\tablenotetext{1}{Standard normal residual. See \citet{lav92}.}
\tablenotetext{2}{Kaplan-Meier residual. See \citet{lav92}.}
\tablenotetext{3}{Spearman rank correlation coefficient $\rho_{\rm s}$ with confidence level $P_0$.}
\tablenotetext{4}{J07 sample after excluding high luminosity quasars with $\log{l_\nu(2500\mbox{~\AA})}>31.5$.}
\end{deluxetable}
\clearpage